\documentclass[traditabstract]{aa}

\usepackage{amsmath}
\usepackage{amssymb}
\usepackage{amsfonts}

\usepackage{fixltx2e}
\usepackage{multirow}
\usepackage[latin1]{inputenc}
\usepackage{txfonts}
\usepackage{calc}
\usepackage[dvips]{graphicx}
\usepackage[english]{babel}
\usepackage{fixltx2e}
\usepackage{tikz}
\usepackage{bbm}
\usepackage{siunitx}
\usepackage{natbib}

\newcommand{\pb}{\textsc{Polarbear}}
\newcommand{\xpure}{\textsc{X$^2$Pure}}
\newcommand{\bh}[1]{\b{\hat{#1}}}
\renewcommand{\eqref}[1]{Eq.~(\ref{#1})}
\newcommand{\secref}[1]{Sect.~\ref{#1}}
\newcommand{\figref}[1]{Fig.~\ref{#1}}
\let\b\mathbf
\let\c\mathcal
\newcommand{\T}{\top}
\newcommand{\AtFA}{\b{A}^\T \F{T} \b{A}}
\newcommand{\AtNA}{\b{A}^\T \b{M} \b{A}}
\newcommand{\N}[1]{\mathcal{N}_{\rm #1}}
\renewcommand{\O}[1]{\mathcal{O}({#1})}
\newcommand{\C}[1]{\b{C}_\b{#1}}

\newcommand{\F}[1]{\b{F}_\b{#1}}
\newcommand{\spanp}[1]{{\rm span}(\b{#1})}

\def\simlt{\lower.5ex\hbox{$\; \buildrel < \over \sim \;$}}
\def\simgt{\lower.5ex\hbox{$\; \buildrel > \over \sim \;$}}

\begin{document}

\title{Making maps of cosmic microwave background polarization for $B$-mode studies: the \pb{} example}
\titlerunning{Making maps of CMB polarization for $B$-mode studies}
\author{ {Davide Poletti} \inst{\ref{apc}}
\and {Giulio Fabbian} \inst{\ref{sissa}, \ref{infn}}
\and {Maude Le Jeune} \inst{\ref{apc}}
\and {Julien Peloton} \inst{\ref{sussex}}
\and {Kam Arnold} \inst{\ref{wisc}}
\and {Carlo Baccigalupi} \inst{\ref{sissa}, \ref{infn}}
\and {Darcy Barron} \inst{\ref{ucb}}
\and {Shawn Beckman} \inst{\ref{ucb}}
\and {Julian Borrill} \inst{\ref{c3lbl}, \ref{ssl}}
\and {Scott Chapman} \inst{\ref{dal}}
\and {Yuji Chinone} \inst{\ref{ucb}, \ref{ipmu}}
\and {Ari Cukierman} \inst{\ref{ucb}}
\and {Anne Ducout} \inst{\ref{imperial}}
\and {Tucker Elleflot} \inst{\ref{ucsd}}
\and {Josquin Errard} \inst{\ref{ilp}, \ref{lpnhe}}
\and {Stephen Feeney} \inst{\ref{imperial}}
\and {Neil Goeckner-Wald} \inst{\ref{ucb}}
\and {John Groh} \inst{\ref{ucb}}
\and {Grantland Hall} \inst{\ref{hsca}}
\and {Masaya Hasegawa} \inst{\ref{kek}, \ref{soken}}
\and {Masashi Hazumi} \inst{\ref{kek}, \ref{ipmu}, \ref{soken}}
\and{Charles Hill} \inst{\ref{ucb}}
\and {Logan Howe} \inst{\ref{ucsd}}
\and {Yuki Inoue} \inst{\ref{sinica}, \ref{kek}}
\and {Andrew H. Jaffe} \inst{\ref{imperial}}
\and {Oliver Jeong} \inst{\ref{ucb}}
\and {Nobuhiko Katayama} \inst{\ref{ipmu}}
\and {Brian Keating} \inst{\ref{ucsd}}
\and {Reijo Keskitalo} \inst{\ref{c3lbl}, \ref{ssl}}
\and {Theodore Kisner} \inst{\ref{c3lbl}, \ref{ssl}}
\and {Akito Kusaka} \inst{\ref{physlbl}}
\and {Adrian T. Lee} \inst{\ref{ucb}, \ref{physlbl}}
\and {David Leon} \inst{\ref{ucsd}}
\and {Eric Linder} \inst{\ref{ssl}, \ref{physlbl}}
\and {Lindsay Lowry} \inst{\ref{ucsd}}
\and {Frederick Matsuda} \inst{\ref{ucsd}}
\and {Martin Navaroli} \inst{\ref{ucsd}}
\and {Hans Paar} \inst{\ref{ucsd}}
\and {Giuseppe Puglisi} \inst{\ref{sissa}, \ref{infn}}
\and {Christian L. Reichardt} \inst{\ref{melb}}
\and {Colin Ross} \inst{\ref{dal}}
\and {Praween Siritanasak} \inst{\ref{ucsd}}
\and {Nathan Stebor} \inst{\ref{ucsd}}
\and {Bryan Steinbach} \inst{\ref{caltech}}
\and\\ {Radek Stompor} \inst{\ref{apc}}
\and{Aritoki Suzuki} \inst{\ref{ucb}, \ref{ral}}
\and {Osamu Tajima} \inst{\ref{kek}}
\and {Grant Teply} \inst{\ref{ucsd}}
\and {Nathan Whitehorn} \inst{\ref{ucb}}
}
\institute{ {AstroParticule et Cosmologie, Univ Paris Diderot, CNRS/IN2P3, CEA/Irfu, Obs de Paris, Sorbonne Paris Cit\'e, France}\label{apc}
\and {International School for Advanced Studies (SISSA), Via Bonomea 265, 34136, Trieste, Italy}\label{sissa}
\and {INFN, Sezione di Trieste, Via Valerio 2, 34127 Trieste, Italy}\label{infn}
\and {Department of Physics \& Astronomy, University of Sussex, Brighton BN1 9QH, UK}\label{sussex}
\and {Department of Physics, University of Wisconsin, Madison WI 53706, USA}\label{wisc}
\and {Department of Physics, University of California, Berkeley, CA 94720, USA} \label{ucb}
\and {Computational Cosmology Center, Lawrence Berkeley National Laboratory, Berkeley, CA 94720, USA}\label{c3lbl}
\and {Space Sciences Laboratory, University of California, Berkeley, CA 94720, USA}\label{ssl}
\and {Department of Physics and Atmospheric Science, Dalhousie University, Halifax, NS, B3H 4R2, Canada}\label{dal}
\and {Kavli IPMU (WPI), UTIAS, The University of Tokyo, Kashiwa, Chiba 277-8583, Japan}\label{ipmu}
\and {Department of Physics, Blackett Laboratory, Imperial College London, London SW7 2AZ, United Kingdom} \label{imperial}
\and {Department of Physics, University of California, San Diego, CA 92093-0424, USA} \label{ucsd}
\and {Sorbonne Universit\'es, Institut Lagrange de Paris (ILP), 98 bis Boulevard Arago, 75014 Paris, France} \label{ilp}
\and {LPNHE, CNRS-IN2P3 and Universit\'es Paris 6 \& 7, 4 place Jussieu, F-75252 Paris Cedex 05, France} \label{lpnhe}
\and {Harvard-Smithsonian Center for Astrophysics, 60 Garden Street, Cambridge, MA 02138, USA} \label{hsca}
\and {High Energy Accelerator Research Organization (KEK), Tsukuba, Ibaraki 305-0801, Japan}\label{kek}
\and {SOKENDAI (The Graduate University for Advanced Studies), Hayama, Miura District, Kanagawa 240-0115, Japan}\label{soken}
\and {Institute of Physics, Academia Sinica, 128, Sec.2, Academia Road, Nankang, Taiwan.}\label{sinica}
\and {Physics Division, Lawrence Berkeley National Laboratory, Berkeley, CA 94720, USA}\label{physlbl}
\and {School of Physics, University of Melbourne, Parkville, VIC 3010, Australia}\label{melb}
\and {Department of Physics, California Institute of Technology, Pasadena, CA 91125, USA}\label{caltech}
\and {Radio Astronomy Laboratory, University of California, Berkeley, CA 94720, USA} \label{ral}
}

\authorrunning{ D. Poletti et al.} 

\date{\today}

\abstract{
Analysis of cosmic microwave background (CMB) datasets typically requires some
filtering of the raw time-ordered data. For instance, in the context of
ground-based observations, filtering is frequently used to minimize the impact
of low frequency noise, atmospheric contributions and/or scan synchronous
signals on the resulting maps. In this work we have explicitly constructed a
general filtering operator, which can unambiguously remove any set of unwanted
modes in the data, and then amend the map-making procedure in order to
incorporate and correct for it. We show that such an approach is mathematically
equivalent to the solution of a problem in which the sky signal and unwanted
modes are estimated simultaneously and the latter are marginalized over. We
investigated the conditions under which this amended map-making procedure can
render an unbiased estimate of the sky signal in realistic circumstances. We
then discuss the potential implications of these observations on the choice of
map-making and power spectrum estimation approaches in the context of $B$-mode
polarization studies. Specifically, we have studied the effects of time-domain
filtering on the noise correlation structure in the map domain, as well as
impact it may have on the performance of the popular pseudo-spectrum estimators.
We conclude that although maps produced by the proposed estimators arguably
provide the most faithful representation of the sky possible given the data,
they may not straightforwardly lead to the best constraints on the power spectra
of the underlying sky signal and special care may need to be taken to ensure
this is the case. By contrast, simplified map-makers which do not explicitly
correct for time-domain filtering, but leave it to subsequent steps in the data
analysis, may perform equally well and be easier and faster to implement. We
focused on polarization-sensitive measurements targeting the $B$-mode component of the CMB signal and apply the proposed methods to realistic simulations based on characteristics of an actual CMB polarization experiment, \pb{}. Our analysis and conclusions are however more generally applicable. } 

\maketitle 

\thanks{\footnotesize{Corresponding author: Davide~Poletti.\\{\hspace*{0.2in}\tt dpoletti@apc.univ-paris7.fr}}}

\section{Introduction}
Cosmic microwave background (CMB) experiments strive to obtain accurate
constraints on cosmological parameters from their surveys. Nowadays, in the
quest for $B$-modes (the curl-like pattern in CMB polarization), their
time-ordered data (TOD) consists of tens of trillions of samples. Map-making,
that is reconstructing a map of the observed sky, is one of the major steps in the data analysis pipeline of any CMB experiment. It aims to compress the volume of the data by many orders of magnitudes, while preserving all relevant cosmological information~\citep[e.g.,][]{Janssen1992, Wright1996, Tegmark1997}.

Map-making is usually couched as a linear problem, and the generalized least square (GLS) technique gives a closed form solution which is unbiased and consistent for any positive definite weight matrix. If the weight matrix is the inverse covariance matrix of the time-domain noise, this solution is also optimal as it has the lowest possible uncertainty. In general the principal difficulty in the GLS estimator is in solving the inverse problem, which requires an inversion of a large system matrix. Performing this inversion explicitly~\citep[e.g.,][]{Borrill1999,Stompor2001} has become very difficult as the number of sky pixels in the surveys increases: the iterative solution has therefore become the standard practice~\citep[e.g.,][]{Wright1996,Dore2001,deGasperis2005, Cantalupo2010, Dunner2013, Szydlarski2014}.

Essentially every real CMB dataset requires some kind of time-domain filtering. The purpose is usually to remove systematic effects such as drifts of often-unknown origin, correlated noise components, or, in the case of ground based experiments, contributions from the ground and atmosphere. Some of these effects can readily be incorporated into the GLS solution~\citep[e.g.,][]{Stompor2001, Cantalupo2010, Dunner2013}, resulting in an unbiased, or nearly unbiased, map estimate. However, it has become common practice to perform the map-making procedure on the filtered data as if the filtering had not been applied~\citep[e.g.,][]{QUAD2010, QUIET2011, SPT2011, BICEP2014, POLARBEAR}. In general this approach is bound to yield an incorrect estimate of the sky signal. However, if the filters are well matched to the data so that the time-domain covariance matrix of the filtered data can be treated as diagonal, this technique usually leads to an enormous simplification of the inverse problem present in the GLS estimator, simplifying the implementation and reducing the computational cost. This approach hinges on the hope that the bias present in the sky signal in the resulting map can be robustly estimated and corrected for at power spectrum level by studying signal-only simulations~\citep{Hivon2002}, and that the loss of statistical precision is minor.

In this work we study the former route and consider map-making procedures that explicitly incorporate and correct for time-domain filtering. We present a general, self-consistent, formalism developed for this purpose and discuss its properties. We point out that even if the time-domain noise is white, such filtering unavoidably leads to correlated noise in the map domain, and in extreme cases to the presence of modes whose amplitude can not be reliably estimated from the data. We then discuss how maps of this kind can be further analyzed using widely popular pseudo power spectrum estimators. Our focus throughout this paper is on the $B$-mode polarization and our principal figure of merit employed here is the uncertainty on the $B$-mode polarization power spectrum obtained by the different approaches.

To demonstrate the formalism in a realistic setting, we use simulations based on the scan strategy and noise typical of the first observational campaign of the \pb{} instrument~\citep{POLARBEAR}. 

This paper is organized as follows. In \secref{framework} we review the map-making formalism and present an extension to the standard procedure that accounts for time-domain filtering. In \secref{degeneracies}, we discuss the effect the filtering may have on the maps reconstructed from the filtered data. In~\secref{templates}, we introduce specific filters typical of ground-based experiments, while
in~\secref{pipeline} we give a worked example demonstrating the effects of such filters on the analysis of mock \pb{}-like data sets. We present the main results of this work in \secref{results} including a performance comparison of the different map estimators,~\secref{comparison}.

\section{Map-making in CMB experiments}
\label{framework}
This section describes the map-making formalism, emphasizing new features related to the presence of time-domain filtering. For the time being, we assume that the problem is algebraically solvable and leave discussion of potential degeneracies to the next section,
\secref{degeneracies}.
\subsection{The standard problem}
The starting point of map-making is the calibrated time-ordered data recorded by the detectors. We collect all these time samples in a vector, $\b{d}$, which contains thus $\N{t}$ elements. The scanning strategy of the telescope and the polarization modulation define how the sky signal contributes to each measured sample, ${d}_t$, which can be then represented as,
\begin{eqnarray} d_t = I_{p_t} + \cos (2\varphi_t) Q_{p_t} + \sin (2\varphi_t) U_{p_t} + n_t. \label{eq:data3stokes}
\end{eqnarray}
Here $n_t$ is the noise; $I$, $Q$ and $U$ are the Stokes parameters of the incoming light for sky pixel $p$ being observed at time $t$; and $\varphi$ is
the orientation of the linear polarization sensitive detector projected on the sky. We assume throughout that the instrumental beams are axially symmetric
and are therefore convolved with the sky signals, which we aim at estimating. There are two important, specific cases of~\eqref{eq:data3stokes} that we will find useful in this paper. One is that of a total intensity measurement,
\begin{eqnarray} d_t = I_{p_t} + n_t, \label{eq:dataIonly}
\end{eqnarray}
and the other of polarization-only measurement,
\begin{eqnarray} d_t = \cos (2\varphi_t) Q_{p_t} + \sin (2\varphi_t) U_{p_t} + n_t. \label{eq:data2stokes}
\end{eqnarray}

All the Stokes parameters characterizing the signal in $\N{p}$ observed sky pixels can be arranged in a single signal $\N{s}$-vector, $\b{s}$, in such a way that the Stokes parameters for one pixel are followed by the Stokes parameter for a subsequent one. The entire data vector, $\b{d}$, can then be represented in a compact way as,
\begin{eqnarray} 
\b{d} = \b{A}\b{s} + \b{n}. \label{data_simple}
\end{eqnarray}
Here, $\b{n}$ is the noise vector with covariance matrix $\C{n}$. The \emph{pointing matrix} $\b{A}$ is a $\N{t}$ by $\N{s}$ known matrix.
Each row of $\b{A}$ defines a linear combination of the signal, which contributes to the measurement at the time corresponding to that row.
A column of $\b{A}$ is a ``time domain signature'' of the corresponding entry of $\b{s}$, telling us when a given sky pixel was observed and with what weight it contributed to the measured signal.

In this form, map-making is a linear statistical problem whose solution is given by the well known GLS estimator,
\begin{eqnarray} 
\bh{s} = (\b{A}^\T \b{M} \b{A})^{-1} \b{A}^\T \b{M} \b{d}, \label{gls_simple}
\end{eqnarray}
which renders an unbiased estimate of the map for any choice of positive
definite weight matrix, $\b{M}$. If $\b{M} = \C{n}^{-1}$ and the noise is
Gaussian, then $\bh{s}$ defines the minimum variance and maximum likelihood solution. The linear independence of the columns of $\b{A}$ is for the time being taken as given. This is a necessary and sufficient requirement to ensure that $\b{A}^\T \b{M} \b{A}$ is invertible and thus \eqref{gls_simple} is uniquely solvable. We discuss this issue in more detail later.

\subsection{Mapmaking with time-domain filtering}
\label{mapmaking}
\subsubsection{The filtering operator}
\label{sec:filteringOp}
Let us collect in a single template matrix, $\b{T}$, all the temporal templates we want to filter from the data, with each template corresponding to one column. For example, a template can be a $\N{t}$-vector equal to one in some time interval and zero everywhere else. Such a template would stand for the removal of a temporal offset in this interval. The size of $\b{T}$ is considerable: in typical ground-based CMB experiments there are hundreds of templates per detector for every scan period (15-90 minutes of observation), easily reaching hundreds of millions of templates. The unwanted contributions in the temporal data can be represented as some linear combinations of the columns of $\b{T}$, 
\begin{eqnarray}
\b{T}\b{x} \equiv \b{\hat T}\b{\hat x},
\end{eqnarray}
where $\b{\hat T}$ denotes a column-orthonormal matrix that spans the same subspace as matrix $\b{T}$ and $\b{x}$ and $\b{\hat x}$ are the corresponding sets of coefficients. In general, we have,
\begin{eqnarray}
\b{\hat T} = \b{T}\,(\b{T}^\T\b{T})^{-1/2},
\end{eqnarray}
so, 
\begin{eqnarray}
\b{\hat T}^\T\,\b{\hat T} = {(\b{T}^\T\b{T})^{-1/2}}^\T \, \b{T}^\T\,\b{T}\,(\b{T}^\T\b{T})^{-1/2} \, = \, \b{1}.
\end{eqnarray}
as required.

We note that the number of columns of $\b{\hat T}$ may be smaller than that of $\b{T}$ if some of the original templates are not linearly independent. In such cases, inverting matrix $(\b{T}^\T\,\b{T})^{1/2}$ would require some pseudo-inverse and the result, ${(\b{T}^\T\b{T})^{-1/2}}$ will be effectively a rectangular matrix.

We can now define the filtering as an operator that projects out from the data all temporal modes defined by the columns of $\b{\hat T}$, 
\begin{eqnarray}
\b{d'} \equiv \b{d} - \b{\hat T}\b{\hat T}^\T\b{d} = ( \b{1} - \b{T}\,(\b{T}^\T\b{T})^{-1}\b{T}^\T)\,\b{d}.
\label{eq:simpleFiltering}
\end{eqnarray}
We note that if our goal is to filter all modes that belong to the subspace spanned by the columns of $\b{T}$, and at the same time to weight all the modes that are orthogonal to this subspace by some symmetric weight matrix, $\b{M}$, then we can generalize the filtering operator in the right hand side of~\eqref{eq:simpleFiltering} so it performs both these functions simultaneously. If we further require that the combined operator, $\b{F}_{\b{T}}$, is also symmetric, such a generalization is unique and the operator reads,
\begin{eqnarray}
\b{F}_{\b{T}} \equiv \b{M} - \b{M}\b{T}\,(\b{T}^\T\b{M}\b{T})^{-1}\b{T}^\T\b{M}.
\label{filter}
\end{eqnarray}
As desired, this operator obviously filters all the modes from the space spanned by $\b{T}$ as,
\begin{eqnarray}
\b{F}_{\b{T}}\b{T} = 0,
\label{eq:filterFilters}
\end{eqnarray}
and weights by $\b{M}$ any mode, $\b{t}_\perp$, that is orthogonal to $\b{T}$,
in the sense of a scalar product weighted by $\b{M}$ (i.e.,
$\b{T}^\T\b{M}\b{t}_\perp = 0$), as
\begin{eqnarray}
\b{F}_{\b{T}}\b{t}_{\perp} = \b{M}\,\b{t}_\perp.
\end{eqnarray}

We emphasize that in general the filtering operator, \eqref{filter} is not
equivalent to filtering the templates one after another, as is often implemented
in practice. Indeed, this would be the case only if the templates happen to be
orthogonal from the outset. In such a case, $\b{T}^\T\b{M}\b{T}$ is diagonal
(i.e., $\b{t}_i^\T\b{M}\b{t}_j \propto \delta_{ij}$, where $\b{t}_i$ denotes the $i$th column of $\b{T}$) and therefore
\begin{eqnarray} 
\F{T} = \b{M} \, \prod_i \b{M}^{-1}\,\b{F}_{\b{t}_i} = \b{M} \, \prod_i \, \left(\b{1} - \b{t}_i\,(\b{t}_i^\T\b{M}\b{t}_i)^{-1}\b{t}_i^\T\b{M}\right).
\end{eqnarray}
If the filters are not orthogonal, and the filtering is performed sequentially, then removing any given template will typically reintroduce some contribution to templates previously filtered, and the final result may depend on the order of the filters. These effects are often small but can sometimes be relevant.

By contrast, the filter proposed in~\eqref{filter} resolves all ambiguities of this kind. With help of this operator we can now generalize the map-making equation, \eqref{gls_simple}, as 
\begin{eqnarray}
\bh{s} &= (\b{A}^\T \F{T} \b{A})^{-1} \b{A}^\T \F{T} \b{d},
\label{eq:gls_filtered}
\end{eqnarray}
where $\F{T}$ acting upon $\b{d}$ removes all the unwanted modes and weights the others as required, while the matrix operator, $(\b{A}^\T \F{T} \b{A})^{-1}$, ensures that the estimator is unbiased. Indeed, inserting~\eqref{data_simple} for $\b{d}$ we get,
\begin{eqnarray}
\bh{s} &= \b{s} + (\b{A}^\T \F{T} \b{A})^{-1} \b{A}^\T \F{T} \b{n},
\label{eqn:estMapSplit}
\end{eqnarray}
and thus,
\begin{eqnarray}
\langle \bh{s} - \b{s}\rangle &= \langle(\b{A}^\T \F{T} \b{A})^{-1} \b{A}^\T \F{T} \b{n}\rangle = 0,
\end{eqnarray}
where the average is taken over a statistical ensemble of instrumental noise realizations and $\langle \b{n}\rangle = 0$.

The map-domain noise covariance matrix of the unbiased map estimator is (see, e.g., \cite{Tegmark1997} or \cite{Stompor2001})
\begin{eqnarray} 
\b{C}_\bh{s}  &\equiv& \langle (\bh{s}-\b{s})(\bh{s}-\b{s})^\T \rangle \nonumber \\  &=& (\b{A}^\T \F{T} \b{A})^{-1} \b{A}^\T \F{T} \, \langle \b{n} \b{n}^\T\rangle \, \F{T} \b{A} (\b{A}^\T \F{T} \b{A})^{-1} \nonumber \\ &=& (\b{A}^\T \F{T} \b{A})^{-1} \b{A}^\T \F{T}\, \b{C}_\b{n}\, \F{T} \b{A} \, (\b{A}^\T \F{T} \b{A})^{-1}.
\label{eqn:noiseCovGen}
\end{eqnarray}
If $\b{M} = \b{C}_\b{n}^{-1}$ then this can be rewritten in a compact way as,
\begin{eqnarray} 
\b{C}_\bh{s} &=& (\b{A}^\T \F{T} \b{A})^{-1},  \label{eqn:noiseCovML}
\end{eqnarray}
owing to the fact that, $ \F{T}\, \b{M}^{-1}\, \F{T} = \F{T}$. This simply generalizes the standard expression for the covariance derived in the case of the maximum likelihood map-making with no filtering, which reads,
\begin{eqnarray} 
\b{C}_\bh{s} &=& (\b{A}^\T \b{C}_\b{n}^{-1} \b{A})^{-1}. 
\end{eqnarray}

\subsubsection{The meta-pixel approach}
We note that an equation analogous to~\eqref{eq:gls_filtered} can be derived from the meta-pixel approach~\citep[e.g.,][]{Stompor2001, Cantalupo2010}. 
In order to do so, let us generalize the data model in~\eqref{data_simple} to incorporate the presence of contributions other than the CMB and noise. This can be done as follows,
\begin{eqnarray} 
\b{d} = \b{A}\b{s} + \b{T}\b{y} + \b{w}. \label{data}
\end{eqnarray}
where $\b{T}$ is, in general, the set of time-domain templates that we want to filter out, $\b{y}$ is their unknown amplitude, and $\b{w}$ is the noise.
We can rewrite Eq. (\ref{data}) as,
\begin{eqnarray} 
\b{d} = [\b{A}, \b{T}] \left[ \begin{smallmatrix} \b{s}\\\b{y} \end{smallmatrix} \right] + \b{w}. \label{eq:dataConcise}
\end{eqnarray}
Estimating the unknown parameters for this data model and for the one in~\eqref{data_simple} are formally equivalent. We will assume that $[\b{A}, \b{T}]$ is full rank: the time-domain signature of the unknown parameters are all linearly independent, postponing discussion of degeneracies to \secref{degeneracies}.
The GLS estimator for Eq.~(\ref{data}), written in a block fashion is
\begin{eqnarray} \label{s_estimator} \bh{s} &= (\b{A}^\T \F{T} \b{A})^{-1} \b{A}^\T \F{T} \b{d}\\ \label{y_estimator} \bh{y} &= (\b{T}^\T \F{A} \b{T})^{-1} \b{T}^\T \F{A} \b{d}
\end{eqnarray}
where $\F{T}$ and $\F{A}$ are filtering operators as in~\eqref{filter}.

As \eqref{s_estimator} shows, $\b{\hat{s}}$ can be computed without explicitly solving for the amplitude of the templates $\b{\hat{y}}$. This is the
\emph{direct} approach, adopted in this paper. Moreover, this is exactly the same expression as was derived in the previous section, \eqref{eq:gls_filtered}.
There are however several advantages to this derivation. For instance, it shows
that if the weights are chosen appropriately (i.e., $\b{M} = \b{C_n}^{-1}$),
then the map estimated via~\eqref{eq:gls_filtered} is both maximum likelihood
and minimum variance. Moreover if the Bayesian perspective is adopted, then the
posterior probability distribution for the combined vector of unknowns
in~\eqref{eq:dataConcise} (i.e., $[\begin{smallmatrix}\b{s}\\
  \b{y}\end{smallmatrix}]$) is Gaussian and the second term in the expression
for the filtering operator, $\F{T}$, \eqref{filter}, arises simply as a result
of a marginalization of this posterior over the unknowns contained in $\b{y}$, assuming flat priors \citep{Stompor2001}. 

This derivation of the estimator also suggests that the map can be solved for in two steps rather than one. Indeed, instead of using~\eqref{s_estimator}, if one first gets the estimate $\bh{y}$, then $\bh{s}$ can be estimated as,
\begin{eqnarray} 
\b{\hat{s}} &= (\b{A}^\T \b{M} \b{A})^{-1} \b{A}^\T \b{M}(\b{d} - \b{T}\bh{y}), \label{s_destriper}
\end{eqnarray}
which, for simple $\b{M}$ may be much easier to solve than~\eqref{s_estimator}.

This latter approach provides a basis for the destriping
technique~\citep[e.g.,][]{Poutanen2004, Keihanen2004, Keihanen2010,
Tristram2011}. We emphasize that the one- and two-step methods are formally
equivalent: they lead to the same estimate of the sky signal. Eq.
(\ref{s_destriper}) is usually easy and cheap to implement as the weight matrix,
$\b{M}$, is typically taken to be diagonal in this context. Choosing between the
two approaches is therefore driven by the cost of~\eqref{s_estimator} compared
to~\eqref{y_estimator}. These two equations require handling dense algebraic
objects of a dimension equal to the number of columns of $\b{A}$ and $\b{T}$
respectively. The former is proportional to the number of observed pixels and
the latter to the number of temporal templates. As this last number typically
grows with the number of detectors and the presence of spurious signals in the
observations, the direct method provides a potentially attractive approach for
modern experiments, which employ arrays of thousands to tens of thousands of
detectors. This includes both ground-based observatories, such as
\pb{}~\citep{POLARBEAR}, and planned CMB satellite missions, like
Lite\textsc{bird}~\citep{Matsumura2016} or COrE~\citep{CORE2011}. By contrast, the two-step approach is very well suited to experiments with a limited number of detectors observing large sky areas, which produce maps with large numbers of pixels but have relatively few templates to be removed. For these reasons it has played particularly prominent role in the analysis of the Planck data~\citep{HFImaps2015,LFImaps2015}.

If $\b{M}$ is diagonal then $\b{A}^\T \b{M} \b{A}$ and $\b{T}^\T \b{M} \b{T}$ are usually sparse and structured and therefore easy to compute and invert. Conversely $\b{A}^\T \F{T} \b{A}$ and $\b{T}^\T \F{A} \b{T}$ are dense and potentially large. The size of the former matrix is proportional to the number of pixels, ranging from $10^5$ for ground-based experiments to $10^9$ for satellites covering the whole sky. The size of the latter matrix also typically exceeds $10^8$ for kilo-pixel arrays. These matrices are not only computationally expensive to manipulate but also often too large to explicitly computate and/or store. Consequently, the solution of the inverse problem either in \eqref{s_estimator} or in \eqref{y_estimator} would ideally be found using some iterative linear equations solvers such the preconditioned conjugate gradient method (PCG)~\citep[e.g.,][]{deGasperis2005, Cantalupo2010,Szydlarski2014}. However, convergence of these iterative solvers may be hard to attain if the matrices are not well conditioned. We discuss
the PCG approach in the context of the extended map-making equation, \eqref{eq:gls_filtered}, in \secref{degeneracyAT}, from the formal point of view,
and in \secref{unbiasedImplementation} and \ref{reconstructedMaps}, in the specific context of \pb.

\subsubsection{The biased map estimator}
Performed directly or iteratively, the inversion of $\AtFA$ (or  $\b{T}^\T \F{A} \b{T}$) is the bottleneck in both implementation and execution time. This is why many experiments prefer to use instead the biased map estimator
\begin{eqnarray} 
\label{s_biased} \b{\hat{s}} &= (\b{A}^\T \b{M} \b{A})^{-1} \b{A}^\T \F{T} \b{d}.
\end{eqnarray}
As $\F{T}$ still acts upon the data vector, $\b{d}$, the templates are still explicitly filtered out of the data. However, this is not accounted for in the system matrix, $(\b{A}^\T \b{M} \b{A})^{-1}$, which therefore does not correct for the filtering but only for the weighting. If $\b{M}$ is diagonal, this choice enormously simplifies the implementation and drastically reduces the computational cost. The price to pay is a bias in the estimator. This bias is usually evaluated and removed at the power spectrum level using Monte Carlo simulations and typically requires some additional assumptions~\citep[e.g., ][]{Hivon2002}. This approach is thus most frequently considered to be a part of the power spectrum estimation pipeline. 

The presence of bias in the map is apparent as we have,
\begin{eqnarray} 
\text{Bias} = \langle \bh{s}-\b{s} \rangle = \left[ (\AtNA)^{-1} \b{A}^\T \F{T} \b{A} - \b{1} \right] \b{s}, \label{bias}
\end{eqnarray}
which does not vanish in general. The covariance of the estimated map then reads,
\begin{eqnarray} 
\b{C}_\bh{s}  &\equiv& \langle (\bh{s}- \langle\bh{s}\rangle)(\bh{s} - \langle\bh{s}\rangle)^\T \rangle \nonumber \\  
&=& (\AtNA)^{-1} \b{A}^\T \F{T} \, \langle \b{n} \b{n}^\T\rangle \, \F{T} \b{A} (\AtNA)^{-1} \nonumber \\ 
&= &(\AtNA)^{-1} \b{A}^\T \F{T}\, \b{C}_\b{n}\, \F{T} \b{A} \, (\AtNA)^{-1}. 
\end{eqnarray}
Therefore, if the filters are chosen such that the matrix $ \F{T}\, \b{C}_\b{n}\, \F{T}$ is nearly diagonal for any diagonal weights, $\b{M}$, then we can 
take them to be, $\b{M} = {\rm diag} \, (\F{T}\, \b{C}_\b{n}\, \F{T})$, yielding,
\begin{eqnarray} 
\b{C}_\bh{s} \approx (\AtNA)^{-1},
\end{eqnarray}
and therefore the covariance can have a particularly simple structure. However this is rarely the case, and instead even if the true noise is uncorrelated and $\b{M} = \b{C}_\b{n}^{-1}$, the covariance reads
\begin{eqnarray} 
\b{C}_\bh{s} = (\AtNA)^{-1} \b{A}^\T \F{T} \b{A} \, (\AtNA)^{-1},
\end{eqnarray}
and thus is a dense matrix with potentially non-negligible, off-diagonal correlations.

We note that the information content of both the unbiased and biased maps is the same, since one can be derived from the other via an invertible linear operation. Indeed, 
\begin{eqnarray} 
\bh{s}_{\rm unbiased} = \b{R}\, \bh{s}_{\rm biased} \label{s_estim_class}
\end{eqnarray}
where $\b{R} \equiv (\b{A}^\T \F{T} \b{A})^{-1}\, (\AtNA)$ is an invertible matrix. However, the key to the biased approach is that no attempt is ever made to estimate the matrix $\b{R}$. Consequently, although the same information is contained in both maps it is encoded differently in each of them, and whenever equivalent biased and unbiased maps are subsequently processed their information is compressed differently, giving rise to different statistical properties in the resulting estimator.

\section{Map-making in the presence of degeneracies}
\label{degeneracies}
So far we have assumed that the generalized map-making equation can be robustly solved, implicitly assuming that the system matrix, $\b{A}^\T\F{T}\b{A}$, is invertible. However this may not always be the case, and in this section we elaborate on this and discuss what can be done in such circumstances.

We first note that invertibility of the system matrix is ensured if the matrix $[\b{A}, \b{T}]$ is full column rank. This can be seen immediately by noting that $(\b{A}^\T\F{T}\b{A})^{-1}$ stands for an upper left diagonal block of the inverse of the matrix,
\begin{eqnarray}
\left[ \begin{array}{c} 
{\displaystyle \b{A}^\T}\\
{\displaystyle \b{T}^\T} \end{array} \right] \, \b{M}\, [\b{A}, \b{T}] = \left[  \begin{array}{c c}
\displaystyle{\AtNA}, & \displaystyle{\b{A}^\T \b{M} \b{T}}\\
\displaystyle{ \b{T}^\T \b{M} \b{A}}, & \displaystyle{\b{T}^\T \b{M} \b{T}}
\end{array}\right],
\label{eqn:matFull}
\end{eqnarray}
which is invertible only when $[\b{A}, \b{T}]$ is full column rank. In practice, since all the operations have to be performed numerically, what really matters is not strict linear independence in the mathematical sense but rather linear independence sufficient to ensure stable and robust, finite-precision numerical calculations, as exemplified by Eqs.~(\ref{s_estimator}), (\ref{y_estimator}), (\ref{filter}), (\ref{s_destriper}) and (\ref{s_biased}).

In general, given a matrix $\b{B}$ and vector $\b{z}$ laying in its
range, if $\b{B}$ is singular we can solve the equation $\b{B}\b{x}=\b{z}$ for $\b{x}$ only down to an unknown contribution from the nullspace of $\b{B}$. Typically, the component of the solution parallel to the nullspace will be arbitrarily set to zero and its true value unavoidably lost.
In practice it could be obtained by regularizating the matrix, which involves first calculating its inverse via computing and inverting its eigenvalues. The regularization is then applied to all eigenvalues that are smaller then some predefined threshold by setting to zero the corresponding eigenvalue of the inverse.

$[\b{A}, \b{T}]$ may not be full column rank for three different reasons, leading to three possible types of degeneracies:

\subsection{The columns of $\b{A}$ are not independent.} 
\label{degeneracyA}
This degeneracy would affect standard map-making as much as our extension, but
we include it for completeness. In this case, the scanning strategy does not
allow the reconstruction of some sky mode. A typical example is 
a polarization
pixel that was not observed with sufficient redundancy in the polarization
angle.

This case is easy to avoid because $\b{A}^\T \b{M} \b{A}$ is easy to build. If $\b{M}$ is diagonal the cost is $\c{O}(\N{t})$ operations and $\b{A}^\T \b{M} \b{A}$ is block diagonal (a block for each pixel); its eigendecomposition (costing $\c{O}(\N{p})$ operations) then enables us to evaluate the condition number of each of the blocks. After pixel selection based on their condition numbers (and the removal of the corresponding samples from the TOD) the new $\b{A}^\T \b{M} \b{A}$ can be safely inverted.

\subsection{The columns of $\b{T}$ are not independent.} 
\label{degeneracyT}
This reflects the fact that there is redundancy in the templates, and basically
corresponds to an attempt to filter the same template twice. For example, this
happens in practice when two different sets of templates (e.g., the polynomial and the ground template filters) both remove the global offset of the TOD.

Since the final goal is to estimate $\bh{s}$ and not $\bh{y}$, Eqs.~(\ref{s_estimator}) and~(\ref{y_estimator}), this degeneracy does not pose any fundamental problem. We merely need to construct a basis of $\spanp{T}$ and use it to define a new (smaller) set of independent templates $\bh{T}$ as described in~\secref{sec:filteringOp}. 

In practice, the situation is also quite straightforward. By construction, there are usually many known orthogonality relations between the templates. As a consequence, $\b{T}^\T \b{M} \b{T}$ is typically reasonably easy to compute and is sparse and structured. Its inverse can then be computed explicitly using standard matrix inversion techniques. The condition number of this matrix provides an easy test of the linear independence of the templates, and if it is too high the matrix has to be regularized while being inverted. Once such a regularized inverse $(\b{T}^\T \b{M} \b{T})^{-1}$ is computed, it should be used 
in the projector of Eqs.~(\ref{s_estimator}) and (\ref{s_biased}), to take care of the redundancies and therefore the degeneracies.

\subsection{Some columns of $\b{A}$ are not independent from the columns of $\b{T}$.}
\label{degeneracyAT}
This is the most insidious type of degeneracy in the map-making problem, and its presence reflects a degeneracy between the sky signal and the
templates. In this case the reconstruction of some sky component is not possible if the templates have been filtered. As a trivial example, by systematically filtering the mean of the total intensity TOD we create a degeneracy with the global offset of the temperature map.

This type of the degeneracy manifests itself as a singularity of both $\b{A}^\T \F{T} \b{A}$ and $\b{T}^\T \F{A} \b{T}$. This can be seen immediately by noting that, if the columns of $\b{A}$ and $\b{T}$ are not independent, then there exits at least two modes, one in the map domain, $\b{\tilde s}$, and one in the template domain, $\b{\tilde y}$, such as,
\begin{eqnarray}
\b{A}\b{\tilde s} = \b{T}\b{\tilde y},
\label{eq:degModes}
\end{eqnarray}
and therefore
\begin{eqnarray}
\b{A}^\T \F{T} \b{A}\b{\tilde s} = \b{A}^\T \F{T} \b{T}\b{\tilde y} = \b{0},\\
\b{T}^\T \F{A} \b{T}\b{\tilde y} = \b{T}^\T \F{A} \b{A}\b{ \tilde s} = \b{0},
\end{eqnarray}
given that $ \F{T} \b{T} = \F{A} \b{A} \equiv 0$, \eqref{eq:filterFilters}. The two modes, $\b{\tilde s}$ and $\b{\tilde y}$, constitute a pair of degenerate modes, which, while residing in different domains, lead to the same effects in the time-domain data and therefore can not be distinguished from each other.

In this case the best one can do to solve the map-making problem is to regularize the inversion of the singular matrix and to compute all the modes of
the map for which the solution can be obtained and determine the modes for which it cannot. These latter modes will be missing from the reconstructed map. We note that this is not due to the regularization procedure but because these modes are removed from the data by the filters. Indeed, 
\begin{eqnarray}
\b{A}\F{T}\b{d} = \b{A}\F{T}\b{A}\b{s} + \b{A}\F{T}\b{n} \, = \, \b{A}\F{T}\b{A}\b{s}_\perp + \b{A}\F{T}\b{n},
\label{eq:sFiltering}
\end{eqnarray}
where the subscript $\perp$ denotes the part of the sky signal orthogonal to the nullspace of $ \b{A}\F{T}\b{A}$, as the part parallel to it is unrecoverably lost. The information about these removed modes then needs to be propagated to next steps of the analysis and properly taken into account to ensure that the final results are statistically meaningful.

This route is only straightforward in practice if all the matrices appearing in Eq.~(\ref{eqn:matFull}), can be constructed and decomposed explicitly.  However, because of their sizes this may be a formidable and often unfeasible task, even with help of the largest massively parallel platforms and parallel software packages. If this is indeed the case, and the solution can be only derived using some iterative technique, then the singular modes may not only be impossible to compute and correct for, but indeed it may not be clear from the outset whether the matrices are singular or not. In such cases this may need to be inferred {\em post hoc} from the behavior of the solver. We discuss these issues in more depth in~\secref{reconstructedMaps}.

We also emphasize that, in the presence of this kind of degeneracy, the maps computed by the direct and two-step approaches may not be identical. Indeed, in the direct case the unconstrained sky modes will be missing from the estimated map, while in the two-step case the situation is different as the regularized inversion has to happen when the estimate of the template amplitudes is performed. Consequently, these will be degenerate ``modes'' of the templates, which will be missing in $\bh{y}$, while the template-corrected data vector, $\b{d} - \b{T}\bh{y}$, will retain a time-domain component that
should have been filtered out. Hence, the map estimated in the second step,
\eqref{s_destriper}, will contain some of the degenerate sky modes, which were
set to zero in the direct approach. Obviously, if the fact that these modes can
not be estimated is properly taken into account in the covariance of the maps,
both maps will be statistically equivalent. We also note that to compute which
map modes are non-constrainable one may use the singular modes of $\b{T}^\T
\F{A} \b{T}$, found while performing the first step of the two-step method, and
use them in the second step by replacing the template-corrected data vector,
$\b{d} - \b{T}\bh{y}$ by $\b{T}\b{z}$, where $\b{z}$ stands for one of the
singular eigenvectors. Map-domain templates resulting from this calculation will
have to be then orthogonalized using, for example, the Gram-Schmidt procedure. 

In the case of the biased map estimator, \eqref{s_biased}, this degeneracy does not pose any numerical issue. As in the unbiased cases the map estimate will have no contribution from the sky signal modes residing in the nullspace of $\b{A}^\T\F{T}\b{A}$ due to the filtering applied in the time domain, \eqref{eq:sFiltering}. Nevertheless, the amplitudes of the filtered modes in the biased map will usually be non-zero as a result of the power leaking to them from the other pixel-domain modes.

\section{The case of ground-based experiments}
\label{templates}
As a specific application of the above formalism let us consider the map-making problem for a modern, ground-based, CMB experiment, which scans the sky with a kilo-pixel array of polarization sensitive detectors in the presence of both atmospheric and ground emission. Commonly, in order to minimize the impact of the atmospheric contribution, the scans are performed at constant elevation for relatively short periods of time. The elevation is then changed to track the sky patch, which moves due to the Earth's rotation. The scan amplitude, the choice of the scan elevations, and the elevation dwelling time are specific to each observation. In the following, we consider only cases that conform to this general description.

The data of such an experiment are typically more complex than the simple model exemplified by \eqref{data_simple}, whether due to the atmospheric, instrumental, or ground contamination. Below we describe common contributions of this type and explain how they fit within the extended data model, \eqref{data}, and how they can be treated with the map-making technique introduced earlier.

\subsection{Noise correlations} 
\label{correlatedNoiseTemplates}
The noise is usually correlated between different detectors and typically displays significant low frequency excess, dubbed $1/f$ noise, which arises either as an instrumental effect, or from atmospheric emmission, or from some other effects. As a result the time-domain noise correlation matrix, $\C{n}$, is dense, making its inversion and multiplication computationally demanding. Moreover, the matrix, $\C{n}$, is unknown and has to be estimated from the data themselves~\citep[e.g.,][] {FerreiraJaffe2000}. This procedure is usually difficult, especially as far as the low frequency modes and detector-detector
correlated modes are concerned. In addition, the low frequency contributions may not be even stationary or Gaussian, and thus can not be properly described merely by a covariance. Consequently, using the optimal weight, $\C{n}^{-1}$, in the map-making process may be difficult. Using a simpler weight matrix, $\b{M}$, does not lead to a bias in the standard map estimator, Eq.~(\ref{gls_simple}), however its choice does affect 
the quality of the final map. This is because for different weights, the time-domain data, $\b{d}$, are coadded differently on the first step of the map-making
procedure, when the right hand side of the map-making equation, $\b{A}^\T \b{M} \b{d}$, is calculated. For instance, diagonal weights (in the map domain) can not selectively suppress some temporal frequency bands over others. Consequently, if $1/f$ noise is present, even if it is Gaussian and stationary, the low frequency modes will not be properly down-weighted as compared to the high frequency ones. This may result in stripes appearing in the direction of the scans. The effect is particularly apparent if the scanning strategy does not provide good cross-linking. Non-Gaussian/non-stationary features can be even more difficult to deal with.

Instead of down-weighting such long modes one may prefer to filter them out, as in the filtered map-making, Eq.~(\ref{eq:gls_filtered}). The time-domain data are then explicitly filtered while being compressed to the pixel-domain object, $\b{A}^\T \F{T} \b{d}$. The long term trends present in the time-ordered data can be removed at this stage. Such removal is blind to the origin and nature of the trends, potentially removing the true sky signal together with the noise and some unwanted spurious contributions. However, the signal-to-noise ratio for these modes is usually very low, as the $1/f$ noise quickly dominates, and the information loss due to the potential removal of the sky signal is typically negligible. If this is the case, then filtering can provide a useful alternative to weighting.

The modes to be filtered out are typically assumed to be arbitrary linear combinations of some sufficiently rich family of temporal templates, which has to be defined case-by-case. For long term trends, the templates are often taken to be piece-wise low order polynomials or harmonic functions defined for all samples of the data, and are represented by columns of some template matrix, $\b{B}$. If the templates are well matched to the problem at hand, then the residual noise, $\b{w}$, defined as,
\begin{eqnarray}
\b{w} \equiv \b{n} - \b{B}\b{x},
\end{eqnarray}
is expected to be 'whiter' than the actual time-domain noise, $\b{n}$, and thus approximately characterized by a diagonal noise covariance matrix.
Here, $\b{x}$ stands for a set of {\em a priori} arbitrary parameters to be determined, similar to the sky signal, $\b{s}$, which define the amplitudes of the corresponding templates~\citep{Stompor2001}. We point out that in general one could introduce some additional information about the long term modes of the noise by setting some constraints on $\b{x}$.  As such constraints are typically hard to identify and might not lead to a significant improvement in the sensitivity, we will not consider them in this paper~\citep[see, e.g.,][for an implementation of this idea in the context of the two-step map-making]{Keihanen2010}. As we will see below, in their absence the map-making will discard all the information in the data that matches the time-domain signature of the templates, effectively filtering all template-like modes out of the data. As previously explained, this causes a loss of a sky mode, $\b{\tilde s}$, only if its time-domain projection $\b{A}\b{\tilde s}$ is a template-like mode. In general, subsequent observations break the degeneracy that a template might have with a subset of the dataset. There are two well-known exceptions. First, the global offset of the time streams is always filtered, as a consequence the global offset of the temperature map is unconstrained. Second, when observing from the Earth's poles, scans at constant elevation always probe constant declination stripes, with different elevations corresponding to different declinations. Filtering the offset from the time streams of each of these scans prevents the reconstruction of the offset of each of these constant declination stripes. These can be partially recovered if the stripes share some
of the sky pixels owing to the assumption that the sky signal is constant across a pixel. The resulting constraints would typically be weak, in particular for the relative offset of two stripes that are not directly adjacent, which could be constrained only via the intermediary ones. Consequently, in such cases one should expect poorly constrained long modes in declination.

\subsection{Ground pickup}
\label{ground_pickup}
Ground-based experiments usually have non negligible ground-synchronous signal contaminating their TOD. The most common source is ground pickup in the far side-lobes of the beam, although the magnetic response of the experiment to the Earth's magnetic field may also be a concern. Experiments try to minimize these side lobes but, as the ground is very bright by CMB standards, their contribution typically cannot be neglected. This ground-synchronous signal could be thought of as a two-dimensional template in Earth-bound coordinates. However, in practice the situation is more complex as the signal can vary in time and may be different for different detectors, as the level and structure of their side-lobes may be very different (and poorly known). Therefore we typically model the ground signal as a one-dimensional template specific to each constant elevation scan and to each detector, or at the very least to a group of detectors located sufficiently close together in the focal plane.

Such a one-dimensional ground template can be parametrized with the azimuth of the observation and represented by one dimensional discretized map with entries standing for the amplitude of the ground signal in each of the disjoint, consecutive azimuth bins. Following our previous argument,
we need to introduce such a template for each detector and each constant elevation scan separately. These, concatenated together, are then denoted as 
$\b{g}$. In the presence of ground pickup, the time-domain data can be then modeled as a sum of three terms: the sky signal term, $\b{A}\b{s},$
the ground-pickup term, $\b{G}\b{g},$ and the noise, $\b{n}$. We can then write,
\begin{eqnarray}
\b{d} = \b{A}\b{s} + \b{G}\b{g} + \b{n}.
\label{eq:dataWithGround}
\end{eqnarray}
This is merely a specialized version of~\eqref{data}. The role played by the matrix $\b{G}$ is analogous to that played by the sky pointing matrix, $\b{A}$. Adopting the simple binned ground template model introduced above, each column of $\b{G}$ is associated with some specific azimuthal bin assigned to some specific detector and some constant elevation scan. This column will be composed of ones and zeros, with $1$ indicating that the given measurement was made by the specific detector and was performed within the specific scan, when the observation's azimuth falls within the specific azimuthal bin.Therefore applying $\b{G}$ to the template, $\b{g}$, will add the same value of the ground pick up to all these measurements.
\begin{figure}[t] \centering 
\includegraphics[width=0.5\textwidth]{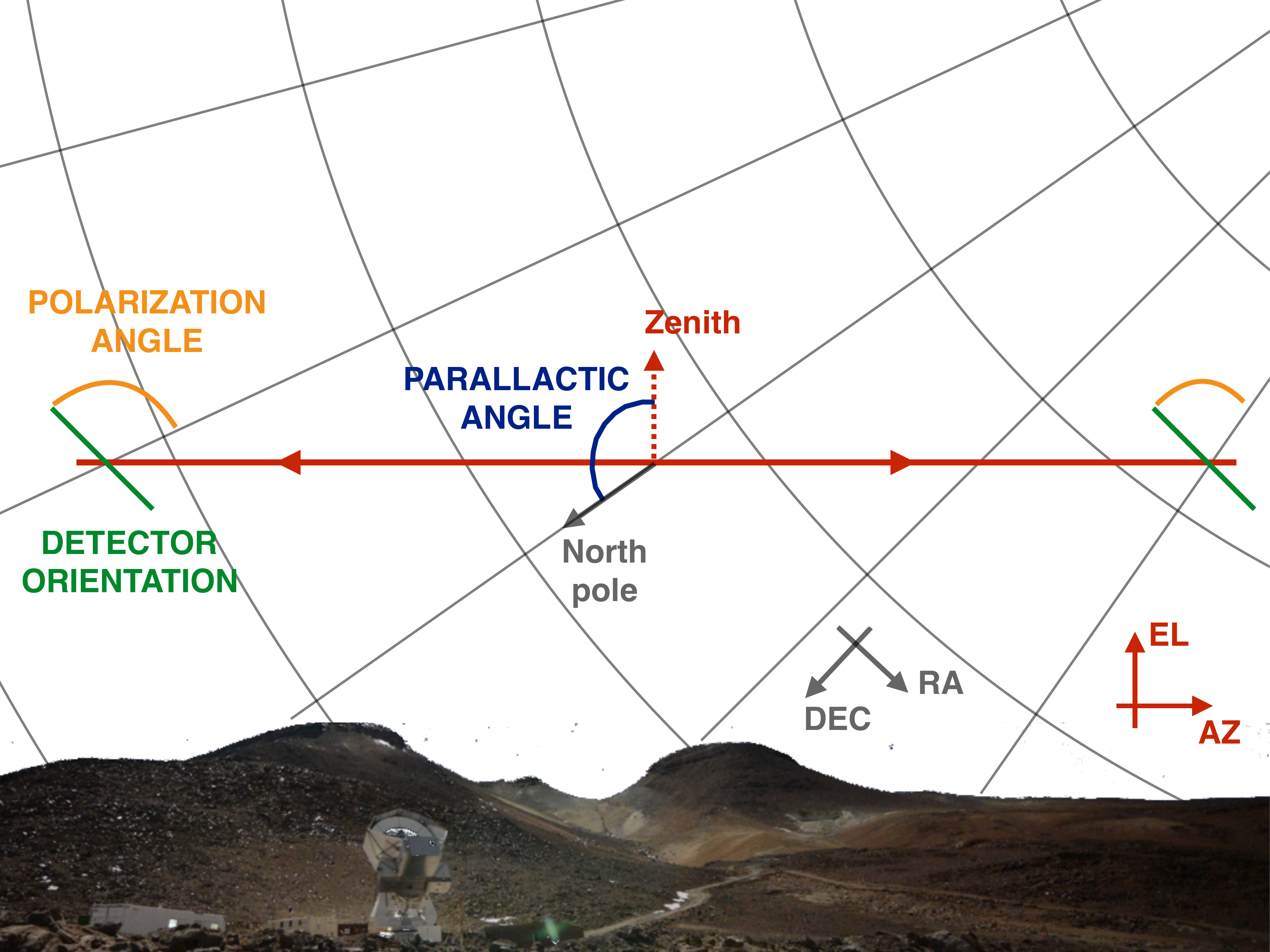} 
\includegraphics[width=0.5\textwidth]{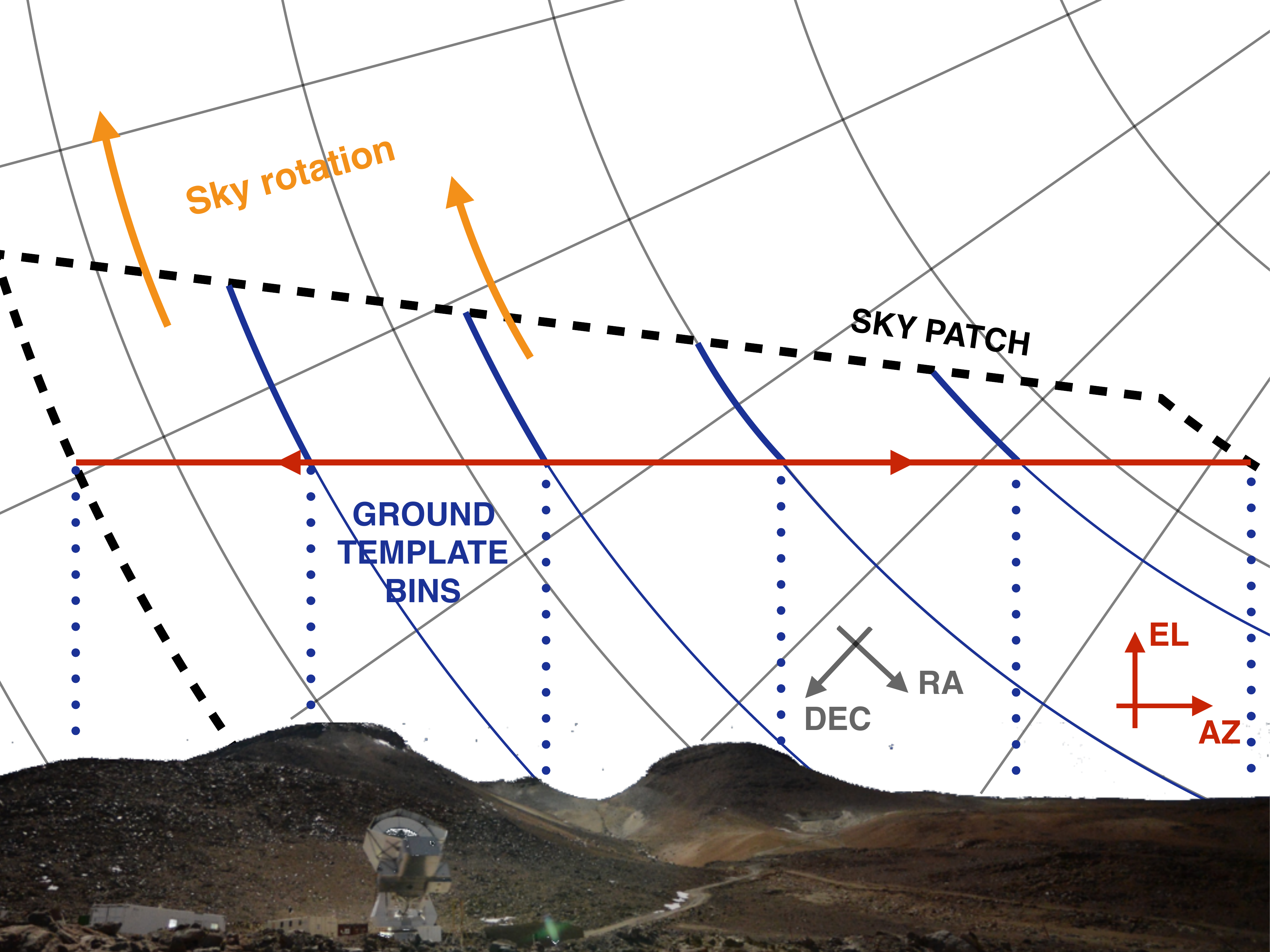} 
\caption{\emph{Top panel} shows the geometry of a constant elevation observation. Grey lines represent the equatorial coordinate system at some fixed time instant. The red line shows the scan in the horizontal coordinate system with the telescope assumed to chop back and forth along a constant elevation direction. The orientation of a polarisation sensitive detector is shown with green lines. It is assumed to be fixed in horizontal coordinates and thus varies somewhat with the observation's azimuth due to changes in the parallactic angle, which is marked in blue. \emph{Bottom panel} emphasizes the effects due to the sky rotation. This is marked with orange arrows. As the sky rotates, the constant elevation scan progressively covers the sky area delineated by a dashed black line. The area above the red line has been already observed. Also marked are its sub-areas, which are observed when the telescope's azimuth falls in one of the azimuthal bins. These for instance may be used to discretize the ground-pickup and are shown here with blue dotted lines. These sub-areas remain disjoint as the sky rotates whenever the azimuthal bins are disjoint, and are separated in the figure with blue solid lines. 
} \label{fig:observationScheme}
\end{figure}
In this section we summarize the effects the modeling of the ground pick-up may have on the quality of the map recovered with the unbiased map estimator, leaving more detailed discussion to Appendix~\ref{groundFilteringDetails}.

We start by considering a site far from the Earth's poles. For a single constant-elevation scan, samples having the same ground pickup will correspond to measurements taken by a single detector with the azimuthal position falling into one of the azimuthal bins. As time progresses and the sky rotates in Earth-based coordinates (see Fig.~\ref{fig:observationScheme} for a sketch of the geometry of the problem) the measurements will cover the sky area extending along the RA direction in the sky coordinates. The size in declination of the area depends on the scan elevation and the azimuth of the bin. The sky areas covered by measurements corresponding to two different azimuthal bins will be disjoint and as both these subsets are affected by a different ground pickup amplitude, their relative offsets will be unconstrained. As a consequence, a single detector map will have multiple degenerate modes, each corresponding to a sky patch swept by a different azimuthal bin. The degeneracies can be in part removed if data of another detector are used, but only if the azimuthal bins of the latter are shifted with respect to those of the first detector in such a way that their corresponding sky patches are also shifted and each patch of the second detector includes pixels from two patches adjacent to the first one. Yet another factor breaking the degeneracy here is the sky pixelization, as the sky pixels crossing the boundary of two adjacent bin sky patches will constrain their relative amplitude. However, in all these cases the degeneracy-breaking may be quite weak because the overlaps typically involve only a limited number of neighboring sky patches corresponding to single azimuthal bins. For observations taken from the poles the situation is different. Since the sky movement is in the azimuthal direction the same sky can be measured in different azimuthal bins. Consequently, there is a significant constraining power on the relative offsets, leaving the overall offset of the observed constant declination stripe as the only truly degenerate mode. We already encountered this degenerate mode in  \secref{correlatedNoiseTemplates}: the offset of the TOD is filtered also when removing correlated noise. The consequences of this degenerate mode and the possible degeneracy breaking effects were already discussed at the end of \secref{correlatedNoiseTemplates}.  

\subsection{Recap} 

The following useful conclusions have been drawn out in this section (and are borne out further by a more detailed analysis in Appendix \ref{groundFilteringDetails}). In the presence of time-domain filtering of the kinds discussed here, only relatively few sky modes are expected to be genuinely degenerate. Nevertheless, a few extra ill-conditioned modes with large variance should also be expected. They are mainly related to the filtering of the ground signal, which leaves poorly constrained modes in the declination direction. The details of these modes will depend on choices made regarding pixelization, definition of the ground template bins, and scanning strategy, but they will be more prominent for observations conducted from the poles. Given this, biased map-making -- by construction oblivious to the presence or absence of such modes -- may be seen as providing a more  convenient and adaptable way to perform the analysis. Indeed, it has been the approach of choice for multiple past analyses of this kind of data sets~\citep[e.g.,][]{QUAD2010, SPT2011, BICEP2014}. However a relevant question, which we discuss in more detail in the reminder of this paper, is whether it is feasible for a ground-based experiment to produce an unbiased estimate of the sky signal, and if so, with what fidelity. 

\section{Worked example: the \pb{} experiment}
\label{pipeline}
This section applies the map-making formalism to realistic simulations broadly based on the \pb{} experiment. We begin by reviewing the experimental
characteristics of \pb{}, and then describe details of the map-making process. We conclude by considering how these different approaches impact the final map properties and, most importantly, the measurement of the $B$-mode polarization power spectrum.

Though the results and conclusions are strictly speaking specific to the \pb{} experiment, they are expected to be qualitatively relevant for the other on-going and planned ground-based experiments, which typically implement a similar data reduction procedure.

\subsection{\pb{} experiment and its observations}
\pb{} is a dedicated CMB $B$-mode experiment operating from the James Ax Observatory in the Atacama Desert in Northern Chile. It is composed of a
cryogenic receiver mounted on the Huan Tran telescope (HTT) \citep{Tran2008}. HTT is a two mirror off-axis Gregorian telescope designed to ensure low
cross polarization. A \SI{4}{\K} aperture stop in the receiver creates a \SI{2.5}{\m} illumination pattern on the primary mirror, resulting in a beam size of
\SI{3.5}{\arcmin} FWHM. The receiver hosts 637 focal plane pixels (1274 transition edge bolometers) sensitive to a spectral band centred at \SI{148}{\GHz} with 26\% fractional bandwidth \citep{Arnold:2012sr}. Each focal plane pixel contains two detectors, henceforth called a ``detector pair'', with each detector sensitive to a different orthogonal polarization direction. The diameter of the field of view of the focal plane is \SI{2.4}{\degree}. The optics of the receiver includes a stepped half-wave plate that can modulate the polarization state of the incoming radiation. Since the optics and the receiver are installed on a moving mount, the azimuth and elevation pointing direction can be controlled during the observations.

The observation considered in the following analysis are based on the first \pb{} campaign, providing a high level of realism to the pointing and polarisation information in our simulations. However, the simulations do not reflect all of the properties of the data sets used in the actual data analysis. Though \pb{} observed three patches, we restrict ourselves to the RA23 patch (R.A. $23^{h}1^{m}48^{s}$ Decl. $-32^\circ48'$). The nominal area of the patch, as used in the analysis of \citet{POLARBEAR} is \SI{8.8}{deg\squared}, but the area actually observed and considered here is much larger, amounting to approximately \SI{43}{deg\squared}. Due to the sky rotation, the patch rises and sets, reaching a maximum elevation of \SI{82}{deg} while the minimum
elevation for observation is \SI{30}{deg}. The scanning consists of 15-minute \emph{constant elevation scans} (CES) ($\N{t}^\text{CES} \sim 25000$
measurements at a rate of \SI{31.8}{\Hz}) in which the telescope moves at a
constant velocity of \SI{0.75}{deg\per\s} in an azimuth range of
\SI{3}{\degree}. We refer to a single left-to-right sweep of the telescope as
\emph{subscan}; there are typically $\N{sub}\sim 150$ subscans in a CES. The CES
ends when the patch leaves the field of view of the telescope, both the azimuth
and the elevation are then modified and a new CES is performed at the new
position of the patch. The half-wave plate position is constant during a CES.
During the first season, its orientation was rotated by \SI{11.25}{deg} every
one-two days during the first half and occasionally during the second half. More details about the observation can be found in \cite[][]{POLARBEAR}.

\subsection{Time domain data model and simulations}
In this section we illustrate in detail our data model, providing a concrete example for the general considerations presented earlier in \secref{framework} and \secref{templates}. Our goal is to investigate the effects that are inherently due to the filtering, rather then the ones that stem from a failure of the applied filtering to remove unwanted contributions. Consequently, our simulations employ an idealized data model, neglecting important properties that real data usually have: imperfect polarization efficiency, noise correlated between time samples and/or detectors, temperature to polarization leakage due to differential gain, beam or bandpass between the two orthogonal detectors, etc., implicitly assuming that all such effects can be removed by the filters. We adopt the filters defined by the proposed data model and used for the actual \pb{} data.

For convenience, we first consider data collected during a single CES by a single pair of detectors in the same focal plane pixel. This is the fundamental
unit of our data model and we have a total of $\N{pair} \times \N{CES} \sim1.5\cdot10 ^{6}$ of them. The two detectors in such a pair are sensitive to two orthogonal polarizations and are referred to as $\parallel$ and $\bot$. We model their TOD to include signal, ground pickup, and templates related to the correlated noise and an actual noise term, 
\begin{eqnarray} 
\left[ \begin{array}[c]{c} \b{d}^{\parallel}\\ \b{d}^{\bot}  \end{array} \right] &= &\b{A} \b{s} + \b{G} \b{g} + \b{B} \b{x} + \b{w}\nonumber\\ &= &\left[ \begin{array}[c]{c} \b{A}^\parallel\\ \b{A}^\bot \end{array} \right] \b{s} + \left[ \begin{array}{cc} \b{G}^d & \b{0}\\ \b{0} & \b{G}^d \end{array} \right] \left[ \begin{array}[c]{c} \b{g}^{\parallel}\\ \b{g}^{\bot}  \end{array} \right] + \b{B} \b{x} + \left[ \begin{array}[c]{c} \b{w}^{\parallel}\\ \b{w}^{\bot}  \end{array} \right], \label{dataModeTopBottom}
\end{eqnarray}
where we have arranged the data vector so that all the measurements of the first detector of the pair are gathered together and followed by the measurements taken by the other one.

The pointing matrices, $\b{A}^{\parallel / \bot}$, are as given
by~\eqref{eq:data3stokes}, and thus have only three non-zero elements for each
row, which correspond to three Stokes parameters of a given sky pixel, $p$,
observed at the time assigned to the row. These are equal to one, $\cos(2\varphi^\parallel_t)$ and $\sin(2\varphi^\parallel_t)$ for detector $\parallel$ and $1$, $\cos(2 \varphi^\bot_t) = -\cos(2\varphi^\parallel_t)$ and $\sin(2\varphi^\bot_t) = - \sin(2\varphi^\parallel_t)$ for detector $\bot$, and for the $I$, $Q$ and $U$ signal components, respectively. The block-diagonal structure of matrix $\b{G}$ is due to the fact that we have introduced two independent ground templates, one for each detector of the pair. Since we will always use the boresight azimuth to define the azimuthal bins, each of the blocks is the same for each detector and all focal plane pixels. Matrix $\b{B}$ describes the time-domain filtering and thus can have more complex structure. In particular it needs to account for two types of contributions: the ones that are correlated between detectors and the ones that are independent. This can be achieved by assuming that matrix $\b{B}$ has the following structure,
\begin{eqnarray}
\b{B} \equiv \left[
\begin{array}{c c c}
\b{B}_{corr} & \b{B}_{uncorr}& \b{0}\\
\b{B}_{corr} & \b{0} & \b{B}_{uncorr}
\end{array}
\right],
\label{eqn:BCorrUncorrStructure}
\end{eqnarray}
where we assumed that we use the same filtering of the uncorrelated part for each of the detectors. This corresponds to the following breakdown of vector $\b{x}$,
\begin{eqnarray}
\b{x} \equiv 
\left[
\begin{array}{l}
\b{x}_{corr}\\
\b{x}_{uncorr}^\parallel\\
\b{x}_{uncorr}^\perp
\end{array}
\right].
\end{eqnarray}
Here, $\b{x}_{corr}$ collects the amplitudes of all the time-domain modes common to both the detectors, while $\b{x}_{uncorr}^{\parallel/\perp}$ those specific to only one of them.

Owing to the orthogonality of the two polarization directions for the two detectors in a pair, we can represent their data with summed and differenced
data streams, $\b{d}^+, \b{d}^-$, which contain the information about total intensity and polarized sky signals, respectively. These are defined as,
\begin{eqnarray} 
\b{d}^{+} & \equiv & \frac{1}{2}( \b{d}^{\parallel}+ \b{d}^{\perp})\\ \b{d}^{-} & \equiv & \frac{1}{2}( \b{d}^{\parallel}- \b{d}^{\perp}).
\end{eqnarray}
Using~\eqref{dataModeTopBottom} and introducing quantities specific to each of the new data streams, defined as,
\begin{eqnarray}
\b{A}^\pm & \equiv & \frac{1}{2}(\b{A}^\parallel\pm\b{A}^\perp)\\
\b{B}^+& \equiv & 
\left[
\begin{array}{c c}
\b{B}_{corr}, & \b{B}_{uncorr}
\end{array}
\right],
\label{eqn:bPlusDef}
\\
\b{B}^-& \equiv &\b{B}_{uncorr}\\
\b{x}^+ & \equiv & 
\left[
\begin{array}{c}
\b{x}_{corr}\\ \frac{1}{2}(\b{x}_{uncorr}^\parallel+\b{x}_{uncorr}^\perp)
\end{array}
\right],
\label{eqn:xPlusDef}
\\
\b{x}^- & \equiv & \frac{1}{2}(\b{x}_{uncorr}^\parallel - \b{x}_{uncorr}^\perp),\\
\b{g}^\pm & \equiv & \frac{1}{2}(\b{g}^\parallel\pm\b{g}^\perp),\\
\b{w}^\pm & \equiv & \frac{1}{2}(\b{w}^\parallel\pm\b{w}^\perp),
\end{eqnarray}
we can express these new data in a concise way as,
\begin{eqnarray} 
\b{d}^\pm = \b{A}^\pm\b{s}_{T/QU} + \b{G}^d \b{g}^\pm + \b{B}^\pm \b{x}^\pm + \b{w}^\pm. \label{dataModeSumDif}
\end{eqnarray}
Here $\b{s}_T$ and $\b{s}_{QU}$ denote sky signal vectors made of the total intensity and interleaved, pixel-by-pixel, $Q$ and $U$ Stokes parameters, respectively. These expressions emphasize that as intended each of the new data streams contains information either about the total intensity, $\b{d}^+$, or polarization, $\b{d}^-$. Moreover, as all the amplitudes appearing on the right hand side of these equations are specific for each data set, each of these two data sets can be analyzed completely separately and, under the assumptions specified earlier, without any loss of accuracy. Specifically, the maps of the total intensity on the one hand, and the Q and U Stokes parameters on the other can be estimated independently. This is the approach we follow in this work.

We point out that a perfect separation of the total intensity and polarization information is strictly speaking only possible if the two detectors of each
pixel pair are perfectly calibrated and have identical beams. Otherwise, some residual total intensity contribution may be present in $\b{d}^-$ and,
less harmfully, some polarization in $\b{d}^+$. The beams of two detectors are more likely to be similar if they belong the same focal plane pixel. Therefore,
using $\b{d}^-$ to constrain the polarization may be less susceptible to total intensity leakage due to beam differences than performing global separation of the three Stokes parameters directly using $\b{d}^{\parallel/\perp}$ as inputs. In any case, if needed, leakage from the total intensity to the differenced data, $\b{d}^-$, can be modeled as a total intensity-like template and used in the map-making process. Though such tests were indeed performed as part of the analysis of the actual \pb{} data set~\citep[][]{POLARBEAR}, we do not consider them in the present work. Leaving aside this kind of systematic effect, it is mathematically equivalent whether we use one or the other data representation, as long as the filtered temporal templates, defined by $\b{B}_{corr}$ and $\b{B}_{uncorr}$, are used consistently.

In~\eqref{dataModeSumDif} the pointing matrices, $\b{A}^+$ and $\b{A}^-$, are given as in Eqs.~(\ref{eq:dataIonly}) and~(\ref{eq:data2stokes}) and therefore are composed of zeros and ones for the summed data and have two non-zeros per row given by $\cos(2 \varphi^\parallel_t)$ and $\sin(2 \varphi^\parallel_t)$ for the differenced data. The ground template operator, $\b{G}^d$, is the same for the summed and differenced data. As an independent ground template is used for each detector pair and for each CES, the $\b{G}^d$ matrix has as many columns as the number of bins used to discretize the ground signal and as many rows as the number of samples in a given CES. We bin the observed azimuths in intervals of the width of $0.08$ deg and thus
have $\N{\b{G}}^{CES,p} \sim$ 100 bins per template. At any given time $t$ the
corresponding row of $\b{G}^d$ has only one non-zero entry (equal to one) in a column corresponding to the ground bin observed at this time.

The $\b{B}^\pm$ matrices define the time-domain filtering applied to both data streams in order to suppress long term correlations. In the \pb{} case the filtering is done subscan-by-subscan~\citep[][]{POLARBEAR}. Consequently the $\b{B}^\pm$ are block diagonal with one block per subscan, and each block displaying the same structure as in Eqs.~(\ref{eqn:BCorrUncorrStructure}) and~(\ref{eqn:bPlusDef}). We denote such an elemental block with a subscript, $s$, to emphasize that we are referring to a single subscan. Each of these blocks removes from a given subscan time-domain trends given by
time domain templates defined as polynomials up to some order, selected to
ensure that the noise after filtering is nearly white. In our analysis,
$\b{B}_{corr, \; s}$ contains four templates (the Legendre polynomials up to the
3rd order, appropriately rescaled to become orthonormal over the time interval given by the
subscan) and $\b{B}_{uncorr,s}$ contains only the constant and linear templates.
Consequently, the columns of $\b{B}_{uncorr,\; s}$ are linearly dependent on
those of $\b{B}_{corr,\; s}$ and we restrict $\b{B}^+$ to the latter ones
(i.e., $\b{B}^+ = \b{B}_{corr,\; s}$) without any loss of generality, as discussed
in~\secref{degeneracyT}. 

$\b{w}^+$ and $\b{w}^-$ are the noise terms, describing the noise in the data after the ground template and temporal trends removal. These noise terms
are expected to be ``prewhitened'' with respect to the actual noise in the data. In the simulations these vectors are modeled as white with inverse variances given by $\omega^+$ and $\omega^-$ respectively. We allow for different weights for each CES and detector pair. These weights are evaluated from the actual data as the inverse of the average of the power spectral density of the real data sum and difference, taken between \SI{1.04}{\Hz} and \SI{3.13}{\Hz}.

It is now straightforward to generalize these considerations to multiple CESs and multiple detector pairs. In both cases we stack all the TOD for every
detector pair and every CES together and, for concreteness, we do so for the summed and differenced data separately. The form of~\eqref{dataModeSumDif} for the concatenated summed and differenced data remains the same but the data objects and operators on its right hand side need to be appropriately redefined. In particular, as we define a different ground template for each detector pair and each CES the global $\mathbf{G}_{\,\rm all}^\pm$ matrix, will be block diagonal, with each block given by the detector-pair and CES specific matrix, $\mathbf{G}^d$. The vector of the ground template amplitudes, $\mathbf{g}^\pm_{\,\rm all}$ will accordingly be made of the detector-pair and CES specific vectors, $\mathbf{g}^\pm$.  Similar generalizations also apply to the temporal drifts term, however in this case one may need, or want, to account for effects that would be correlated between different detector pairs. Such effects could for instance be a result of contributions to the summed data due to atmospheric fluctuations. Consequently, the ultimate filtering operator, $\mathbf{B}^\pm_{\,\rm all}$, may not be strictly block diagonal but have rather a form resembling that of~\eqref{eqn:BCorrUncorrStructure}. In the example studied however we do not include this possibility but introduce a separate template for each detector, each subscan and each polynomial order. This adds some flexibility that may permit better accounting for systematic effects, but it may not be always
advantageous as far as statistical uncertainties are concerned due to the significant number of extra independent degrees of freedom this choice implies.
The number of polynomial templates per CES and detector pair is
$\N{\b{B}}^{CES,p} = \N{sub} \times \N{poly} \sim 150 \times \N{poly}$, where
$\N{poly}$ is four or two if we are considering the sum or the difference of the detector pair. Consequently the total number of templates per CES and detector pair sum (resp. difference) is $\N{\b{T}}^{CES,p} = \N{\b{G}}^{CES,p} + \N{\b{B}}^{CES,p} \sim $ 700 (resp. 400). 

\subsection{Map-making: estimators and implementation}
\label{map_estimators}
We estimate the sky signals using both the unbiased, Eq.~(\ref{s_estimator}), and the biased, Eq.~(\ref{s_biased}), estimators. 
The weight matrices, $\b{M}$, are assumed to be block diagonal, with blocks corresponding to different CESs for every detector pair. The blocks are in turn taken to be diagonal and proportional to a unit matrix with the proportionality coefficient given by
the noise weights as introduced at the end of the previous section. Consequently, the weight matrix block corresponding to the $i$th detector pair and $c$th CES reads,
\begin{eqnarray} 
\b{M}_{c,i}^\pm = \omega_{c,i}^\pm \b{1}. \label{blockWightMatrix}
\end{eqnarray}

We define sky pixels for which the signal estimation is performed prior to
map-making. This is done in two steps. First, we remove pixels for which a
two-by-two diagonal block of $\AtNA$ matrix has a condition number higher than $10^6$. This ensures that we retain only the pixels for which there is sufficient observation redundancy to allow for numerically disentangling two Stokes parameters, Q and U. In addition, we also remove pixels that have not been observed sufficiently frequently. The threshold is chosen as a minimal number of CESs during which the pixel was observed, taken to be $120$ here. We have found empirically that this criterion helps avoid strongly degenerate modes localized at the lightly-observed patch boundaries. As a result our sky maps are composed of roughly \num{5.2e+4}, nside=2048 HEALPix pixels covering approximately \SI{43}{deg\squared}. Once the pixels are designated for removal we propagate this information back to the TOD, flagging out all the samples which fall in one of those pixels.

Map-making requires an application of the filtering operator, $\F{T}$, to the
time-ordered data vectors, $\b{d}^\pm$. This has to be preceded by a computation
of the filtering operator itself. All the matrix operators involved in the
computation (i.e., $\b{B}^\pm$, $\b{G}$, $\b{T} \equiv [\b{B}^\pm, \b{G}]$, and $\b{M}$) are block diagonal, as is $\F{T}$. We pre-compute the orthonormalization kernel $\b{K} \equiv (\b{T}^\T\b{M}\b{T})^{-1}$ via its explicit construction and inversion. We first build the template coupling kernel, $\b{T}^\T\b{M}\b{T}$, which requires $\N{t} \times \N{poly}^2 \sim 10^{12} - 10^{13}$ operations thanks to the sparsity and structure of the time-domain templates. The cost of the inversion (with regularization) of the kernel is $(\N{T}^{CES,p})^3 \times \N{CES} \times \N{pair} \sim 10^{13} - 10^{14}$ operations. This number is considerable, but performing a large number of small matrix inversions is very efficient because of the locality of the data to be processed by the CPUs, resulting in a considerable advantage on modern massively parallel computing systems. The kernel's rank is approximately equal to the total number of templates, $\N{\b{T}}^{CES,p} \times \N{CES} \times \N{pair} \sim10^{8}-10^{9}$. Storing this object in the memory is demanding, even when its 
block diagonal structure is explicitly taken into account, since it amounts to $(\N{\b{T}}^{CES,p})^2 \times \N{CES}
\times \N{pair} \sim 10^{11}-10^{12}$ double precision numbers.

We note that by construction $\b{B}^\pm$ and $\b{G}$ are individually column-orthogonal, but when considered together their columns are in general not orthogonal and possibly not even linearly independent. Indeed, one degeneracy of each CES and detector-pair block is readily expected. This is the one between the constant mode filtered by the time-domain templates and the constant offset of the ground template. Consequently, the kernel, $\b{K}$, is in general non-trivial and its inversion needs to be regularized, see  \secref{degeneracyT}. In doing so, we set a threshold of $10^{6}$ on the condition number of each diagonal block of $\b{T}^\T\b{M}\b{T}$. With the kernel precomputed and stored in the computer memory, we apply the filter to the data without ever computing it explicitly. Instead, we perform the operations included in its implicit form,~\eqref{filter}, from the right to left,
 computing first $\b{M} \b{d}$, followed by $\b{T}^\T (\b{M} \b{d})$, and then $\b{K} (\b{T}^\T \b{M} \b{d})$. We then loop over the TOD again to compute $\b{A}^\T \F{T} \b{d}$. Both these operations have a $\O{\N{t}}$ cost. This approach facilitates the entire operation: storing the filtering matrix, $\F{T}$, would require a prohibitive amount of memory and its application to a time-domain vector would take too much computational time.

\subsubsection{Unbiased map estimator}
\label{unbiasedImplementation}
We have implemented the unbiased map estimator Eq.~(\ref{s_estimator}) in two different ways. In the first case, we perform an explicit construction and inversion of the $\AtFA$ matrix, while in the second we employ an iterative solver instead.

In the explicit implementation, we first compute $\AtFA$. This is done as follows. Consider a time sample $t$ and call $p$ the observed sky pixel. Since the only non-zero entries of the $t$-th row of $\b{A}$ are the columns corresponding to pixel $p$, the $t$-th column of $\F{T}$ contributes only to the $p$-th column of $\F{T}\b{A}$. Analogously the $t'$-th row of $\F{T}$ contributes only to the $p'$-th row of $\b{A}^\T\F{T}$. Therefore, in order to build the $\AtFA$ matrix we loop over the elements of $\F{T}$: for the $(t',t)$ entry we compute its contribution to the $(p',p)$ block of $\AtFA$. As mentioned earlier the $\F{T}$ matrix is not stored in the memory, but rather its elements are computed on the fly from the matrices $\b{K}$ and $\b{T}$ (stored in the memory in a compressed form). The non-zero entries of $\F{T}$ are the $\N{CES} \times \N{pair}$ blocks ($\N{t}^{CES}$ by $\N{t}^{CES}$) on the diagonal. These blocks are scattered across a number of processors that ranges between about 1300 and 5800, depending on the computational platform we used. Each processor is responsible for computing the contribution of its blocks to $\AtFA$ and the result is reduced at the end. Since the matrix, $\AtFA$, can not be stored in the memory of a processor (and not even on an
entire compute node) we divide it into blocks (typically 576) and we compute them one by one, at each step only considering entries $(t',t)$ of $\F{T}$ such that $(p',p)$ is inside the block of $\AtFA$ being computed. The computational cost scales as the number of non zero entries of $\F{T}$: $\N{CES} \times \N{pair} \times (\N{t}^\text{CES})^{2} \sim 10^{15}$ operations.

We then perform the eigendecomposition of $\AtFA$ representing it as, 
\begin{eqnarray} 
\label{eigendecomposition}
\AtFA = \b{V} \text{diag} (\b{e})\b{V}^\T.
\end{eqnarray}
This is done with help of a ScaLAPACK routine, \texttt{pdsyevr}, \cite{scalapack}. The numerical cost is $\O{\N{p}^3} \sim 10^{15}$
operations. This scaling relation is the main obstacle in the application of the explicit implementation to maps with a larger number of pixels. By construction the eigenvalues, $\b{e}$, are all non-negative numbers though numerically some small eigenvalues may
turn out to be negative. The inversion of this matrix is then performed by inverting its eigenvalues. Since the condition number of the matrix is typically very large, the inversion needs to be regularized by employing a (pseudo)inverse defined as,
\begin{eqnarray}
(\AtFA)^{-1} = \b{V} \text{diag} (\b{\tilde e})\b{V}^\T,
\label{eqn:svdAtFA}
\end{eqnarray}
where,
\begin{eqnarray}
\b{\tilde e}_i \equiv
\left\{
\begin{array}{l l}\medskip
\b{e}_i^{-1}, & \hbox{\ \ \ \ \ \ \ if $\b{e}_i > 10^{-6} \max_j \b{e}_j$;}\\
0,    & \hbox{\ \ \ \ \ \ \ otherwise.}
\end{array}
\right.
\end{eqnarray}
One of the advantages of this estimator is that once the computationally heavy objects are evaluated, we can efficiently produce multiple simulated realizations of the reconstructed sky maps directly in the map domain, see Sect.~\ref{sect:sims} for more
details. We emphasise that the construction and inversion of the system matrix are the most complex and expensive parts of the algorithm: their memory requirements force an intrinsically parallel implementation and a non-negligible fraction of the execution time is spent in communication between the compute nodes.

In the case of the iterative solver, the system matrix, $\AtFA$, is never explicitly constructed. Instead, we follow the general blue-print of maximum likelihood map-making code implementations~\citep[e.g.][]{Cantalupo2010} and apply the factors defining the matrix from right to left. The filtering operator is applied as described above, again without being ever explicitly constructed. We use a preconditioned conjugate gradient (PCG) technique~\citep[e.g.,][]{GolubvanLoan} with the preconditioner set to be $(\AtNA)^{-1}$, which is either diagonal, for the total intensity, or block-diagonal, for the polarization, and can therefore be straightforwardly computed, stored in the memory, and applied to a vector whenever needed. This is again the standard choice~\citep[e.g.][]{Cantalupo2010}. To quantify the convergence one typically uses the norm of the map level residuals,
\begin{eqnarray}
r^{\left(i\right)} \equiv \frac{|\b{A}^\T \F{T} \b{A} \b{m}^{\left(i\right)} - \b{A}^\T \F{T} \b{d}|}{|\b{A}^\T \F{T} \b{d}|},
\label{eqn:pcgResDef}
\end{eqnarray}
where $\b{m}^{\left(i\right)}$ stands for the solution after the $i$th
iteration. We notice that $r$ is a dimensionless quantity. A typical requirement for convergence would then be $r^{\left(i\right)} \le 10^{-6}$.

Unlike the explicit implementation, the iterative solver is not applied to the full dataset. The constant elevation scans are divided into 267 nearly even groups and, using the iterative solver, one map $\bh{s}_\alpha$ is obtained independently for each group $\alpha$. Afterwards we coadd the maps as follows
\begin{eqnarray} 
\bh{s} = \left[ \sum_\alpha (\AtNA) \big|_\alpha \right]^{-1} \sum_\alpha (\AtNA)\big|_\alpha\; \bh{s}|_\alpha,
\end{eqnarray}
where as before $|_\alpha$ denotes a quantity computed using only the data belonging to subset $\alpha$.

\subsubsection{Biased map estimator}
The biased map estimator Eq. (\ref{s_biased}) requires a construction of $\AtNA$, its inversion and its multiplication by a vector, $\b{A}^\T \F{T} \b{d}$. All these operations pose no issues given the block-diagonal structure of the matrix and the pixel selection procedure applied to the data beforehand, which ensures that each of its blocks is invertible. The computational cost is then driven by the construction of the kernel $\b{K}$ ($\sim 10^{13} - 10^{14}$ operations). As mentioned earlier, this kernel is block-diagonal and can be constructed and inverted efficiently on a single modern processing unit. Consequently, the entire estimator can be implemented and executed using serial or embarrassingly parallel programming models.

\subsection{Simulations}
\label{sect:sims}
In our analysis we use signal-only, noise-only, and signal and noise simulated map reconstructions. As the time-domain data set is quite large we strive to perform the simulations in the pixel domain whenever possible. We simulate signal- and noise-only maps separately, as described below, and the total maps are then produced by co-addition of these at the map level.

We also produce specialized simulations in order to validate our implementations. These are described in Sect.~\ref{validation}.

\subsubsection{Signal-only maps}
\label{sec:signalSimulations}
For the simulations of the CMB sky signal we assume the power spectrum defined by the Planck best fit parameters
\cite{planck2015parameters}. We set the tensor-to-scalar ratio, $r$ to zero, for definiteness, as the value of $r$ is not relevant in the case considered here, given the focus of the first \pb{} campaigns on sub-degree angular scales. We synthesize the 'true sky' maps, denoted here as $\b{s}^\text{sim}$, using the \texttt{synfast} tool of the HEALPix package~\citep{Gorski2005}. 

To simulate the CMB-only sky maps as reconstructed with the explicit implementation of the unbiased map-maker we take the simulated true sky maps, $\b{s}^{\textrm sim}$ and remove from the simulated realization of the sky maps the eigenvectors corresponding to the eigenvalues set to zero in the $\AtFA$ inversion regularization. We have validated that this is equivalent, algebraically and numerically, to first projecting the sky signal, $\b{s}^\text{sim}$, to the time-domain, $\b{A}\,\b{s}^\text{sim},$ and then running the map-making procedure on the derived time-ordered data. For the biased map estimator we apply the operator, $(\AtNA)^{-1}\AtFA$, directly to the generated, true sky maps, $\b{s}^\text{sim}$.  Again this is algebraically and numerically equivalent to first producing the signal-only data streams, $\b{A} \b{s}^{\textrm{sim}}$, and then applying the biased map-making operator, $(\AtNA)^{-1}\b{A}^\T\b{F}$, to them.

Only for the iterative implementation of the unbiased map-making do we actually run the map-making solver on simulated time-streams, in order to reproduce effects related to the convergence (or otherwise) of the iterative algorithm.

\subsubsection{Noise-only maps}
\label{sec:noiseSimulations}
The noise-only maps reconstructed with the explicit implementation of the
unbiased map estimator are computed directly in the pixel-domain as $\bh{n} =
\b{V} \b{z}$, where $\b{z}$ is a pixel-domain vector of Gaussian random
variables with variance given by the corresponding entry of $\b{\tilde{e}}$, and
$\b{V}$ and $\b{\tilde{e}}$ are defined in Eq.~(\ref{eqn:svdAtFA}). This assumes
that the weight matrix, $\b{M}$, provides a correct description of the TOD noise
(i.e., $\b{M} = \b{C_n}^{-1}$). Were this is not the case, we would have to start from simulating a noise timestream, $\b{n}$, with the desired noise properties and then process it with the map-making algorithm. 

For the biased map-making we follow the latter path even if the noise is uncorrelated. We therefore start by generating a timestream of uncorrelated Gaussian numbers of appropriate variance and projecting it into the pixel-domain using the corresponding map-making procedure.

\begin{figure}[t] 
\includegraphics[width=.5\textwidth]{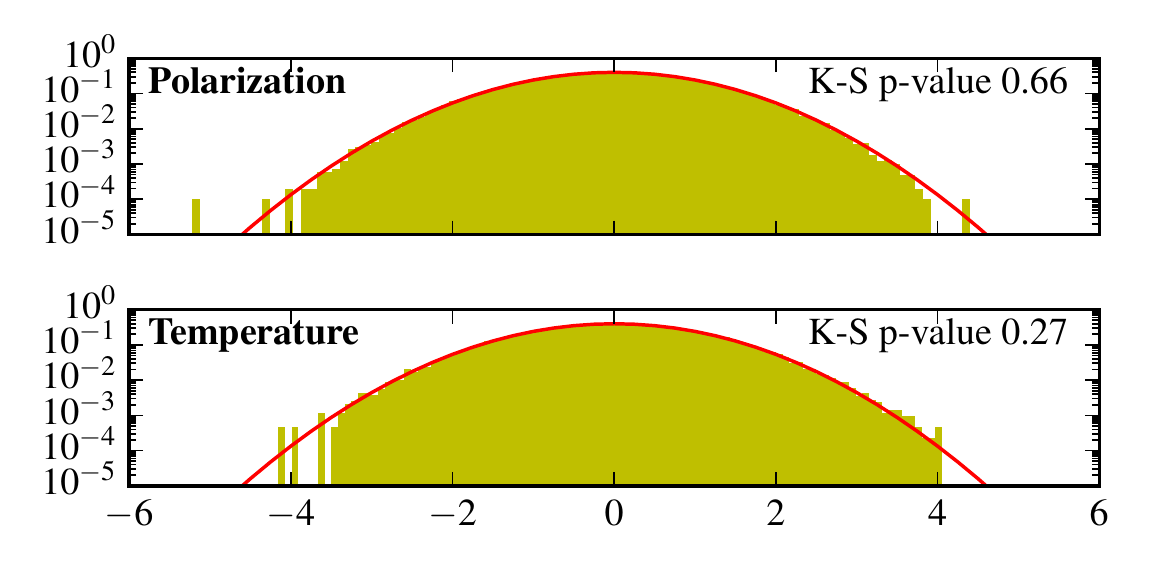} 
\caption{Histogram of the whitened unbiased map: $\b{V}\text{diag}(\sqrt{\b{e}})\b{V}^\T \bh{s}$, where $\bh{s}$ is estimated starting from a time-domain white noise simulation (see \secref{validation} for the details).} \label{fig:prewhitenedDistribution}
\end{figure}
\begin{figure*}[t] \centering 
\includegraphics[width=\textwidth]{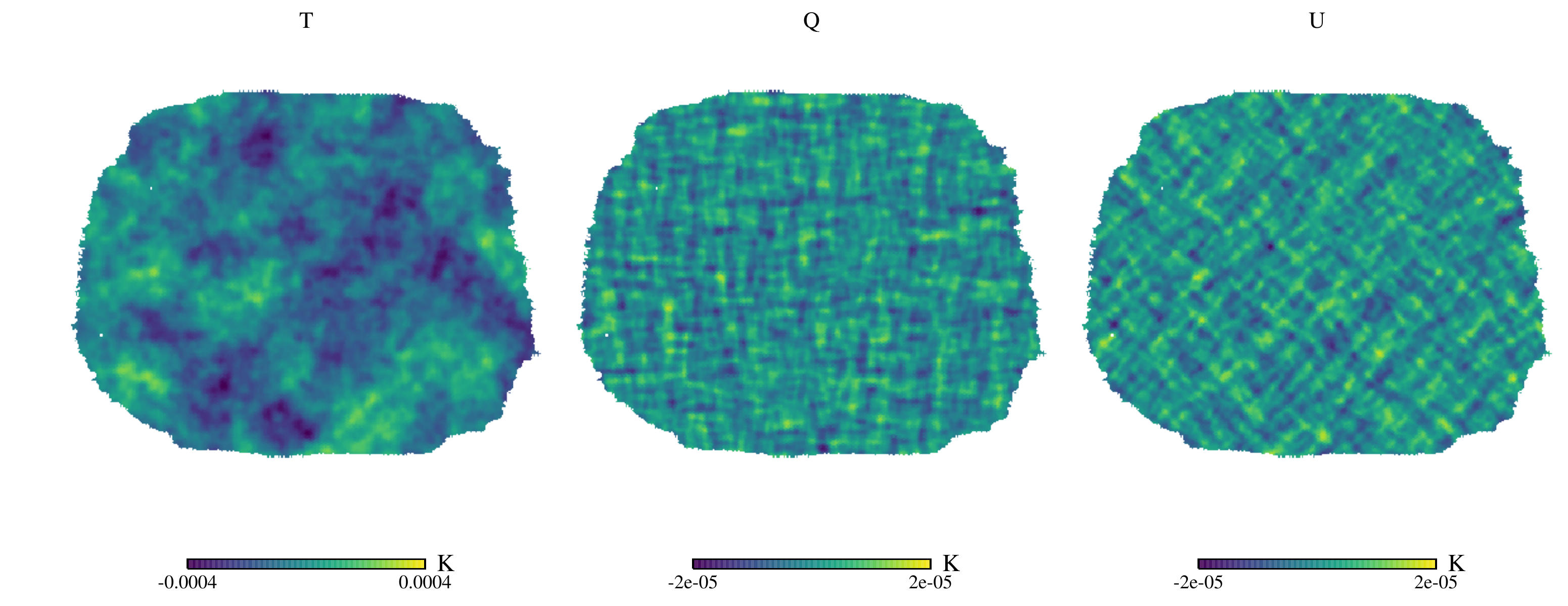} 
\caption{Maps of the input sky used here for the reconstruction comparison of the different map estimators.} \label{fig:inputMap}
\end{figure*}
\begin{figure*}[!ht] \centering 
\includegraphics[width=\textwidth]{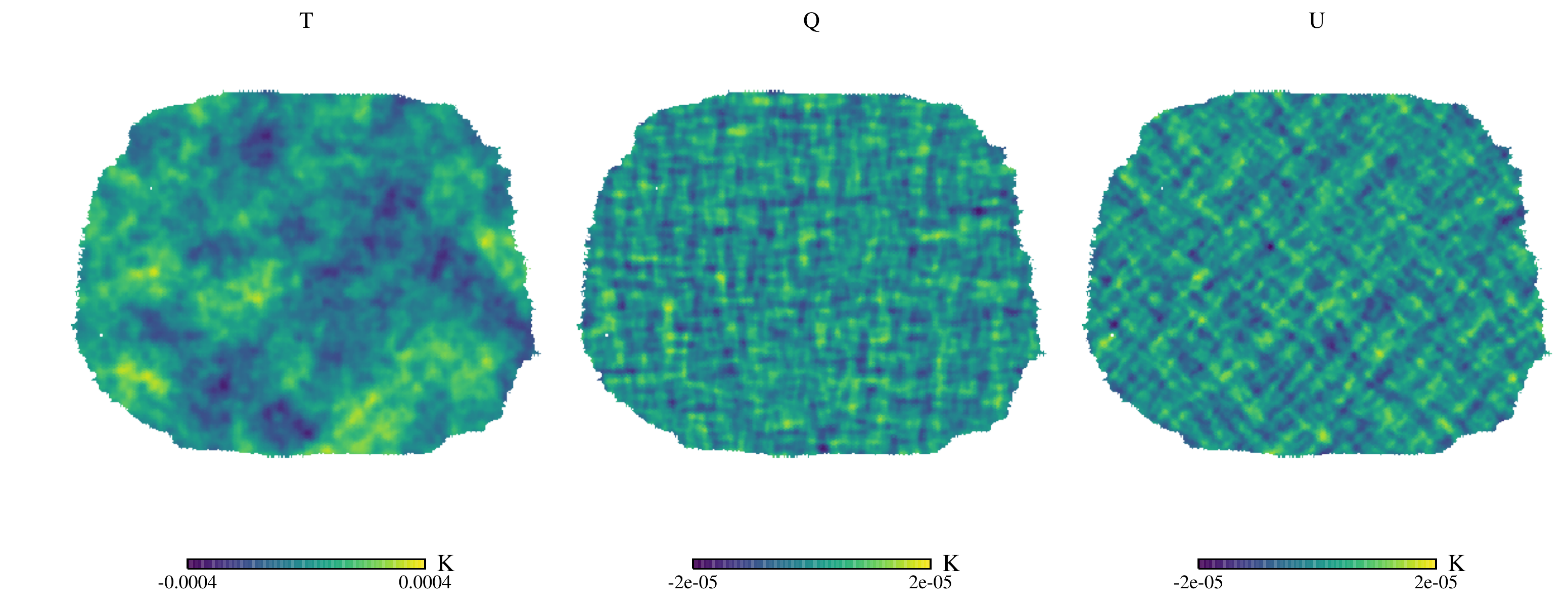}\\ 
\includegraphics[width=\textwidth]{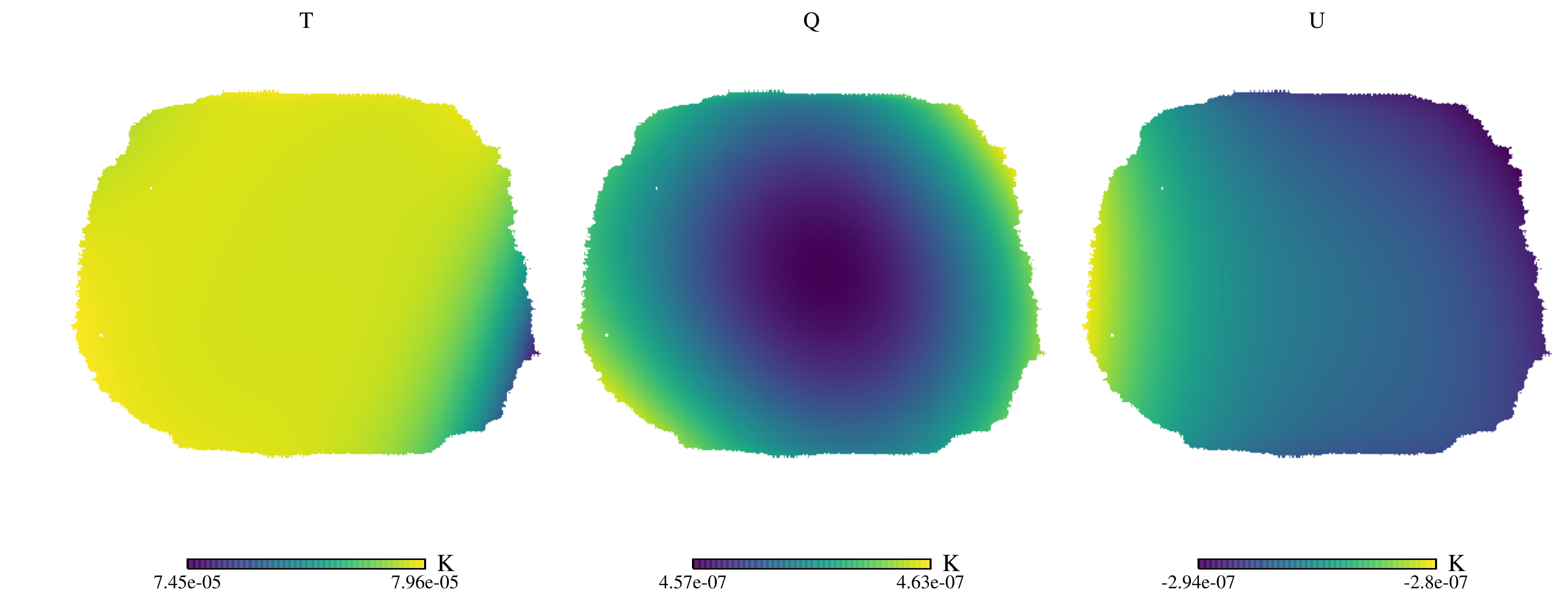}\\ 
\includegraphics[width=\textwidth]{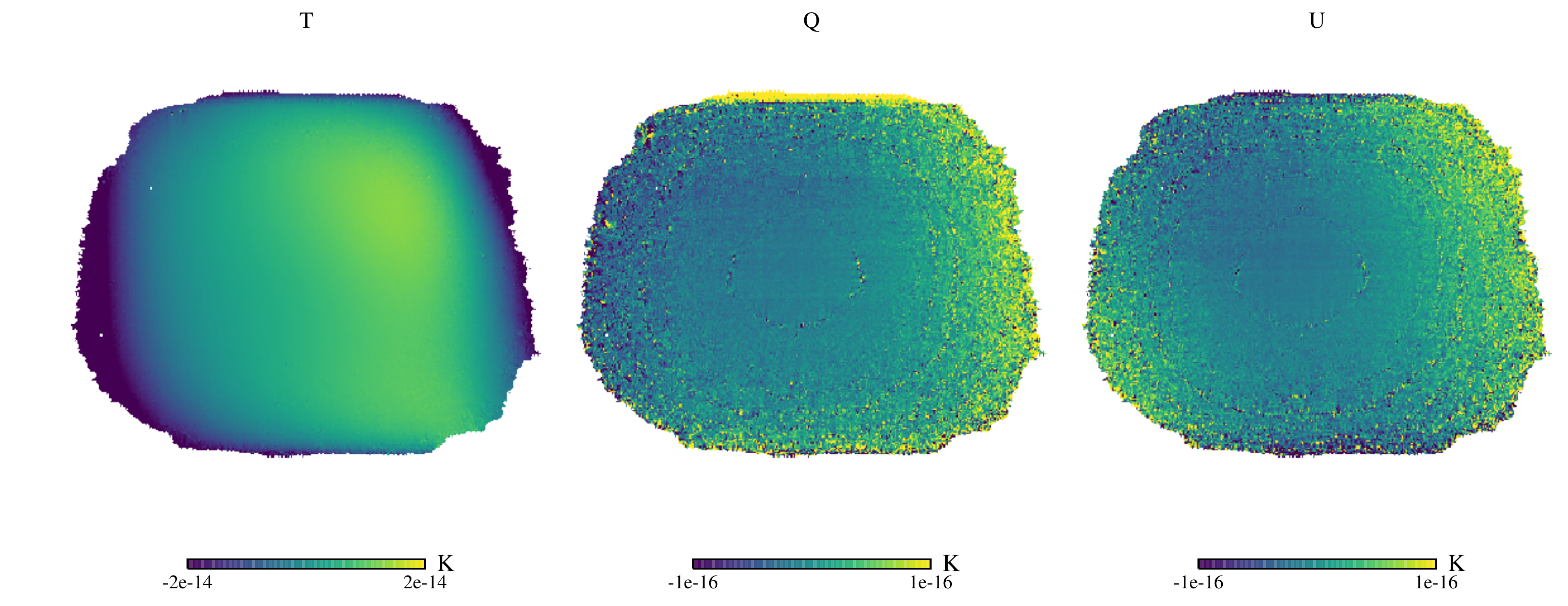} 
\caption{Maps derived with the explicit implementation of the unbiased map estimator. \emph{Top row:} reconstructed maps. \emph{Middle row:} difference between the reconstructed  and the input maps. \emph{Bottom row:} difference between the reconstructed and the input maps with the singular modes removed from the input maps. } \label{fig:unbiasedMap}
\end{figure*}
\subsection{Validation and verification of the map-making code}
\label{validation}
We have performed multiple tests in order to validate and verify the map-making code. For validation, we test whether the filtering operator, $\F{T}$, removes all the unwanted modes as desired; for verification we perform a number of full, end-to-end runs of the code, testing that the various outputs have the expected statistical properties.

Our validation test checks that the filtering operator $\F{T}$ satisfies the relation $\F{T}\b{T} \b{y} = 0$, for some vector of template amplitudes $\b{y}$. Since we are interested in map domain residuals, we actually test whether
\begin{eqnarray}
\b{s}^\text{res} = (\b{A}^\T \b{M} \b{A})^{-1} \b{A}^\T \F{T}\b{T} \b{y} = 0.
\end{eqnarray} 
We separately test the ground pick up filtering, $\b{T} = \b{G}$, and the polynomial filters, $\b{T} = \b{P}$. In the former case, we produce a simulated ground signal timesteam as follows. For every CES, there is an index $i$ for each ground template bin. $i$ ranges between 0 and $\sim 100$. For each CES we set the amplitude of the $i^\text{th}$ ground template to $y^G_i = 1 +
0.01 \times i$, while the entries of $\b{y}$ corresponding to the polynomial templates are set to zero. We obtain a simulated data vector $\b{d}^\text{ground} = \b{T}\b{y}$, where $d^\text{\,ground}_t = y^G_{i_t}$. This construction has been devised in order to ensure that the simulated timestream is a linear combination of all the ground templates, with elements that are of order unity and are always positive. This last condition mimics the worst-case scenario of a ground signal that is coherent in time. In a very similar fashion, we test the filtering operator on the polynomial filters by setting $d^\text{\,poly}_t = 1 + 0.01 \times i$, where now $i$
is the index of the swipe in azimuth within the CES (remember that we have a set of polynomial templates for each constant direction azimuthal glide). We find that the map domain residuals, $\b{s}^\text{res}$, never exceed $10^{-6}$ for temperature and $10^{-8}$ for polarization. These levels are expected given the precision of our filter orthogonalization procedure, which tends to be merely approximate for very short subscans. They are however negligible for any practical purposes.

As part of our end-to-end verification tests we study the statistical properties
of the noise-only maps produced by our map-making code. In this case, we produce
the simulated noise-only stream in time-domain, with properties described by the
diagonal weight matrix, $\b{M}$, and processed it via our code. The output map
was then prewhitened using the square root of the theoretically expected
covariance, $(\AtFA)^{-1}$. The result was then histogrammed,
Fig.~(\ref{fig:prewhitenedDistribution}), and compared to a Gaussian with unit
variance. The agreement was found to be very good, for example the
Kolmogorov-Smirnov test found $p$-values of 0.66 for polarization and 0.27 for temperature, showing that the $(\AtFA)^{-1}$ matrix (explicitly computed) reproduces correctly the covariance properties of the unbiased map, including the correlations due to the time-domain filtering.

Other examples of end-to-end tests, involving a direct comparison of the known input with the output are discussed in ~\secref{reconstructedMaps}. As emphasized there, the overall agreement is found to be excellent.

\subsection{Polarized power spectrum}
\label{power_spectrum}
The sky maps reconstructed from the measurements of a CMB experiment typically serve multiple purposes. They may be the 
end product of the analysis whose goal is a representation of the sky signal in the observed sky area. However, they will often be only a step toward some more profound statistical analysis of the underlying signal.

In the following we will therefore not only look at the reconstructed maps as images of the true sky, but will also compare them from the point of view of the constraints on the power spectra which can be derived from their analysis. In this latter case, we focus specifically on the $B$-mode power spectrum and use the pseudo-power spectrum approach to its estimation~\citep[e.g.][]{Hauser1974, Hivon2002}. This method has gained significant popularity in the field, thanks to its flexibility and relatively straightforward implementation. As our goal is the spectra of the $B$-mode polarization, and the observations we consider cover only a limited sky area, we use a so-called pure pseudo spectrum approach, which explicitly corrects for the bias and variance effects of the so-called $E$-to-$B$ leakage generated by the presence of the observed sky boundary. The technique was first proposed in~\citet{Bunn2003}, and later implemented and elaborated on in~\citet{Smith2006} and~\citet{Smith2007}. In this work,
we use a numerical code, \xpure{}, developed by~\citet{Grain2009}, which has been described, tested, validated and exploited both there and in follow-up work~\citep[e.g.,][]{Grain2012, Ferte2013, Ferte2015, planckXXX, Krachmalnicoff2015}. We refer the reader to these papers for more details.

The pure pseudo-spectrum technique removes the bias due to leakage by estimating the $E$ and $B$ spectra simultaneously and allowing for an off-diagonal $EB$ block of the coupling kernel, which is then used to model and subtract the leaked $E$ power 
from the $B$-mode spectrum. The enhancement of the power spectrum variance due to the leakage is then suppressed with the help of appropriately constructed apodizations. Ideally these have to be estimated separately for every harmonic domain bin for which the spectrum is to be computed~\citep[][]{Smith2007}; again we use the code implemented by~\citet{Grain2009}. Once the apodizations are estimated we use them for all the $B$-mode spectra we estimate, irrespective of the algorithm used to produce the maps.

The pure techniques are only designed to deal with $E$-to-$B$ leakage due to the cut sky. However, other sources of leakage are also usually present. For instance, at small angular scales leakage typically arises as a result of the pixelization adopted for the recovered map. This is typically found to be subdominant to the uncertainty due to the noise on small scales, and thus can typically be left uncorrected with little, if any, impact on the precision of the final results.

Leakage can also be expected if the $Q$ and $U$ maps used for power-spectrum estimation do not faithfully reflect the true underlying sky signal. This is certainly the case for biased map-making, but can also be relevant for the unbiased approach if degeneracies are present. The leakage arising in such cases can bias the estimated spectra on all angular scales of interest, and
thus must be carefully accounted for. Though solutions to this problem have been proposed~\citep[e.g.,][]{Bunn2003, BICEP2014, BICEP2016}, they are computationally very heavy. Here, instead, we use a phenomenological approach based on~\citet{Hivon2002} and model the biased spectra as
\begin{eqnarray} 
\left( \begin{array}[c]{c} \tilde{C}_\ell^{EE}\\ \tilde{C}_\ell^{BB} \end{array} \right) = \left[ \begin{array}[c]{cc} f_\ell^{EE} & f_\ell^{EB} \\ f_\ell^{BE} & f_\ell^{BB}  \end{array} \right] \left( \begin{array}[c]{c} C_\ell^{EE}\\ C_\ell^{BB} \end{array} \right) + \left( \begin{array}[c]{c} N_\ell^{EE}\\ N_\ell^{BB} \end{array} \right). 
\label{spectrum model}
\end{eqnarray}
where $\tilde{C}_\ell^{XX}$ represents the biased power spectrum computed from the map, while $C_\ell^{XX}$ is the true sky power spectrum to be estimated.

The $\tilde{C}_\ell$ spectra are evaluated using \xpure, the bins in $\ell$ are centred at $\ell = 400, 700, 1100, 1500, 1900$ and have a width of $\Delta \ell = 400$, except for the first bin ($\Delta \ell = 200$). The transfer functions, $f_\ell^{XY}$, are evaluated using CMB signal only simulations assuming the power spectrum defined by the Planck best fit parameters \cite{planck2015parameters} and zero tensor-to-scalar ratio, $r=0$.

In order to evaluate $f_\ell^{XY}$ a $Y$-only realization of the sky is performed using the \texttt{synfast} tool of the HEALPix package~\citep{Gorski2005}. Knowing the underlying algebra of the map estimator, the biased map is extracted from the simulated sky (analogously one could project the sky into the time-domain using $\b{A}$ and run the map-maker on these timestreams, see
\secref{sec:signalSimulations} for more details). A biased spectrum $\tilde{C}_\ell$ is computed then using \xpure. This procedure is repeated for $100$ independent sky realizations and the transfer function is evaluated as
\begin{eqnarray} 
f_\ell^{XY} = \frac{\langle \tilde{C}_\ell^{XX} \rangle}{C_\ell^{YY}},
\end{eqnarray}
where $\langle \cdots \rangle$ denotes the average over the 100 simulations.
The variance on this determination of $f_\ell^{XY}$ is
$\text{Var}(\tilde{C}^{XX}_\ell)/(100\, \langle C^{XX} \rangle^2)$, in our case it corresponds
to an uncertainty on $f_\ell^{XY}$ of the order of percent, which is
sufficiently small to ignore possible biases that would arise from an inaccurate
estimation of the transfer functions.
Analogously the noise bias, $N_\ell$, is estimated as the average of spectra produced by \xpure{} on $100$ noise only maps. These maps were produced running each map-maker
on timestreams drawn from a Gaussian distribution with covariance $\C{w}$ (or
through an equivalent procedure, see \secref{sec:noiseSimulations}).

We now have all the ingredients needed for calculating the unbiased power spectrum estimator, which is given by
\begin{eqnarray} 
\left( \begin{array}[c]{c} \hat{C}_\ell^{EE}\\ \hat{C}_\ell^{BB} \end{array} \right) = \left[ \begin{array}[c]{cc} f_\ell^{EE} & f_\ell^{EB} \\ f_\ell^{BE} & f_\ell^{BB}  \end{array} \right]^{-1} \left( \begin{array}[c]{c} \tilde{C}_\ell^{EE} - N_\ell^{EE}\\ \tilde{C}_\ell^{BB} - N_\ell^{BB} \end{array} \right). 
\label{spectrum}
\end{eqnarray}
This extension to polarization of~\cite{Hivon2002} is equivalent to the approach adopted by \cite{POLARBEAR} if $f^{BE},f^{EB} \ll 1$, a condition which is also fulfilled in our case since we are considering a very similar observation.
\begin{figure*}[!ht] \centering 
\includegraphics[width=\textwidth]{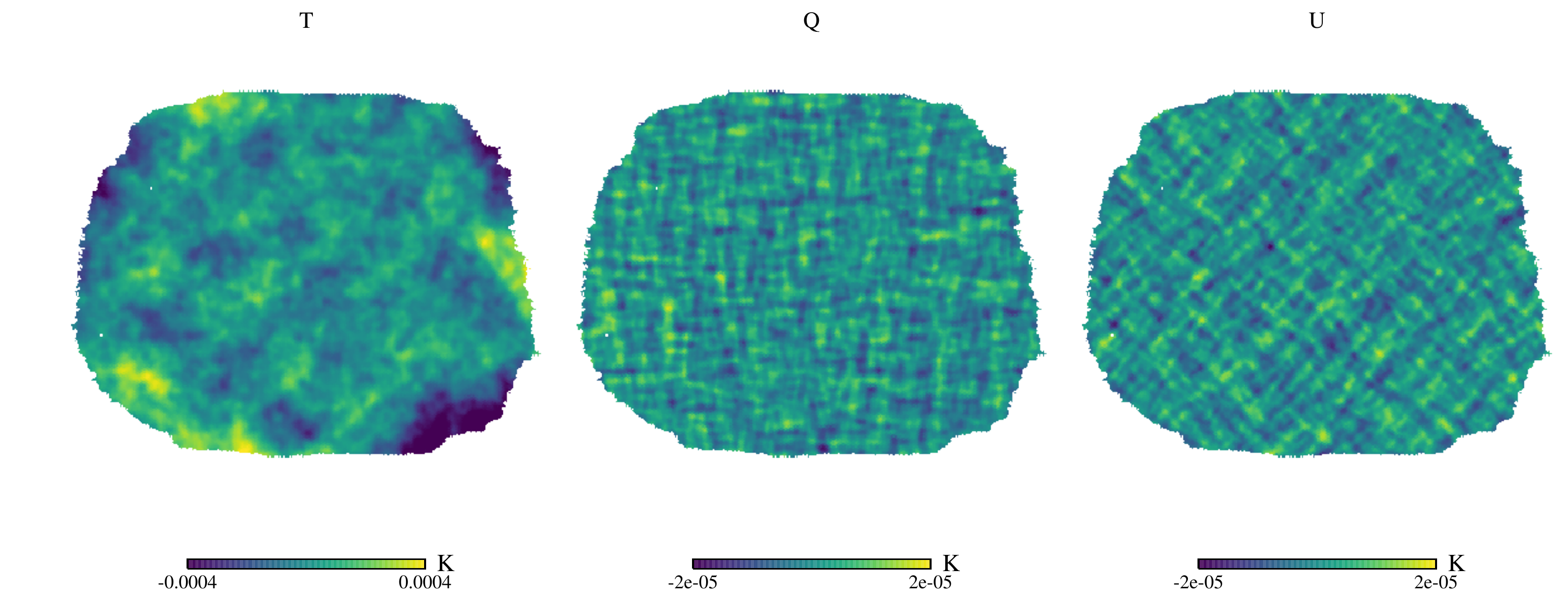}\\ 
\includegraphics[width=\textwidth]{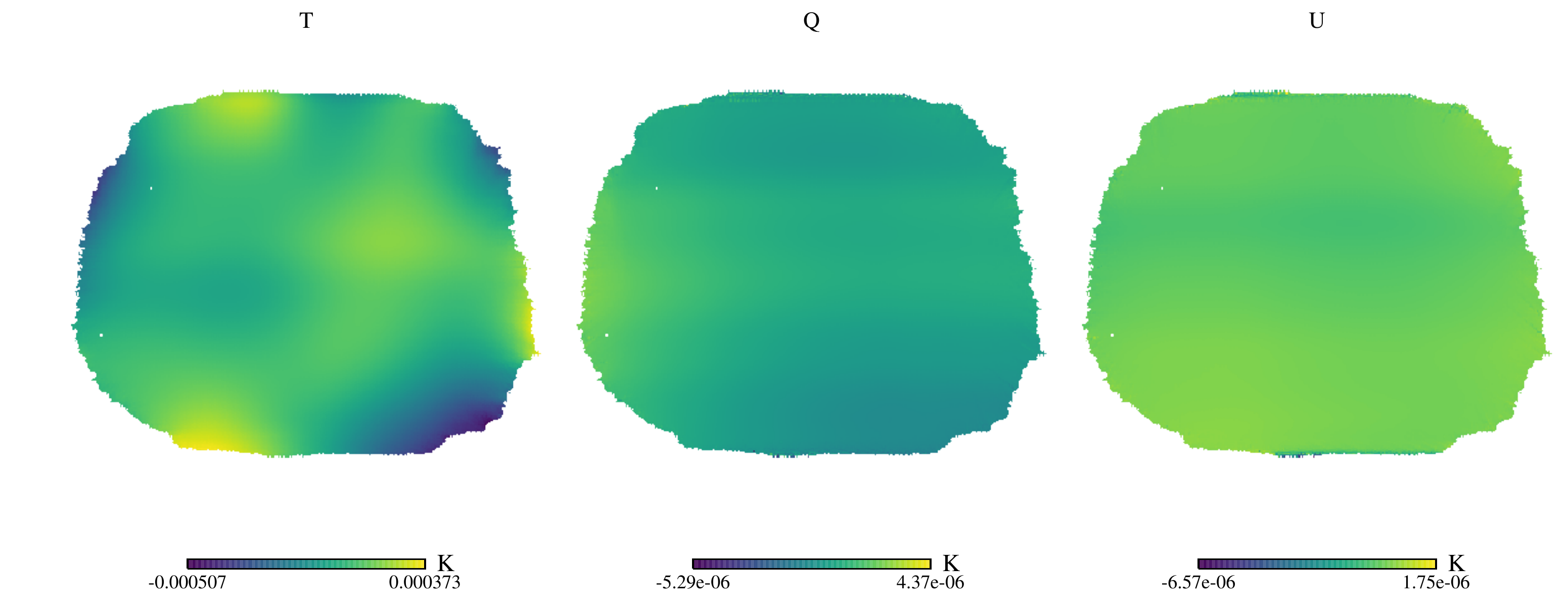}\\ 
\includegraphics[width=.33\textwidth]{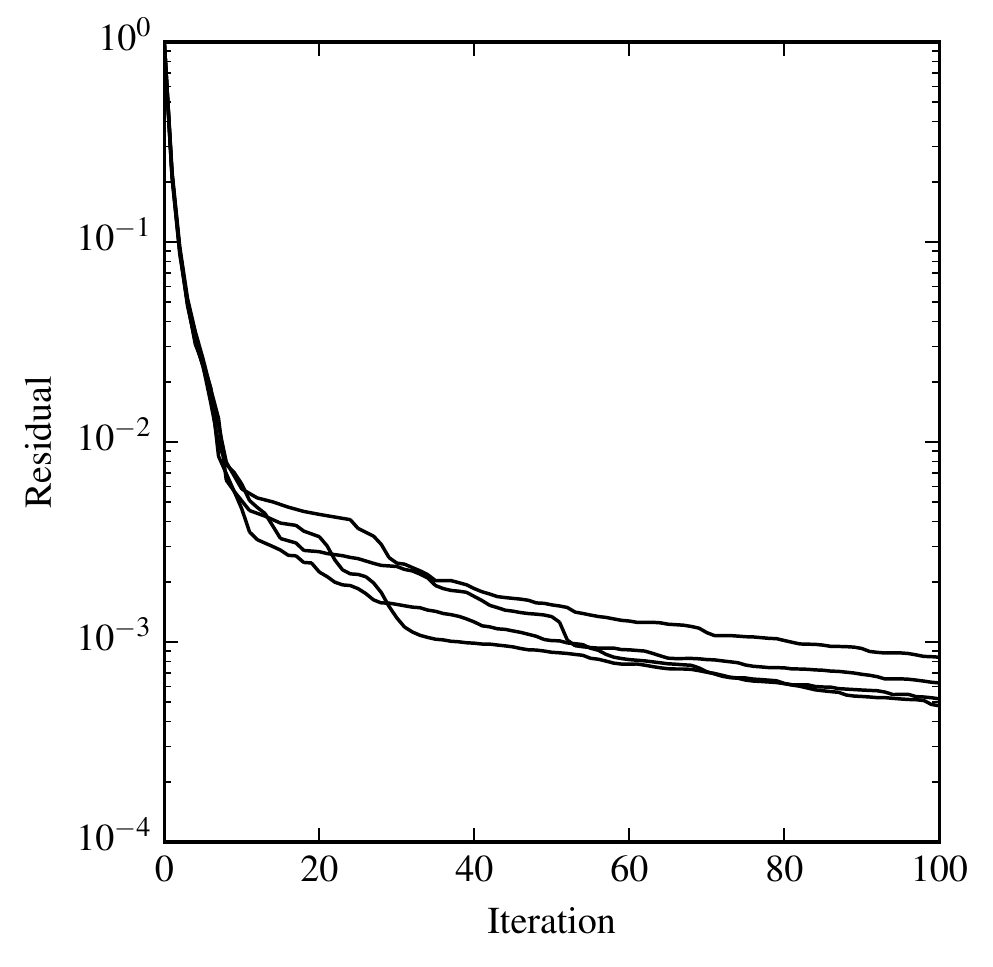} 
\includegraphics[width=.32\textwidth]{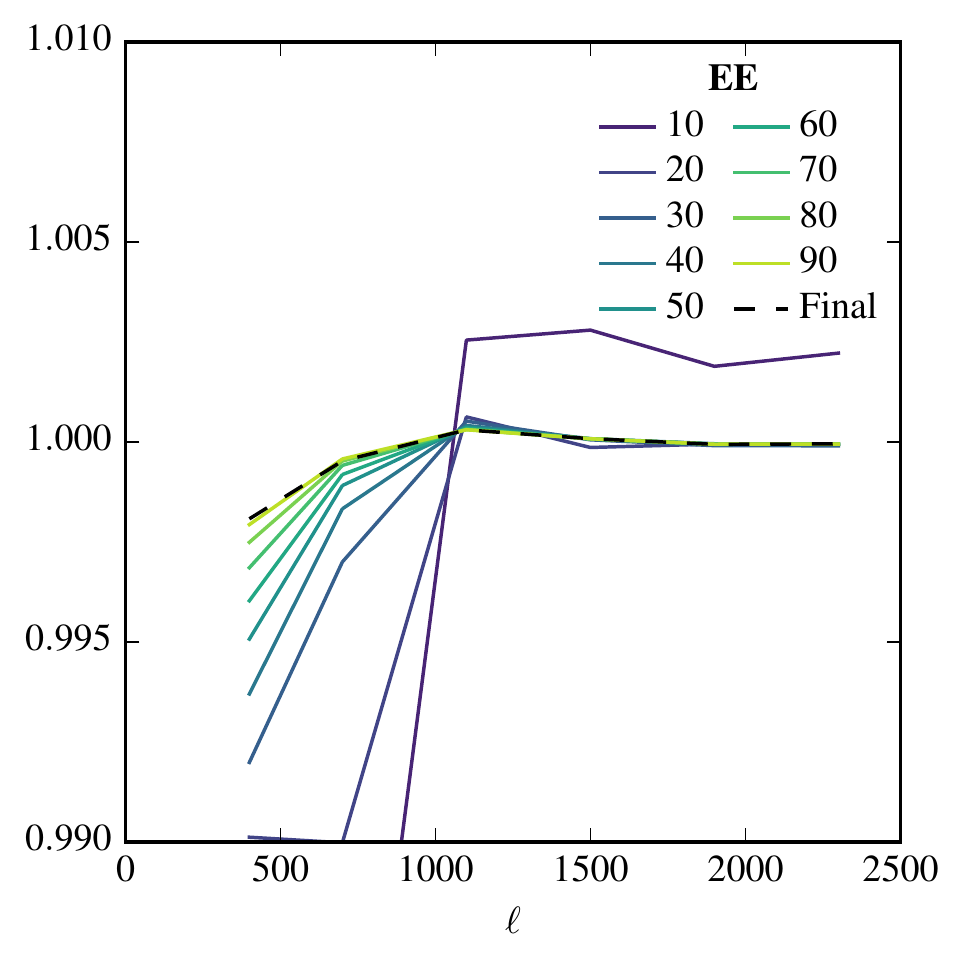} 
\includegraphics[width=.32\textwidth]{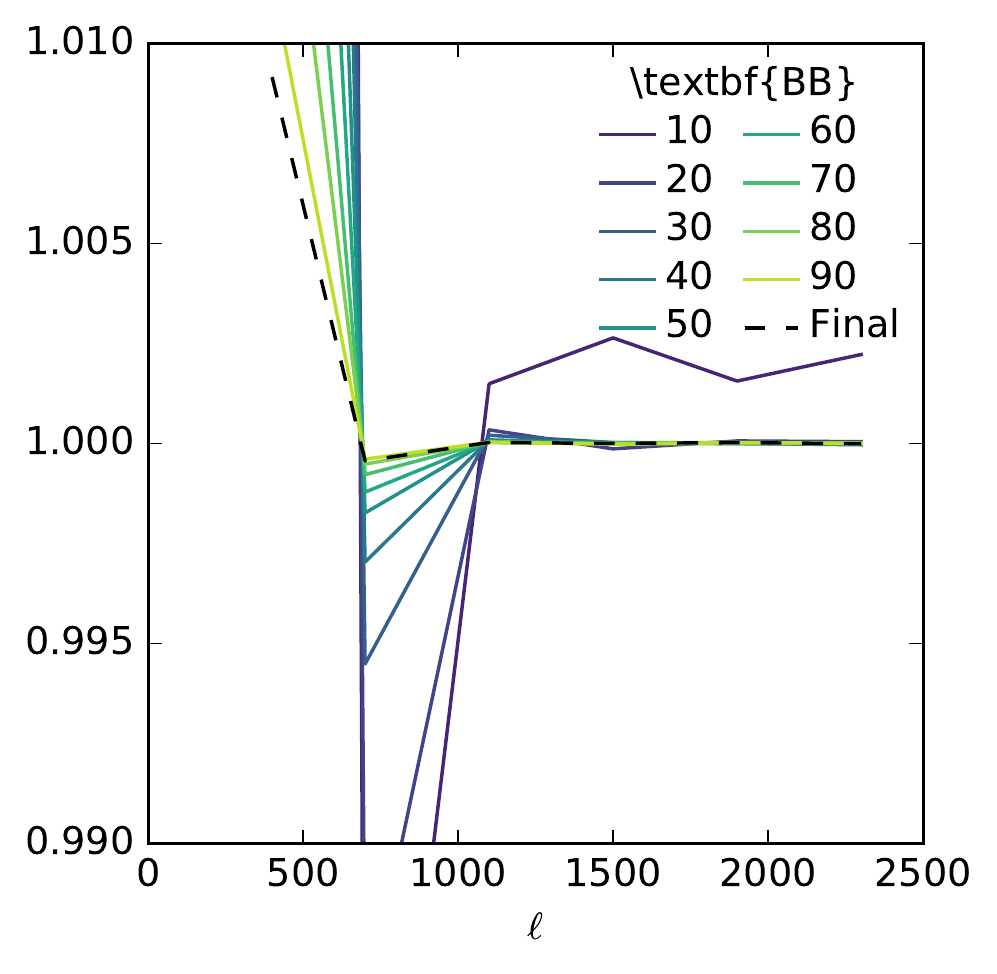} 
\caption{Results from the iterative implementation of the unbiased map
  estimator. \emph{Top row:} reconstructed maps. \emph{Middle row:} difference
  between the reconstructed and the input maps. \emph{Bottom row:} A study of convergence of our iterative solver. The left panel shows the standard residual,
 as defined in  \eqref{eqn:pcgResDef}, which saturates and does not converge to our fiducial
 level of  $10^{-6}$  in as many as $100$ iterations. The middle and right panels show that this lack of convergence is due to the largest angular modes as the
  fractional difference between the power spectrum of the input map and the power spectrum of the $i$th map estimate in the
  multipole range $\ell \in [500, 2100]$ becomes quickly very small  and reaches the level of better than $0.1$\% in fewer than $\sim\hskip -2pt 100$ iterations. 
  This last observation has been used to set the convergence criterion 
  used in the analysis of the first year {\sc Polarbear} data set~\citet{POLARBEAR}.} \label{fig:pcgMap}
\end{figure*}
\begin{figure*}[h!] \centering 
\includegraphics[width=\textwidth]{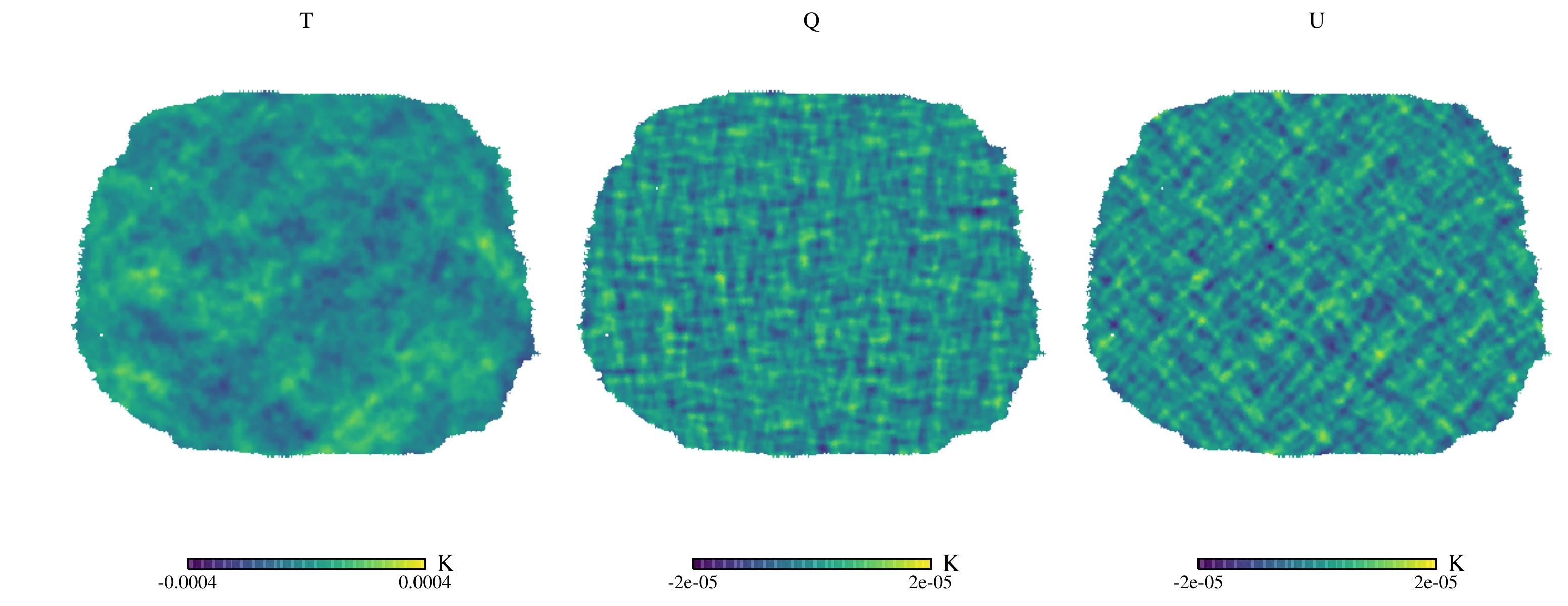}\\ 
\includegraphics[width=\textwidth]{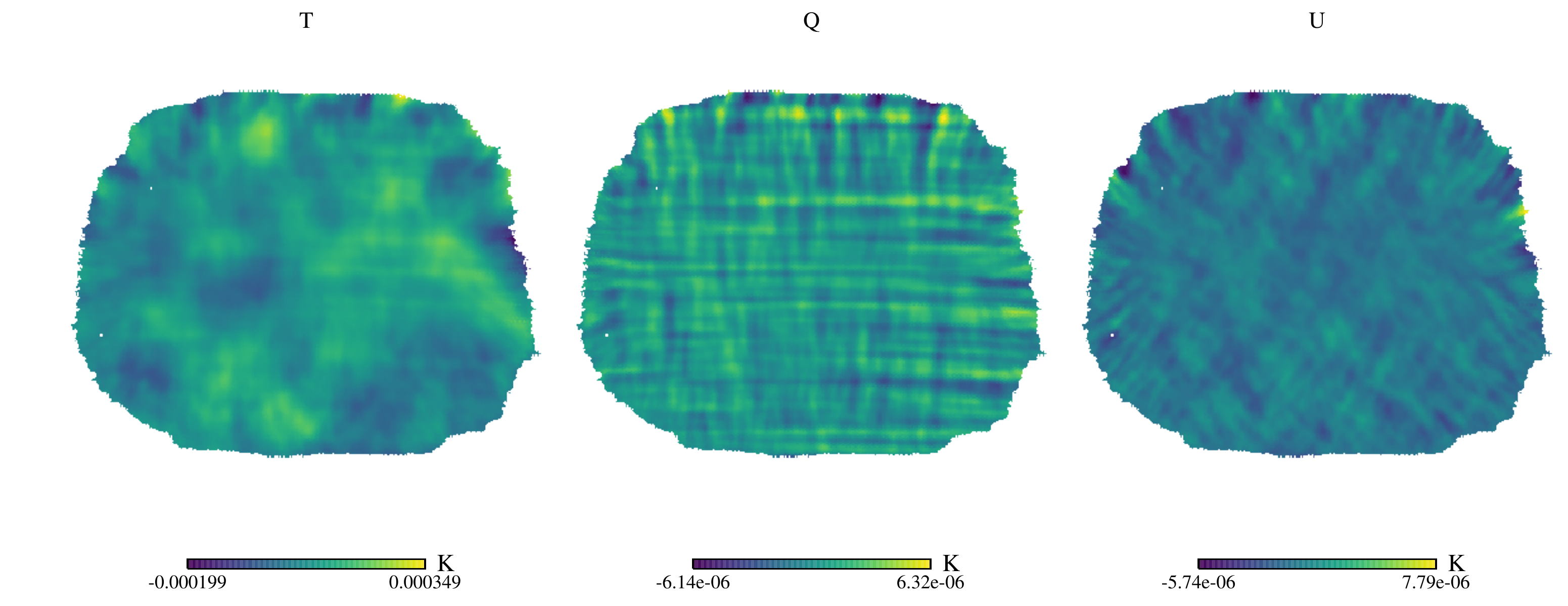}\\ 
\caption{Map estimates derived using the biased map estimator, Eq.~(\ref{s_biased}). \emph{Upper row:} reconstructed maps. \emph{Bottom row:} difference between the reconstructed and the input maps.} \label{fig:biasedMap}
\end{figure*}

\section{Results}
\label{results}
Here we describe the maps produced from \pb-like simulations using the different map-making approaches and their implementations. We assess the maps from the point of view of the fidelity with which they reconstruct the actual sky signal. However, we also look at them as merely a step toward a statistical characterization of the signal. In this latter case, we use the $B$-mode polarization power spectrum as a comparison metric and specifically focus on the power spectrum estimation approach as introduced in the last section.

\subsection{Reconstructed sky maps}
\label{reconstructedMaps}
In Figs. \ref{fig:inputMap}, \ref{fig:unbiasedMap}, \ref{fig:pcgMap}, \ref{fig:biasedMap} we compare the maps reconstructed using
both the biased and unbiased estimators, in the latter case using both the explicit and the iterative solver. The maps were computed 
assuming noiseless data and the same pixel size was adopted for the simulation and reconstruction to avoid any pixel effects. In none of the cases considered does the reconstructed sky correspond exactly to the input. We discuss each of the cases in turn below.
\paragraph{Unbiased maps via the explicit solver.} The residual present in this case, \figref{fig:unbiasedMap}, is due to the presence of singularities in the system matrix, $\AtFA$. As expected from our earlier discussion, \secref{degeneracies}, and confirmed by our
results in~\secref{eigenstructure}, there are two such singular modes for
polarization and one for temperature. The inversion regularization procedure
removes these two modes from the estimated map, even if they contain actual sky
signal. These singular modes result from the interplay between the scanning
strategy
and the filtering. They make the filtered data insensitive to these modes,
which are unavoidably lost. Therefore, the map estimated from the
unbiased estimator using the explicit solver may not be strictly speaking
unbiased but provides an unbiased representation of all the modes that can be
constrained from the available data given the chosen filtering.

\paragraph{Unbiased maps via the iterative solver.} The residual in this case is clearly more pronounced and complex (see \figref{fig:pcgMap}, middle panel). As we mentioned earlier we use a PCG solver and adopt as the preconditioner the matrix 
$(\b{A}^\T\b{M}\b{A})^{-1}$. One might expect that the result of the unbiased
map estimator should be the same, whichever solver is applied. However, the
result shown in the figure corresponds to an incompletely converged iterative
solution. Indeed, we have found that the iterative solution residuals,
~\eqref{eqn:pcgResDef}, do not decay to zero, see the bottom left panel of
\figref{fig:pcgMap}, but instead asymptote to about $10^{-3}$, which is roughly
three orders of magnitude above our fiducial convergence criterion of $10^{-6}$. This is the case even if we allow as many as a few thousand iterations. Such a behavior is indeed expected in linear systems for which the system matrix is (numerically) nearly singular~\citep[][]{Hanke1995, Szydlarski2014}. 

To understand the effects of this lack of convergence of the solver on the estimated maps, we have computed the (pseudo) power spectra of the estimated map after $i$th iteration, bottom right panels of \figref{fig:pcgMap}. We see that although the very low $\ell$ part of the spectrum does indeed fail to converge, convergence is quickly reached in the intermediate and high $\ell$-range. This again is consistent with the singular modes found in the explicit solver having only large angular scales. If the singular modes are known, we could readily remove them from the solution, and thus from the residuals,  at each step of the iteration and restore the proper convergence. However, this typically would require as many computations as the explicit solver, undermining the most important advantage of the iterative one.

We can still use the PCG solver in such circumstances by using this practical workaround: instead of monitoring a single residual as given by~\eqref{eqn:pcgResDef}, we track the behavior of residuals at the scales of interest for the power spectrum, as shown in
\figref{fig:pcgMap}. Nevertheless, we have to be aware of the fact that some of the power in the final solution may be compromised.

In the case of the map shown in the figure, convergence can indeed be reached in fewer than $100$ iterations in the so-called 
science band defined in~\citet{POLARBEAR}, which was the band of interest for the first round of the \pb{} papers. 

\paragraph{Biased maps.} The biased map estimator, as expected, leads to the largest residuals. These are particularly pronounced in the outskirts of the map, where the pixels crossing (and thus the filtering) may be highly anisotropic, but are still readily visible in the central part of the map, where the cross-linking and pixel sampling are better.

\begin{figure}[t] 
\centering 
\includegraphics[width=0.5\textwidth]{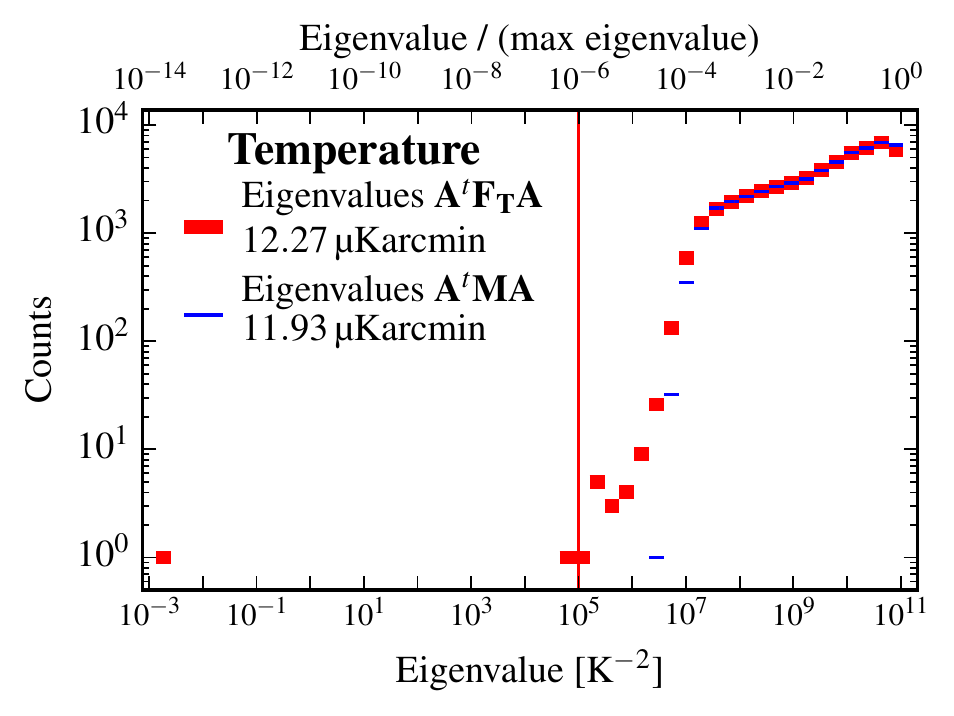} 
\caption{Eigenvalues of the temperature block of the $\AtFA$ and $\AtNA$ matrices. The spectra of the two matrices are very similar. The major difference is a group of poorly constrained modes, including one that is formally degenerate corresponding to the offset of the map.} 
\label{fig:eigenvaluesT}
\end{figure}

\begin{figure}[t] 
\centering 
\includegraphics[width=0.5\textwidth]{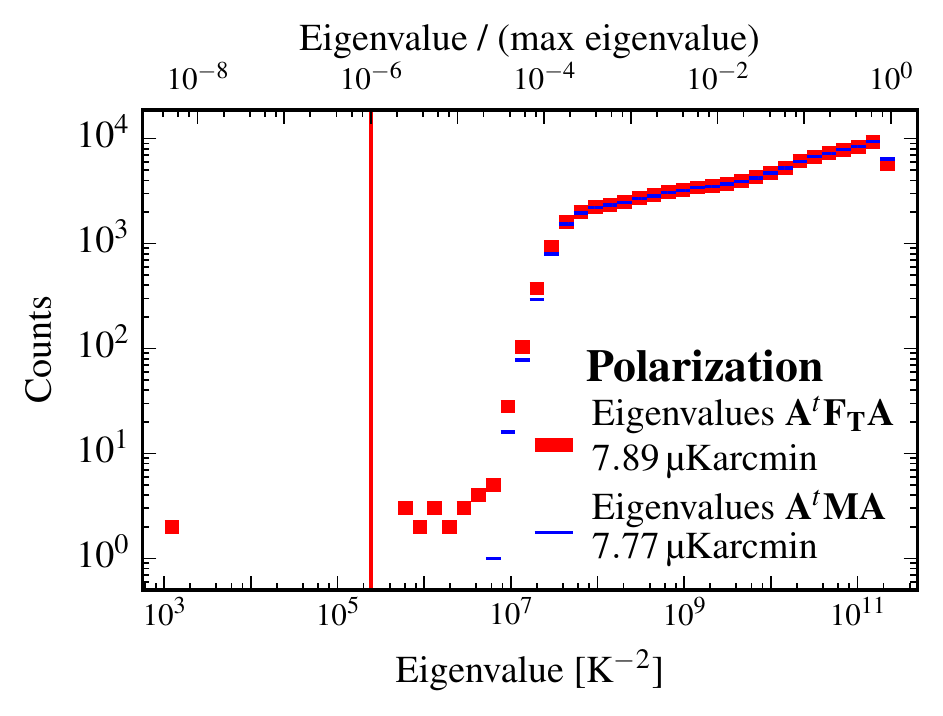} 
\caption{Eigenvalues of the polarization block of the $\AtFA$ and $\AtNA$ matrices. As with temperature, the spectra of the two matrices are very similar. In this case we also have tens of poorly constrained modes. We also expect to have two nearly degenerate modes, one for each Stokes parameter (cf. \figref{fig:eigenvaluesT}. Though they are treated as singular because of numerical reasons, their degeneracy is partially broken, as explained in \secref{ground_pickup}.} \label{fig:eigenvaluesP}
\end{figure}

\begin{figure*}[t] 
\centering 
\includegraphics[width=0.34\textwidth]{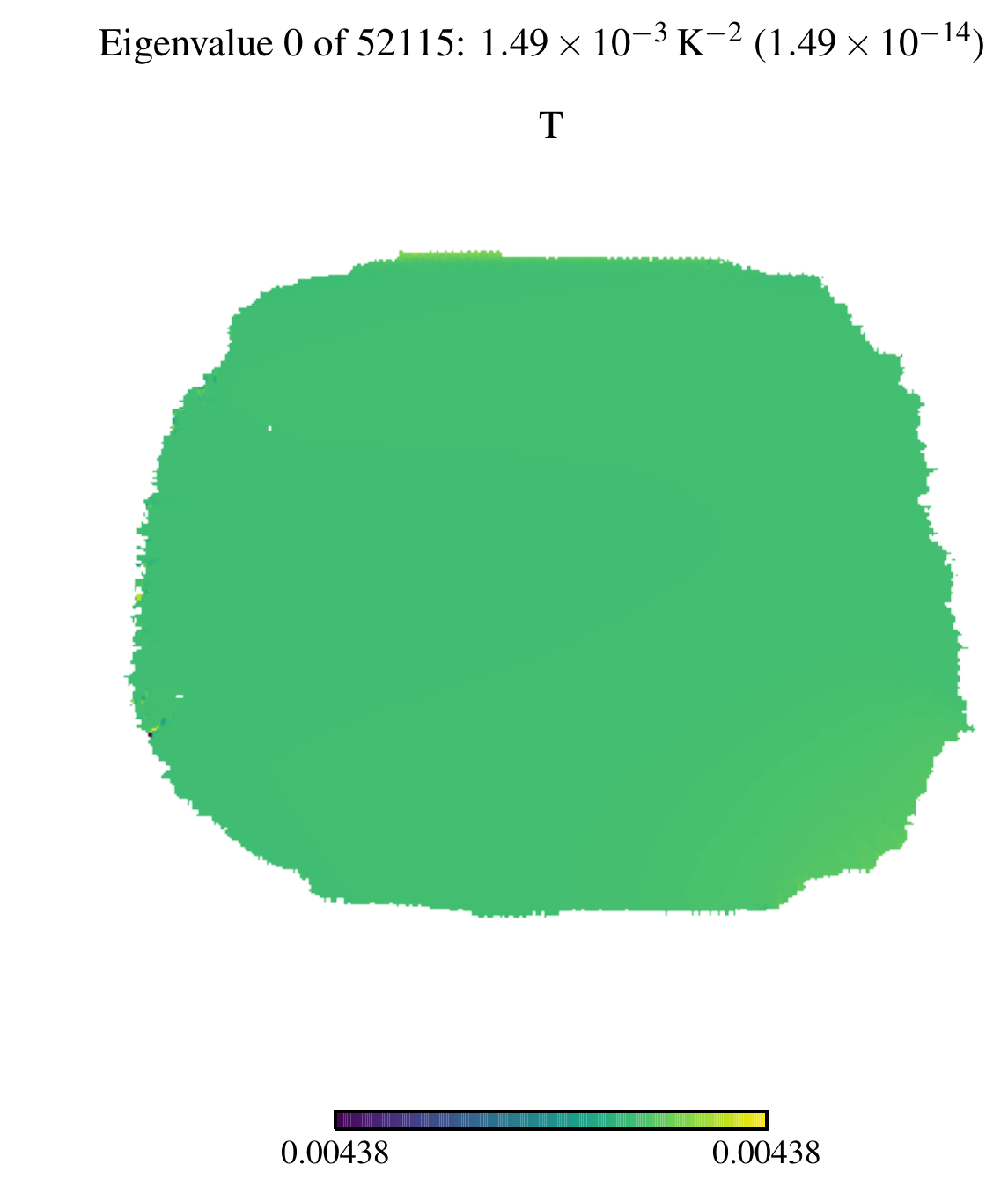} \hspace{0.01\textwidth} \includegraphics[width=0.64\textwidth]{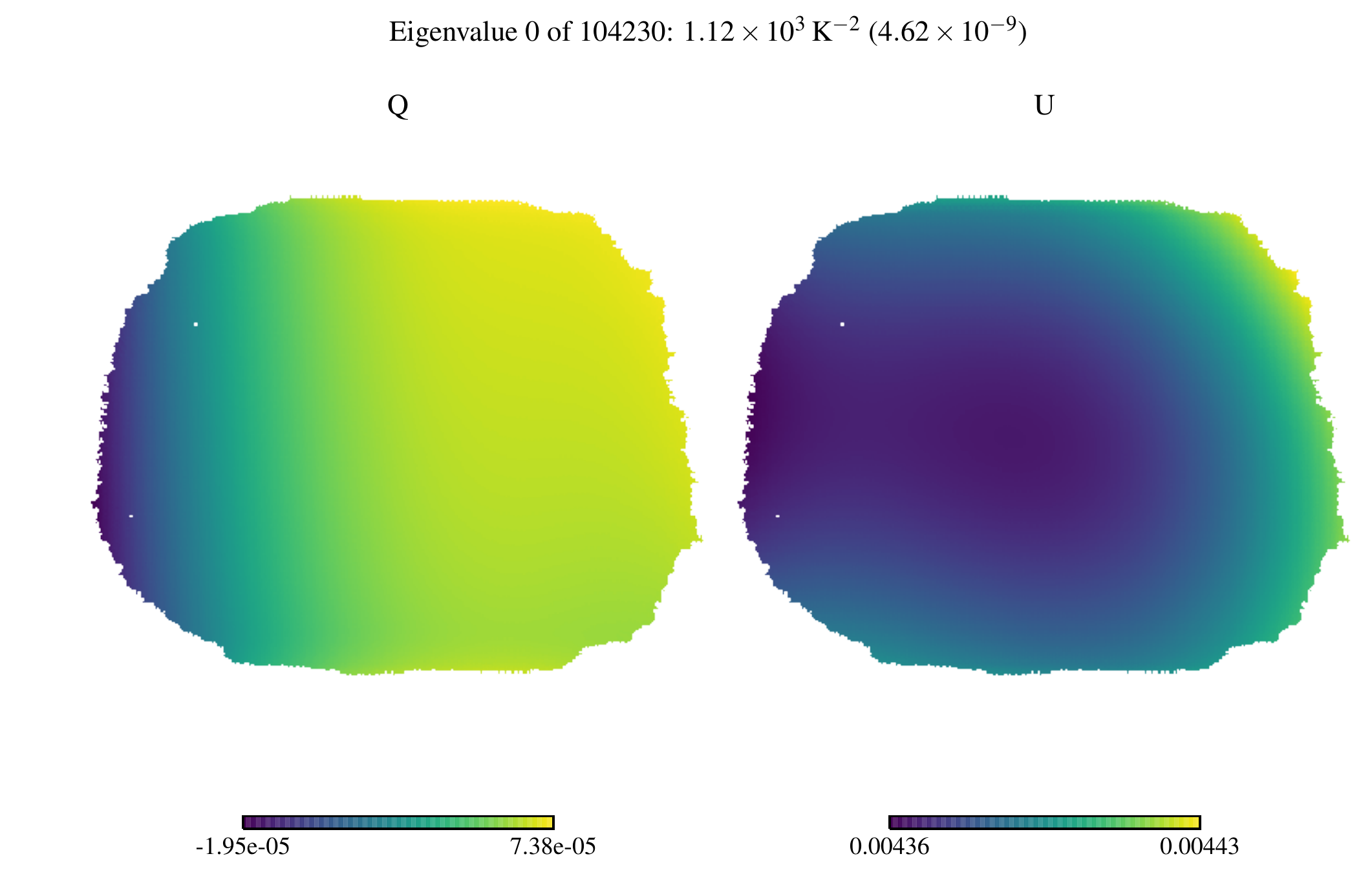} 
\caption{\emph{Degenerate modes} (the parenthesis in the title reports the corresponding eigenvalue divided by the largest eigenvalue). The degenerate mode of the temperature (left) map reconstruction is the global offset. The Q and U maps on the left are one of the two singular modes of polarization, the other mode is very similar: Q and U maps are swapped and the sign of one of them is flipped. The degenerate modes of the polarization are more complex because of the modulation of the polarization angle during the observation (see \secref{ground_pickup}). For the same reason the degeneracy of these modes is partially broken (though they have to be treated as singular in order to preserve the numerical stability of the inversion of the $\AtFA$ matrix).} \label{fig:singularModes}
\end{figure*}

\begin{figure*}[h] 
\centering 
\includegraphics[width=0.34\textwidth]{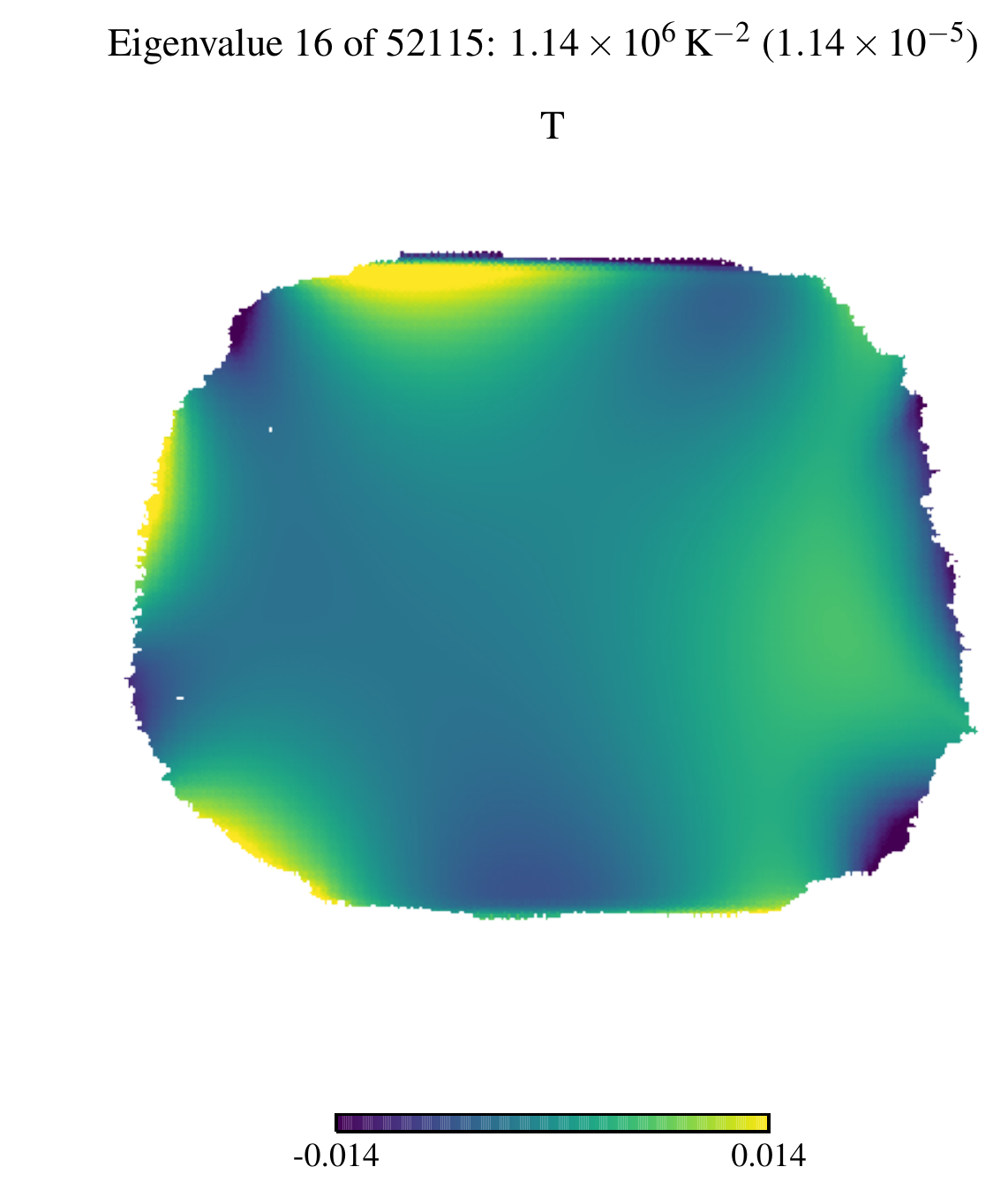} \hspace{0.01\textwidth} 
\includegraphics[width=0.64\textwidth]{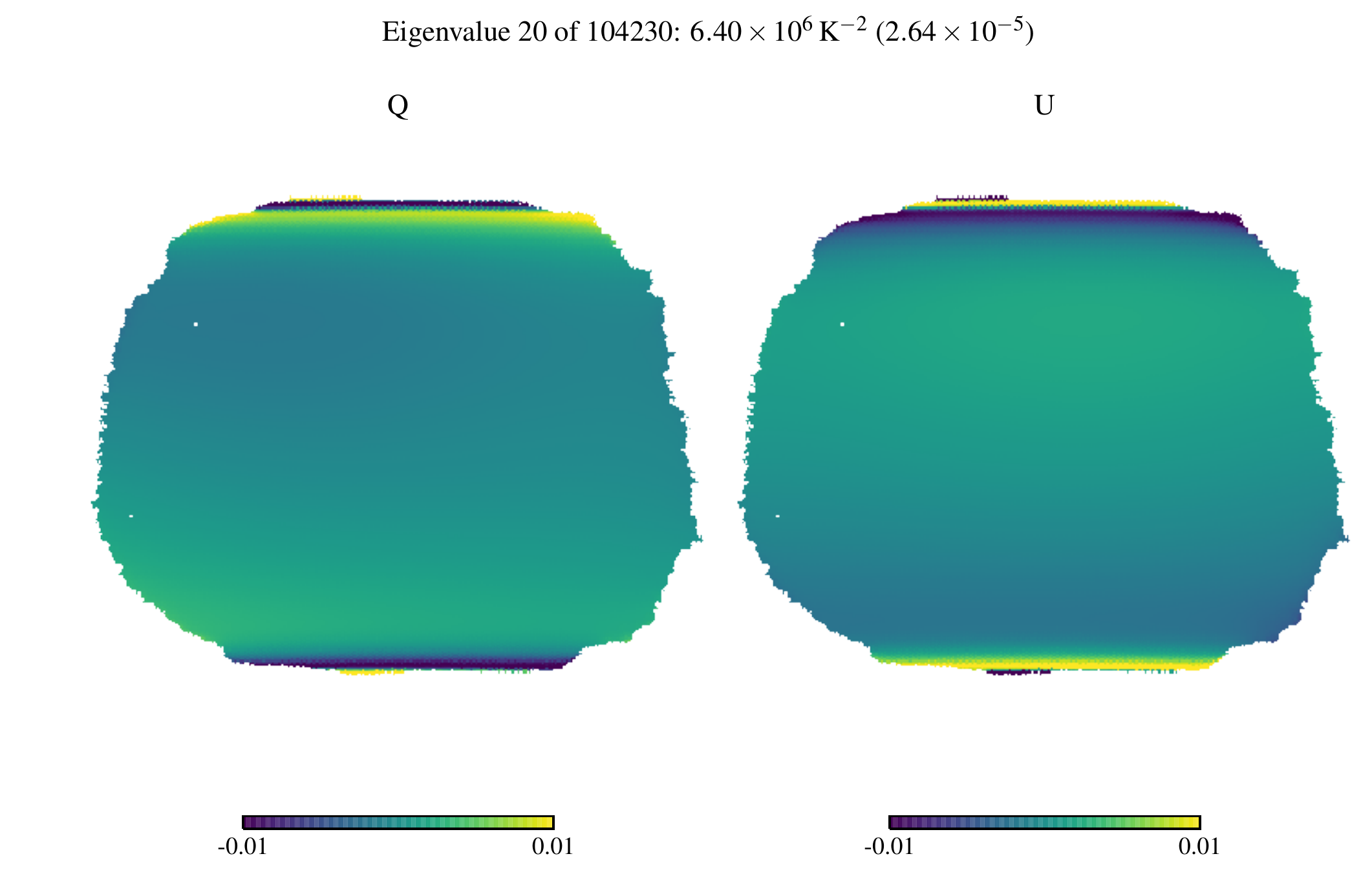} 
\caption{\emph{Examples of nearly degenerate modes} (the parenthesis in the
title reports the corresponding eigenvalue divided by the largest eigenvalue).
These modes are composed of some prominent feature at the boundaries and a
(usually) weaker long mode. The structures at the boundaries correspond to sets
of pixels that are heavily affected by filtering. For polarization the dominant
effect is the ground removal. At the high and low declination ends of the
observed area the redundancy of the observations is low and therefore the
degeneracy breaking effects discussed in \secref{ground_pickup} are mild. For
temperature, the high order of the polynomial filtering plays a significant
role, adding prominent features at the boundaries at intermediate declinations
and increasing the complexity of the long modes. We emphasize that the prominent features at the boundaries saturate the color scale.} \label{fig:nearlySingularModes}
\end{figure*}

\begin{figure*}[t] \centering 
\includegraphics[width=0.34\textwidth]{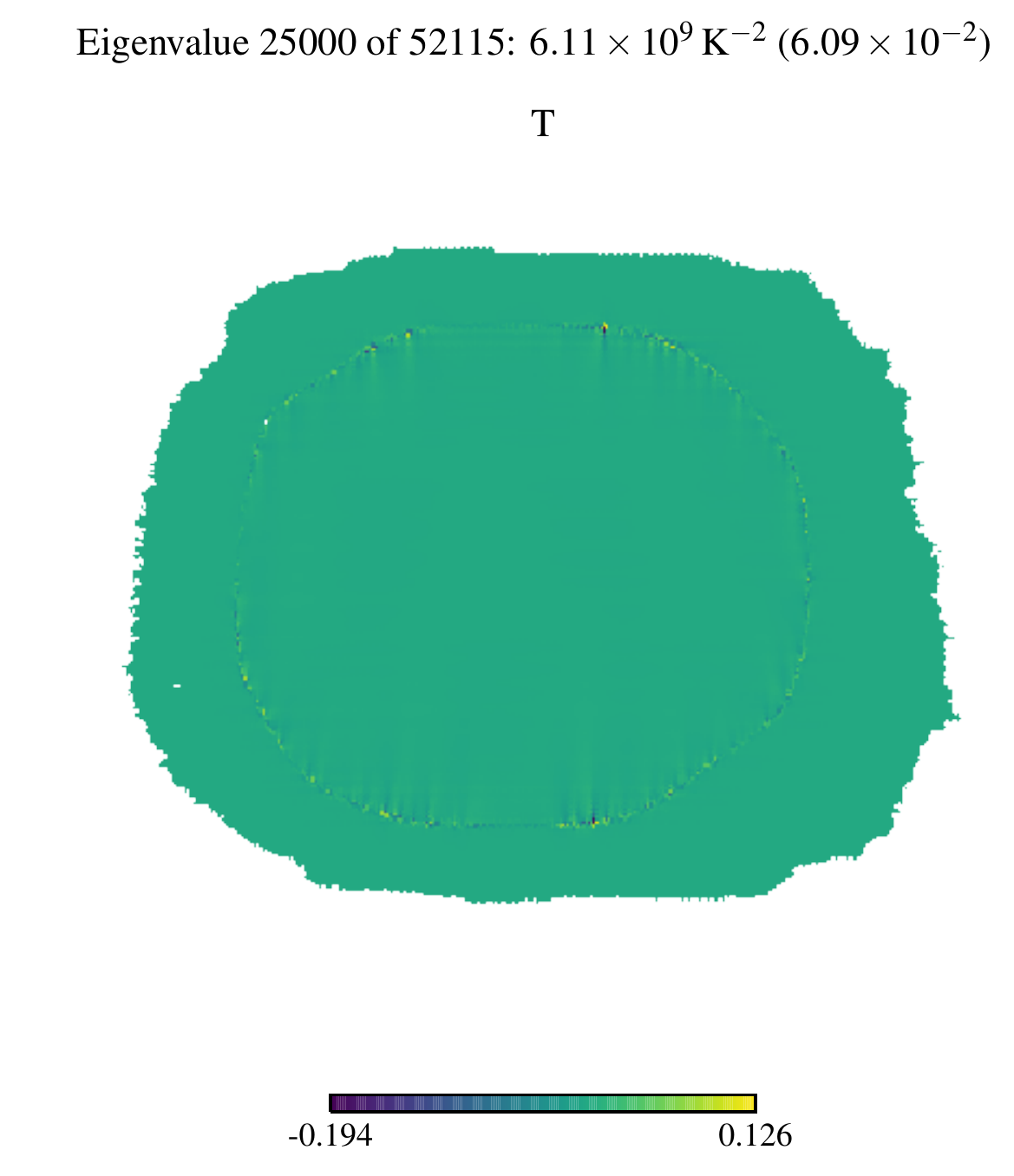} \hspace{0.01\textwidth} 
\includegraphics[width=0.64\textwidth]{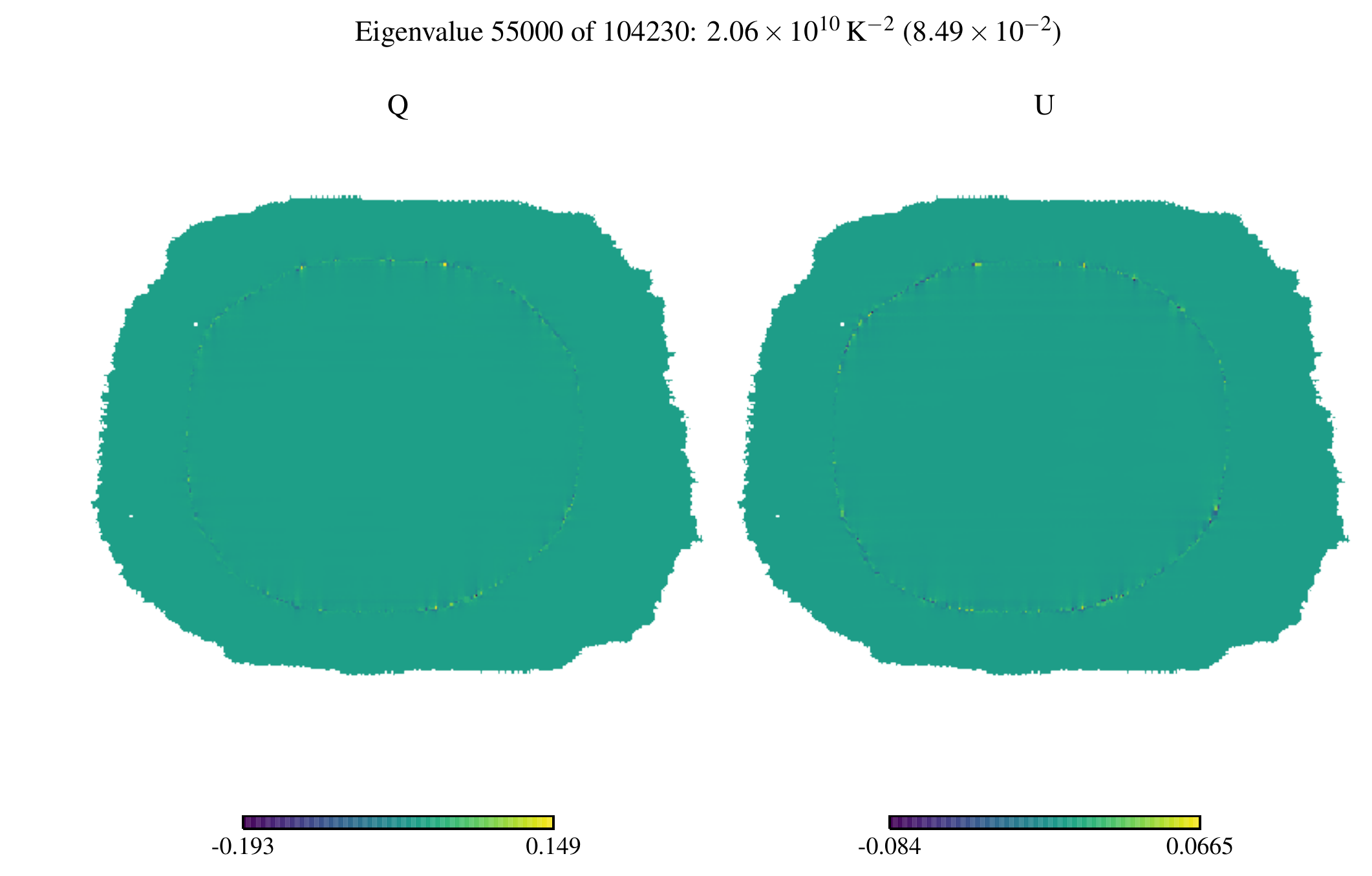} \\ 
\includegraphics[width=0.34\textwidth]{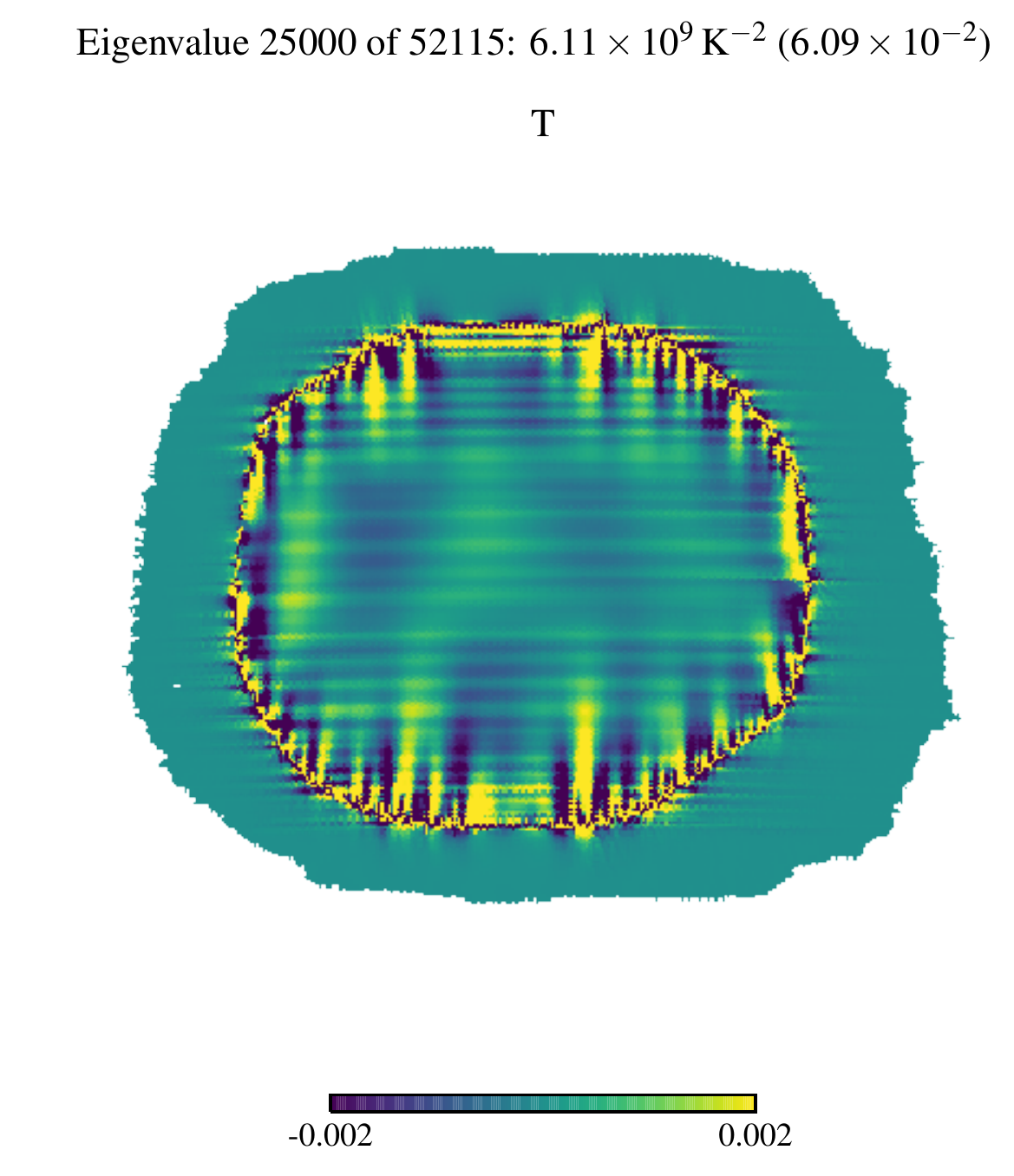} \hspace{0.01\textwidth} 
\includegraphics[width=0.64\textwidth]{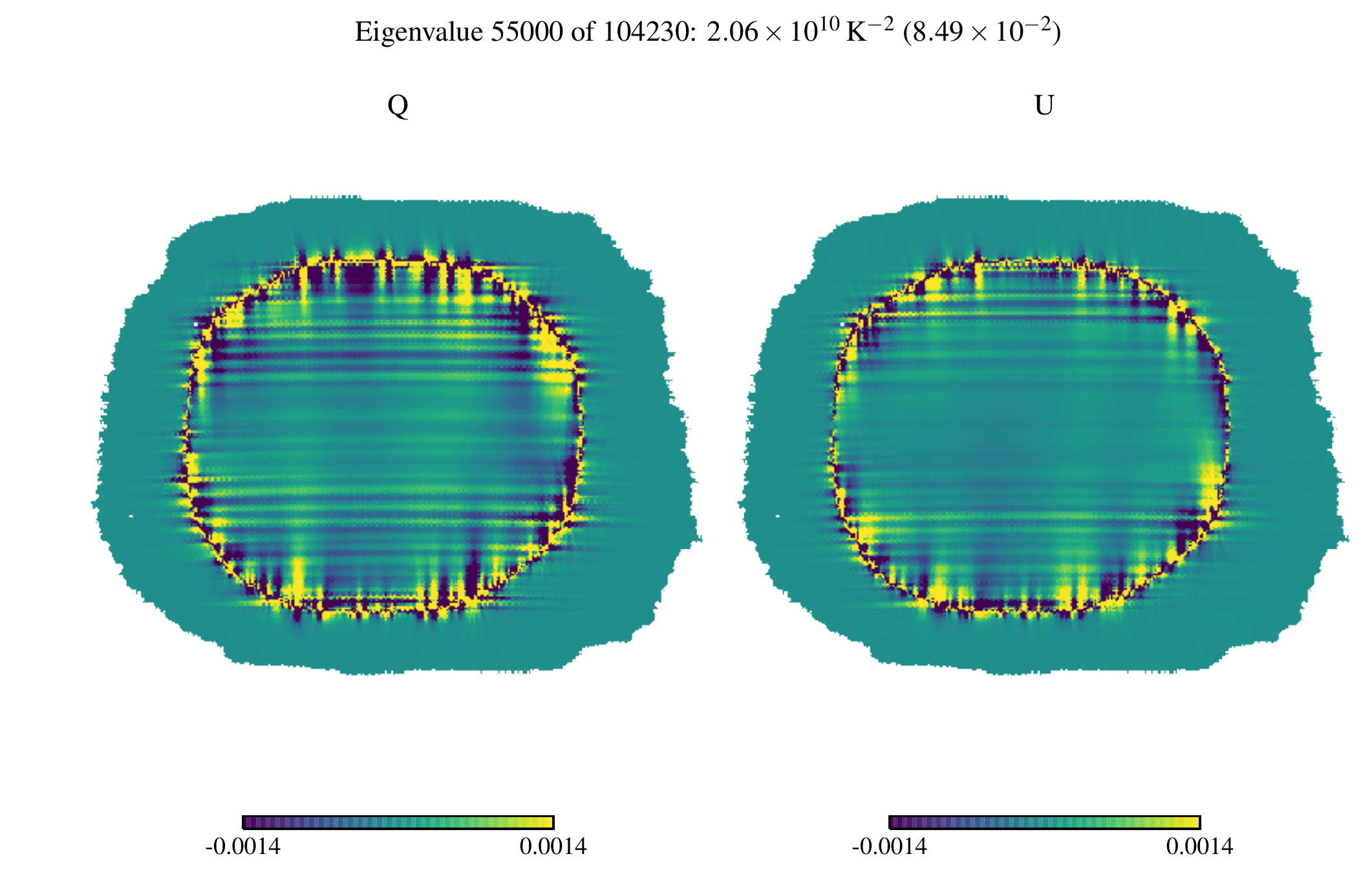} 
\caption{\emph{Examples of pixel-like modes} (the parenthesis in the title reports the corresponding eigenvalue divided by the largest eigenvalue). These modes involve mainly a ``ring'' composed of a very limited number of pixels (see the first row). In the second row, the same eigenvectors are displayed with a color scale squeezed by roughly two orders of magnitude, emphasizing structures inside and outside of the ring. Outside of the ring the structures quickly fade away. Also moving inwards the structures decrease their amplitude. However, compared to the outward structure, their typical length is much larger and, most important, they
do not completely disappear: they have a relevant amplitude in the whole inner region. We emphasize that the ``horizontal'' structures represent the correlations induced by the ground template filtering.} 
\label{fig:pixelLikeModes}
\end{figure*}

\subsection{Eigenstructure of $\AtFA$.}
\label{eigenstructure}

The eigenstructure of $\AtFA$ not only determines which modes are missing from the final unbiased sky estimate, it also provides information about the modes, which, while not singular, are not well constrained by the data. This is because the matrix is closely connected to the noise covariance in the pixel domain, $\b{N_p}$, as shown by Eqs.~(\ref{eqn:noiseCovGen}) and~(\ref{eqn:noiseCovML}).

In the case with no filtering at all the matrix, $\AtFA$, reduces to the
well-known $\AtNA$, which, for the diagonal weights assumed here, is block
diagonal with two-by-two blocks describing the (weighted) coupling between $Q$
and $U$ Stokes parameters in each pixel. The off-diagonal elements of
these blocks will typically be negligible for pixels observed with a
sufficiently homogeneous distribution of polarization angles, while the diagonal
elements will be approximately equal to the eigenvalues of the two-by-two block. These are essentially given by the number of observations per pixel, and their corresponding eigenvectors are spatial modes equal to zero everywhere but in the given pixel. Departure from such behavior would then indicate the presence of strong off-diagonal coupling in some of the pixels.

The spectrum of $\AtNA$ can therefore be used as a good reference for assessing the impact of the filtering on the map domain noise spectrum. We compare the eigenstructure of the matrices, $\AtFA$ and $\AtNA$, in Figs.~\ref{fig:eigenvaluesT}  and~\ref{fig:eigenvaluesP}, for temperature and polarization respectively. In both cases, $\AtFA$ and $\AtNA$ have very similar
spectra with the exception of tens of poorly constrained modes, defined as those with eigenvalues smaller by at least five orders of magnitude than the maximum eigenvalue. The corresponding eigenvectors, \figref{fig:nearlySingularModes}, are long modes and,
in many cases, they exhibit a striped structure close to the boundaries. We
interpret these modes as the result of the ground signal filtering discussed in
\secref{ground_pickup} as their number roughly corresponds to the number of bins
in the ground-bin. Indeed, following the discussion of \secref{ground_pickup},
we expect that with every ground-template bin there should be an associated
ill-constrained mode, corresponding to an offset of the constant declination
strip swept by the azimuth range of the bin during the time of a single constant
elevation scan. The fact that the recovered eigenvalues are not numerically zero
demonstrates that these degeneracies are weakly broken as expected given that
the presence of sky pixels observed with the telescope orientation corresponding
to two different ground-template bins. Consequently, this leads to only one
truly degenerate mode per each Stokes parameter map, see~\figref{fig:singularModes}.

We also see that the two most singular eigenvalues of the polarized case are significantly larger than the most singular eigenvalue in the case of temperature. The difference is at least in part due to the numerical precision of the computations, however it is also consistent with our earlier expectation that the polarizer angle change across the constant elevation sweep can break the degeneracy between the sky signal offset for each strip and the amplitude of the ground template in the corresponding bin. In the case under study, given the limited primary mirror chop, the effect is very weak.

While regularizing the inversion of the matrix, $\AtFA$, we remove these modes from the solution together with all the modes which are smaller than $10^{-6}$ of the maximal eigenvalue. This still leaves significant number of ill-constrained modes in the estimated maps. Although they do not give rise to any discernible artefacts in the caseof the noiseless results as shown~\figref{fig:unbiasedMap}, when noise is included in the time-domain data, these modes may dominate the maps visual appearance. However, despite being noisy these modes are correctly estimated, and are neither artefacts of the estimator or its implementation, nor remnants of any incompletely-filtered parasitic signal such as atmosphere. Rather they reflect the actual uncertainty that the observation and filtering incur. These modes are typically missing in the unbiased maps derived with the iterative solver as well as in the biased maps. This is because these modes are either the most difficult to converge (in the case of the iterative solver) or are explicitly filtered out (in the biased map-maker). Consequently, although these maps may occasionally -- and somewhat deceptively -- look better, they will nonetheless be missing information which is correctly included in the unbiased map computed using the explicit solver.

The main sequence of eigenvectors are ``pixel-like'' modes,
\figref{fig:pixelLikeModes}, in the sense that in each of these modes the most
relevant structure involves a very limited number of pixels. As one intuitively
expects, these pixels move from the boundary toward the center of the patch as
the eigenvalue of the mode grows (i.e., as the mode is better constrained). If there were no filtering, $\AtFA$ would be block diagonal (it would be equal to $\AtNA$) and therefore each eigenvector would correspond to exactly one pixel. Setting a threshold on the magnitude of the eigenvalues would be then equivalent to performing a selection of the best observed (here innermost) pixels. In the presence of filtering, the main effect of setting a threshold is still selecting the innermost pixels. However, because of the correlations that the filtering introduces, signal is also removed from all over the map,
affecting areas relatively far from the boundary pixels removed. This effect is visible at map-level in the lower panel of \figref{fig:pixelLikeModes} and it is investigated at the power-spectrum level in \secref{power_spectrum}: removing modes that mainly involve pixels outside of the power-spectrum mask has an important impact on the power spectrum uncertainty and bias.

\begin{figure*}[t] \centering 
\includegraphics[width=0.49\textwidth]{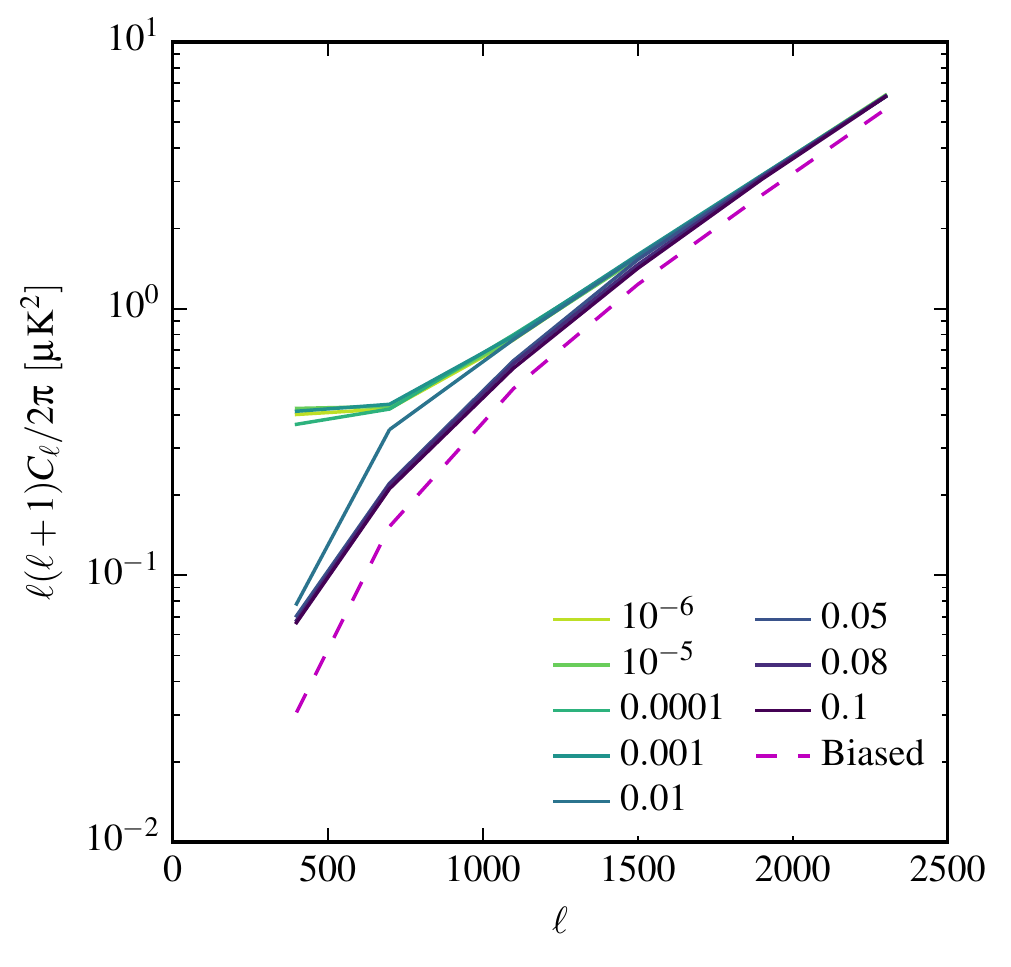} 
\includegraphics[width=0.49\textwidth]{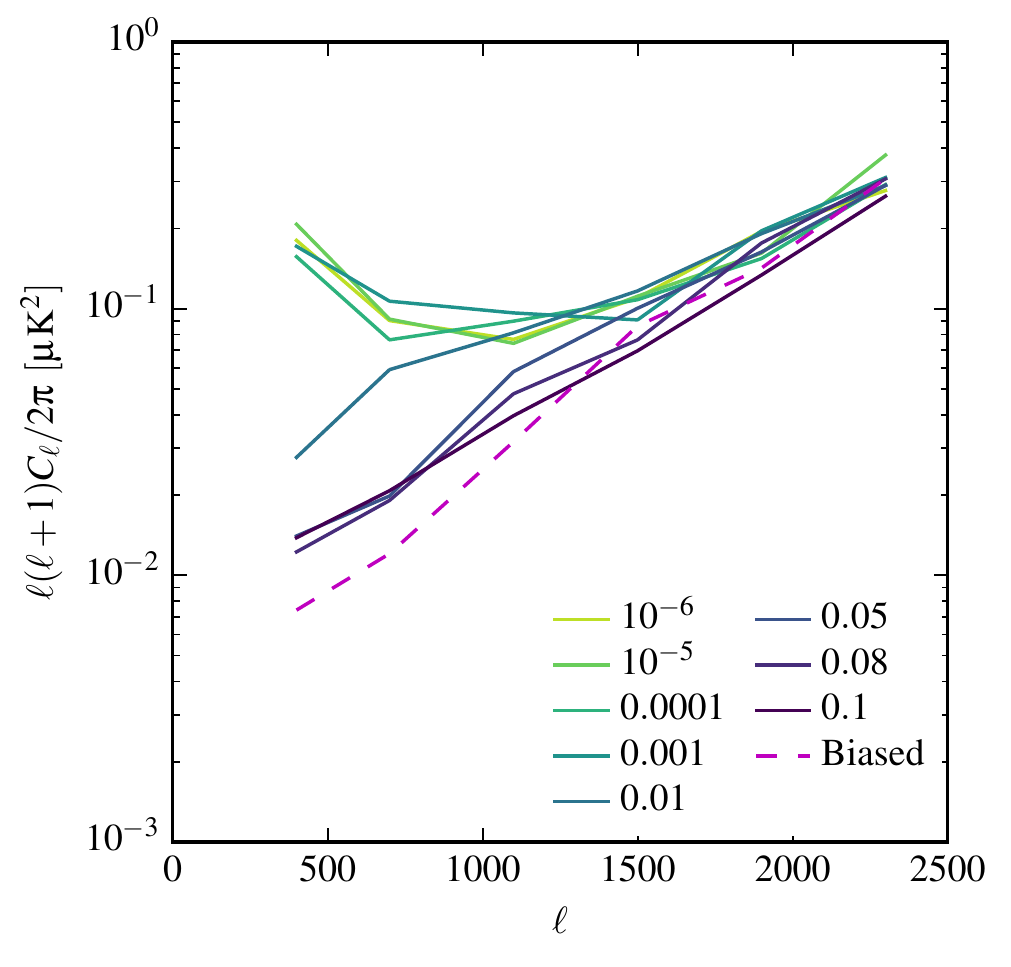} 
\caption{Mean (left) and standard deviation (right) of the $B$-mode power spectra (as computed by \xpure) of 100 noise only simulations for different input maps. ``Biased'' refers to the biased map estimator defined in Eq. (\ref{s_biased}). The other
lines refer to different maps derived from the unbiased map estimator Eq. (\ref{s_estimator}). The value in the legend is the $\alpha$ parameter that quantifies the amount of map domain filtering applied to the unbiased map. Higher values of $\alpha$ correspond to more aggressive filtering, see \secref{noise_bias} for more details} 
\label{fig:noise}
\end{figure*}

\subsection{Power-spectrum analysis}
We now investigate how the map domain properties for the different estimators affect our pure pseudo-power spectrum estimator.

In \secref{noise_bias}, we consider the spectrum of noise-only simulations as computed by \xpure. The qualitative results support filtering the noisiest modes from maps produced with the unbiased map estimator.

Finally, in \secref{comparison}, we compare the unbiased power spectra of the maps produced by the biased and unbiased map estimators (before and after map domain filtering).

\subsubsection{The noise bias}
\label{noise_bias}
The maps generated by the different map-makers have different noise properties, in terms of both the noise amplitude and its correlations.

\figref{fig:noise} shows the mean of the spectra produced by \xpure{} from the noise-only simulations, measuring the noise bias. We then take the noise simulations in pairs and evaluate the uncertainty on the noise bias as the standard deviation of the cross spectrum of the two noise maps within a pair.

In order to interpret these results consider first the unbiased case. The analysis of the eigenstructure of $\AtFA$ - which is the covariance matrix of the map estimator - has shown that the noisiest modes involve large scales. Since the power spectrum estimator can down-weight pixels but not modes, the large noise power carried by these modes can dominate the power
spectrum at large scales. This consideration suggests that the noisiest modes are the cause of the noise increase at large scales compared to the usual $\ell^2$ trend.

We therefore filter our unbiased map, progressively removing the noisiest
eigenvectors of the $\AtFA$ matrix: if the eigenvalue of a given eigenvector is
less than $\alpha$ times the maximum eigenvalue, the mode is filtered out of the
map. We consider several values of $\alpha$: \num{e-6}, \num{e-5}, \num{e-4},
\num{e-3}, \num{e-2}, \num{0.05}, \num{0.08}, \num{e-1}. We note that this
procedure is similar to the rejection of the noisiest pixels in a map, the only difference is that we remove modes, because our noise is correlated.

As the noisiest modes are removed from the unbiased map (i.e., as $\alpha$
increases) the power spectrum slowly converges to $\ell^2$ behavior. We note that this low $\ell$ noise increase is not caused only by the ``long modes'' related to the ground template, it is actually dominated by poorly constrained pixel-like modes: using the $10^{-4}$ threshold the long modes are removed but we still observe an important noise excess. This suggests that the cause is the not the ground template marginalization but the polynomial filtering (or a combination of the two).

On the contrary, in the biased map the way noise is correlated does not cause any noise increase at large scales, both the mean and the standard deviation of the noise power spectra follow the usual $\ell^2$ behavior. We stress that in this section both the spectra derived from the biased map and the ones derived from filtered maps are biased: we have to debias them before making quantitative statements about the uncertainty on their spectra, see next section.

\begin{figure*}[t] \centering 
\includegraphics[width=0.48\textwidth]{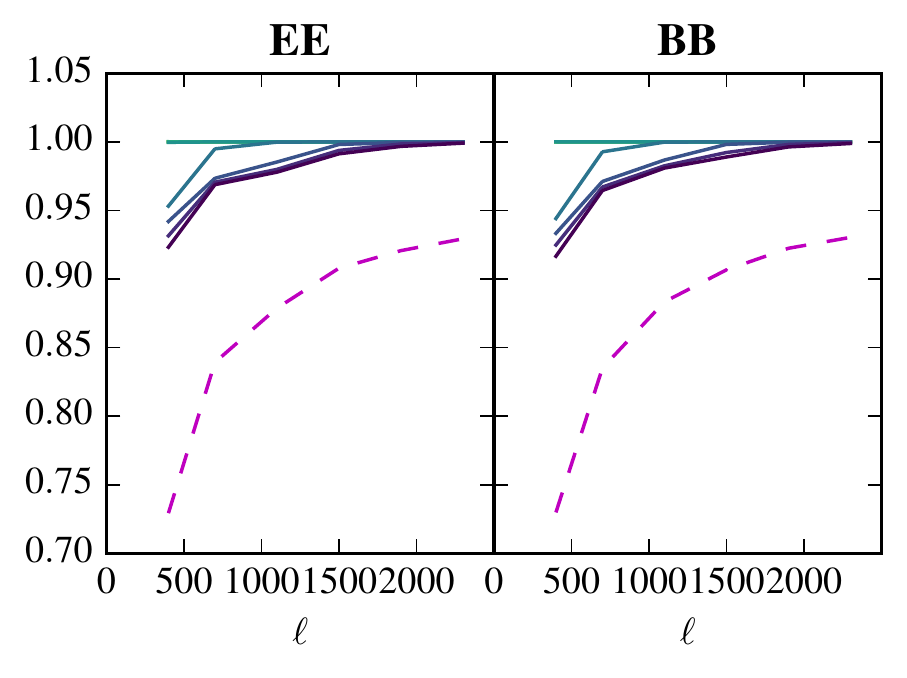} 
\includegraphics[width=0.51\textwidth]{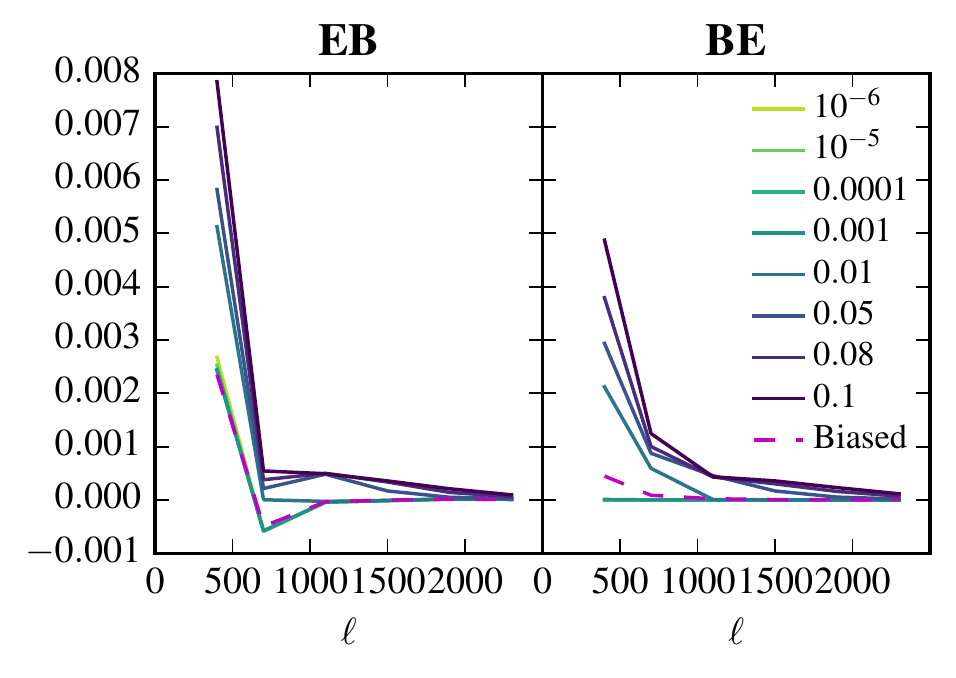} 
\caption{Transfer functions of the different map estimators. For the explanation of the legend, see \figref{fig:noise}} 
\label{fig:transfer_functions}
\end{figure*}

\subsubsection{Performance comparison}
\label{comparison}
As described in \secref{power_spectrum}, in order to debias the spectra, signal only simulations are used to evaluate the transfer functions $f^{XY}$ (\figref{fig:transfer_functions}) and noise only simulations for evaluating the noise bias \figref{fig:noise}.
We use these quantities to get an unbiased power spectrum estimator for each map
estimator and apply it to three sets of 100 simulations:
$E$ modes only for evaluating the uncertainty due to $E$ to $B$ leakage; 
noise only for evaluating the uncertainty due to the noise; 
$E$ and $B$ modes and noise for evaluating the total uncertainty on the $BB$ spectrum ($E$ to $B$ leakage, $BB$ cosmic variance and noise uncertainty).

In the first of these three cases the simulations do not contain noise and
therefore we evaluate the uncertainty as the standard deviation of the auto-spectra
of each simulation.
In \secref{noise_bias} we have shown that, for the
biased power spectrum estimates, the mean value of the noise simulations is comparable
with their dispersions, the auto-spectrum of noisy simulations  would then 
result in an asymmetric distribution. Consequently, in order to evaluate the
uncertainty of noisy simulations we prefer to group them
in pairs and evaluate the uncertainty as the standard deviation of the
cross-spectrum of the two simulations of each pair. We note that, when the
simulations contain also signal, the signal is the same in the two simulations.

\begin{figure*}[!ht] 
\centering 
\includegraphics[width=0.33\textwidth]{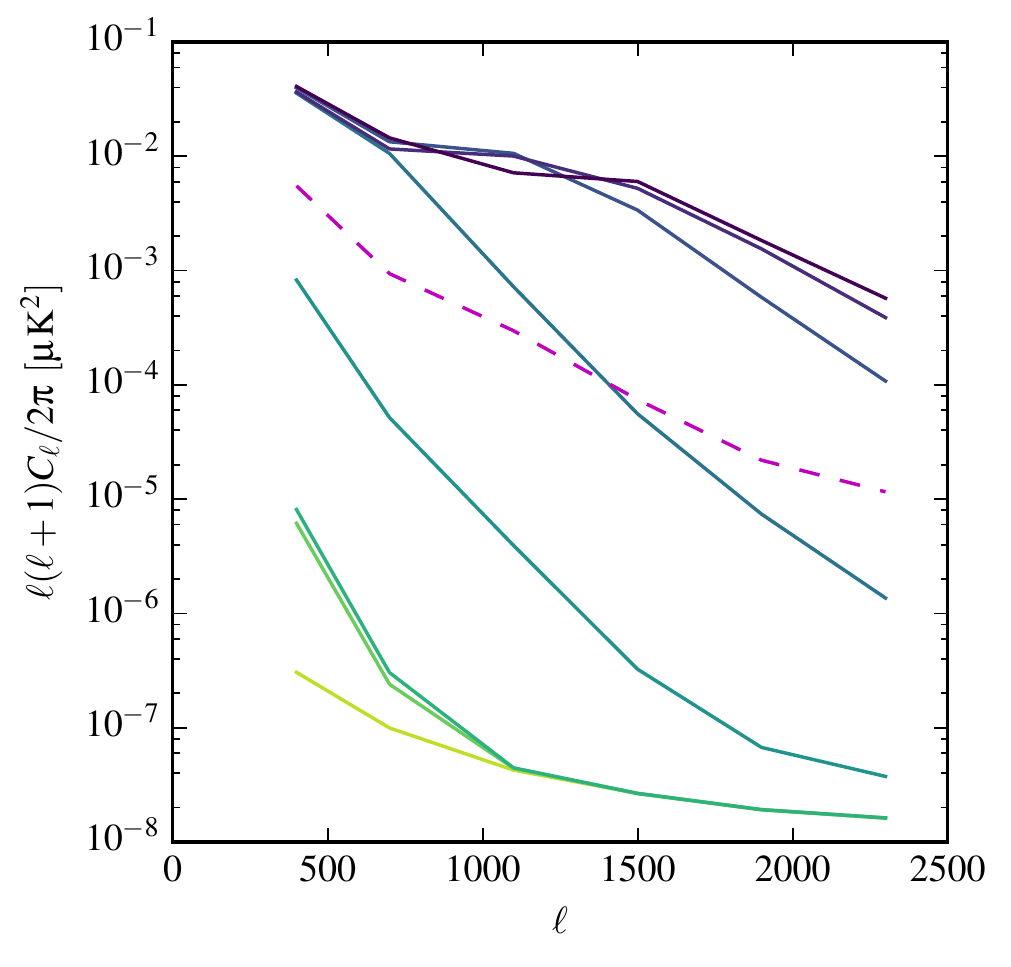} 
\includegraphics[width=0.33\textwidth]{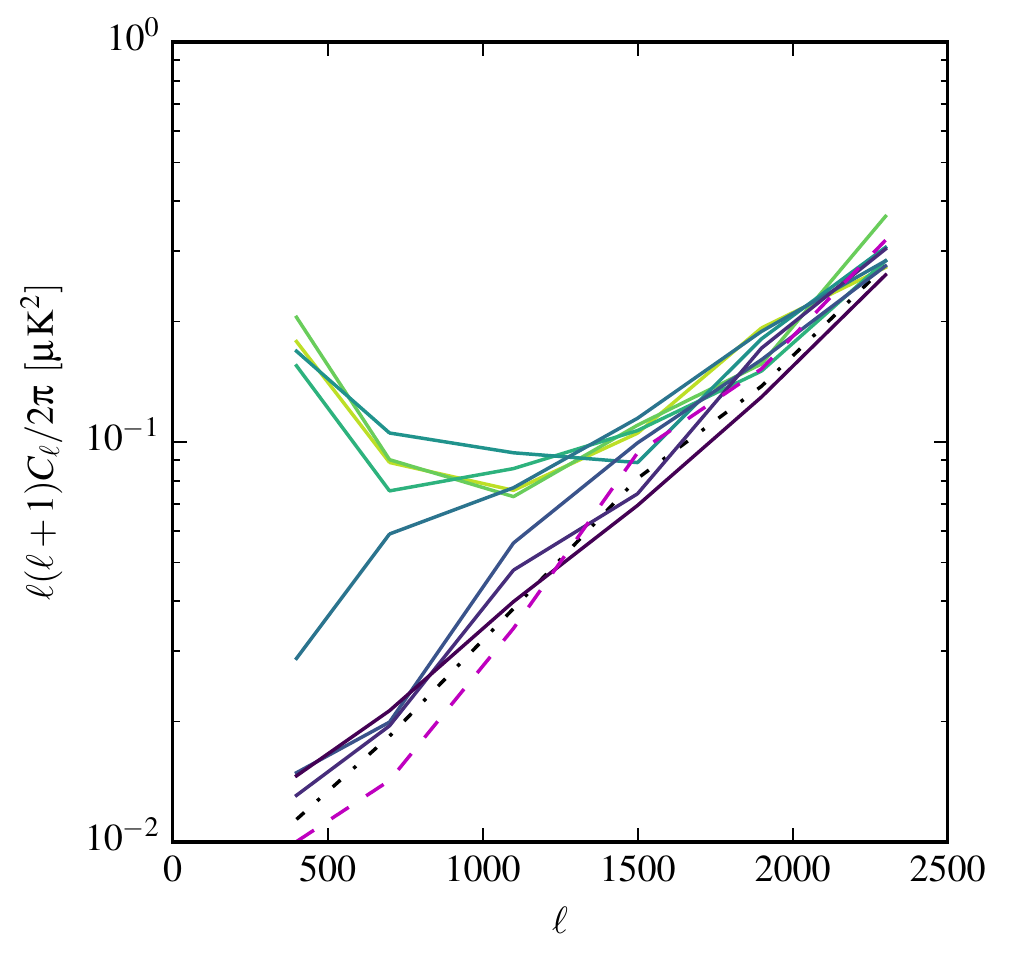} 
\includegraphics[width=0.33\textwidth]{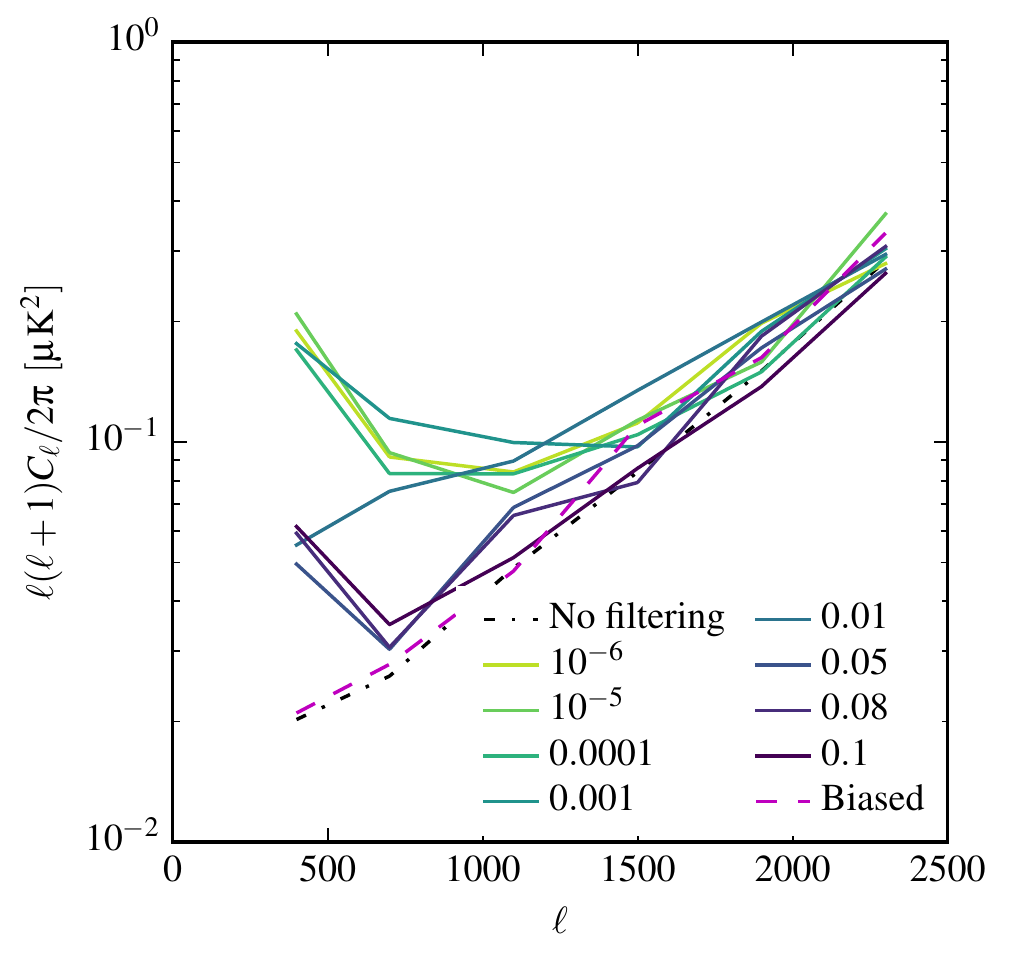} 
\caption{Uncertainty on the unbiased estimation of the $B$-mode power spectra based on 100 simulations containing $E$ only (left panel), noise only (central panel) and $E$, $B$ and noise (right panel), see \secref{comparison} for more details. The last case estimates the total uncertainty on the $BB$ power spectrum estimation. The spectra are estimated with help of cross-spectra of simulated pairs of maps. For the explanation of the legend, see \figref{fig:noise}. In the case of simulations containing noise
we also display the ``No filters'' case, in which the simulated white noise
TOD are not filtered: the simple map-making Eq. (\ref{gls_simple}) is
adopted.} 
\label{fig:uncertainty_BB}
\end{figure*}

\subsubsection*{$E$ to $B$ leakage}
As expected from the pure formalism implemented in \xpure, the unbiased map estimator has basically no uncertainty due to $E$ to $B$ leakage (the leakage is only due to pixelization effects). However the leakage quickly increases as the threshold on the eigenvalues increases and more modes are filtered from the map. The leakage becomes relevant at large scales for any threshold higher then $0.001$. The biased map estimator too has a significant amount of leakage at large scales but performs better
than any threshold greater than $0.001$.

These considerations can also be made by observing the transfer functions in \figref{fig:transfer_functions}. The $BE$ transfer function quantifies the average contribution of the $EE$ power in the sky to the $BB$ power in the reconstructed
map: the departure from zero is relevant for the biased map estimator only for the first bin and it is considerably more pronounced for $\alpha \geq 0.001$.

\subsubsection*{Noise uncertainty}
In \secref{noise_bias} we have shown that, for different map estimators and different thresholds $\alpha$, we get different dispersions of the raw spectra of noise simulations. However, the $BB$ transfer function in \figref{fig:transfer_functions} shows that they also have different loss of $BB$ power. Restoring this power boosts the spectrum of the noise too. The interplay between the two effects can be non-trivial.

However, \figref{fig:uncertainty_BB} shows that the latter effect has minor impact: the dispersion of the unbiased spectrum of noise only simulations is still the higher the lower the threshold $\alpha$ and the lowest for the biased map estimator.

Because of the correlated nature of the noise in the estimated maps, the power spectrum estimator can not properly down-weight the noisy modes. Their large fluctuations dominate the power at large scales, boosting the noise uncertainty. Filtering these modes out of the map alleviates this noise excess at large scales and the lowest noise uncertainty is reached by the most aggressive filtering ($\alpha = 0.1$). The biased map estimator performs extremely well in this respect: despite the fact that its noise is correlated too, the uncertainty due to noise is comparable with the $\ell^2$ trend expected by the uncorrelated noise case. We stress that the unbiased and biased map estimators preserve the same amount of information (we can convert one into the other anytime using an invertible linear operator). The disparity in their power spectrum noise uncertainty is purely due to the fact that we are using a suboptimal weighting for the power spectrum estimation.

\subsubsection*{Total uncertainty on the $BB$ spectrum}
\begin{figure}[t] 
\centering 
\includegraphics[width=0.5\textwidth]{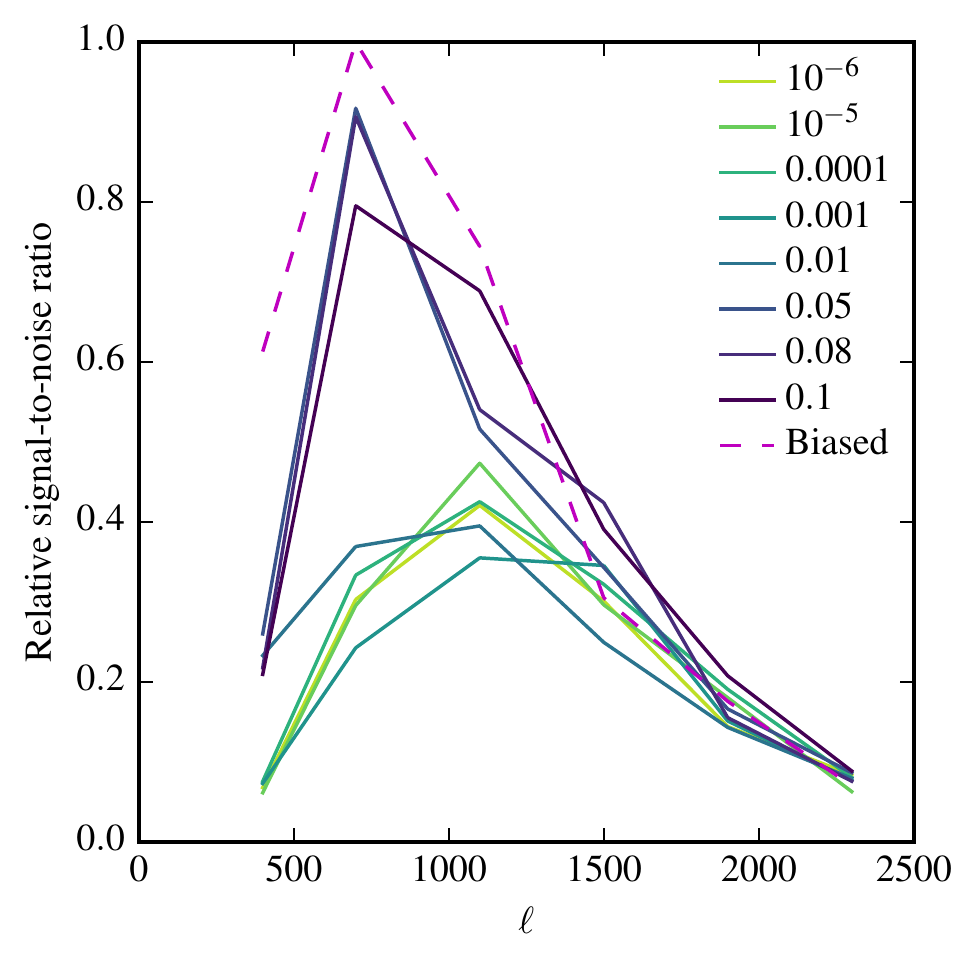} 
\caption{Ratio between the signal and the total uncertainty for $B$-mode power spectra derived with the different power spectrum estimation choices. The results are quoted relative to the highest ratio obtained for the biased map. The loss of precision of the spectra derived from the unbiased map is apparent and related to the strong correlated noise modes present in the map. The loss can be mostly recovered with help of a progressively more aggressive removal of the noisiest modes as expressed by  the increasing value of parameter $\alpha$ as defined in the text and given in the legend.} 
\label{fig:signal_to_uncertainty_BB}
\end{figure}

Finally we consider the total uncertainty on the $BB$ spectrum for the different map estimators. In \figref{fig:uncertainty_BB} we show its spectrum while in \figref{fig:signal_to_uncertainty_BB} we express it as ratio of the lensing $BB$ spectrum and total uncertainty.

As far as the unbiased map estimator is concerned, given the specific noise
level of these simulations, the noise plays a dominant role. Controlling its
large scale excess is more important than controlling the $E$ to $B$ leakage and
consequently we find that the higher the threshold on the eigenvalues the lower
the overall uncertainty is. We note that the most aggressive
threshold removes more than 50\% of the modes but retains more than 90\% of the
information (computed as the sum of the eigenvalues retained over the sum of all
the eigenvalues). The biased map estimator has good noise level over the entire spectrum and, even if the estimator produces $E$ to $B$ leakage at large scales, the resulting uncertainty is below the noise
level. As a result it performs substantially better than any other estimator studied here in the low $\ell$ part of the spectrum.

We also investigate how the situation would change if the noise level was lower by extrapolating the total uncertainty assuming the observation time was $x$ times longer. For each power spectrum bin, this total uncertainty is evaluated as
\begin{eqnarray} 
\sqrt{\sigma^2_S + \left(\frac{\sigma_N}{x}\right)^2 + 4 \frac{N}{x} \,S}, 
\end{eqnarray}
where $S$ and $N$ are the mean power of the signal only ($E$ and $B$) and noise only simulations respectively and the $\sigma$s are their standard deviations. For the biased map estimator the uncertainty due to $E$ to $B$ leakage is smaller than the $BB$ cosmic variance. Therefore, the unbiased map estimator has superior leakage control but it is not the limiting factor in the case we are considering. Consequently, the factor $x$ required for the unbiased map estimator to have better performance then the biased one is very large (about 10). We emphasize that this statement depends strongly on the specific case we are considering. The situation might be very different if the $E$ to $B$ leakage were to provide a more significant contribution to the overall uncertainty, as can happen when larger scales are probed. In such cases the unbiased map-making approach may be more readily favored.

\section{Conclusions}
Time-domain filtering is unavoidable in the analysis of CMB datasets. In this work we have addressed the issue of producing unbiased sky-signal estimates from filtered time-ordered data. We have presented a general map-making formalism that 
strictly accounts for the presence of time-domain filtering and which is capable of producing as faithful estimates of the sky signal as can be ever obtained in such circumstances. We have shown, however, that some modes of such estimates may be unconstrained or only poorly constrained. This happens whenever there are sky signal modes that are degenerate (or nearly degenerate) with the filtered time-domain modes, and which therefore cannot be fully disentangled from these in the map-making.

Subsequently, we have focused on ground-based experiments and two specific classes of filters -- polynomial and ground-synchronous -- which are frequently applied to remove long temporal modes and ground-signal pick-up. We have presented general considerations relevant to this case and then demonstrated them on specific, simulated data sets based on the first observational campaigns of the \pb{} experiment.

In this latter context we have shown that nearly unbiased maps of all three Stokes parameters can indeed be derived. Nonetheless, these contain a number of poorly constrained sky modes which have to be properly accounted for in the ensuing analysis of the maps in order to fully capitalize on the advantages of the proposed map-making. In particular, we have shown that the performance of simple pseudo-spectrum-based methods, commonly used to estimate the power spectra of the derived maps, may be significantly affected by the presence of such modes. This is because this approach is not naturally capable of weighting different sky modes differently and therefore these modes tend to either lead to excess noise at low multipoles if they are left in the map, or to enhance the $E$-to-$B$ leakage if they are excised.
We stress that this excess noise at low multipoles should not be
interpreted as a downside of the map estimation technique but rather a
demonstration that the map-domain correlation induced by popular time-domain
filtering operations can severely affect the performances of power-spectrum
estimators that use only a pixel-by-pixel weighting.

We have developed a practical approach to compensate for such deficiency but found out that, at best, we can only match the performance of the quicker and simpler biased map-making, based on a simple, noise-weighted binning of the filtered time-domain data. Consequently, more involved and resource-consuming techniques, such as those based on maximum likelihood principles, may need to be employed to exploit at the full potential of the unbiased maps. If, on the contrary, pseudo-power spectrum estimators are used, the unbiased maps can lead to lower $E$-to-$B$ leakage than the biased ones. Consequently, some improvements in the uncertainty of the recovered $B$-mode spectrum can be expected whenever $E$-to-$B$ leakage, and not the noise, is a dominant source of uncertainty and aggressive mode removal is unnecessary.

We note that the availability of nearly unbiased maps should also be important whenever multiple maps (e.g., corresponding to different frequency bands or coming from different experiments), need to be combined, as in pixel-based component separation approaches or map-level cross-analyses involving multiple data sets. In some applications, some of the ill-constrained modes may have to be removed from the unbiased map prior to further processing, as in the power spectrum estimation used in this work. Even though this effectively leads to biased maps, the removed sky modes will be known and the effects of their removal can be kept track of. This has to be further investigated in more detail and on a case-by-case basis, and is therefore left for future work.

Since the specific filters studied in this work are common for ground based experiments, the conclusions derived in this work should be of importance for many operating and planned ground-based experiments.


\begin{acknowledgements}
The POLARBEAR project is funded by the National Science Foundation under Grants No. AST- 0618398 and No. AST-1212230.
The  James  Ax  Observatory  operates  in  the  Parque
Astron\'omico Atacama in Northern Chile under the auspices of the Comisi\'on Nacional de Investigaci\'on Cient\'ifica
y Tecnol\'ogica de Chile (CONICYT). 
This research used resources of the National Energy Research Scientific Computing Center, a DOE Office of Science User Facility supported by the Office of Science of the U.S. Department of Energy under Contract No. DE-AC02-05CH11231. 
KEK authors acknowledge the support of MEXT KAKENHI Grant Number JP15H05891 and JSPS KAKENHI Grant Number JP26220709.  
In Japan, this work was supported by JSPS Core-to-Core Program, A. Advanced Research Networks and used computational resources of the HPCI system (Project ID:hp150132). 
In Italy, this work was supported by the RADIOFOREGROUNDS grant of the European Union's Horizon 2020 research and innovation programme (COMPET-05-2015, grant agreement number 687312)
as well as by the INDARK INFN Initiative. 
JP acknowledges support from the Science and Technology Facilities Council [grant number ST/L000652/1] and from the European Research Council under 
the European Union's Seventh Framework Programme (FP/2007-2013) / ERC Grant Agreement No. [616170].
CR acknowledges support from a Australian Research Council's Future Fellowship (FT150100074).

\end{acknowledgements}

\bibliographystyle{aa}
\bibliography{biblio}
\appendix
\section{Filtering the ground-synchronous signal}
\label{groundFilteringDetails}
In this appendix we study in detail the filtering of ground-synchronous signals, elaborating on and justifying the conclusions presented in~\secref{ground_pickup}. 

We start by considering the data recorded from a single detector, $d$, during a constant elevation scan, $s$. According to the data model in Eq. (\ref{eq:dataWithGround}), the data recoded in a given azimuthal bin, $\psi$, is
\begin{eqnarray}
\left.\b{d}\right|_{d, s, \psi} & = & \left.\b{A}\right|_{d, s, \psi} \left.\b{s}\right|_{d, s, \psi} + \left.\b{G}\right|_{d, s, \psi} \left.\b{g}\right|_{d, s, \psi} + \left.\b{n}\right|_{d, s, \psi} \nonumber\\
& = & \left.\b{A}\right|_{d, s, \psi} \left.\b{s}\right|_{d, s, \psi} + \left.\b{1}\right|_{d, s, \psi} g_\psi + \left.\b{n}\right|_{d, s, \psi},
\label{eq:dataScan}
\end{eqnarray}
where $\left.\b{1}\right|_{d, s, \psi}$ is a vector of ones of the appropriate length and $g_\psi$ denotes a ground template amplitude common to all selected samples.

Let us now focus on the shape of the sky patch corresponding to $\left.\b{s}\right|_{d, s, \psi}$. The geometry of the problem is depicted in Fig.~\ref{fig:observationScheme} and, for the time being, we neglect the role of the sky pixels. We consider a small azimuthal change of the pointing direction at a point on the sky at which the parallactic angle is $\eta$. The change in horizontal coordinates $ \Delta (\text{Az}, \text{El}) = (\delta, 0)$ corresponds to an interval in the equatorial ones given by,
\begin{eqnarray} 
\Delta (\text{RA}, \text{Dec}) = (- \delta \cos \eta, - \delta \sin \eta)  \label{RaDecOfAzInterval}
\end{eqnarray}
In particular, at the South Pole the horizontal coordinates correspond to the equatorial ones after flipping the
$y$ (El) axis, thus we have always $\eta = \pi$ and $\Delta (\text{RA},
\text{Dec}) = (\delta, 0)$. For a given elevation, the parallactic angle is
always the same for a given azimuth but depends on the azimuth's value, so
although it changes across a single ground template bin the changes are very
small. Consequently, each constant elevation scan crossing a bin will draw a
line interval on the sky given approximately by Eq.~(\ref{RaDecOfAzInterval}).
Because of the Earth's rotation, if we keep on crossing the bin multiple times
the intervals will cover a trapezoidal shape in sky coordinates, as shown in the
bottom panel of Fig.~\ref{fig:observationScheme}. The size of the trapezoid
depends on the bin width but also on the parallactic angle,
Eq.~(\ref{RaDecOfAzInterval}). However, the lines traced by the azimuthal bin
end points always follow the constant declination direction on the sky. We note that if $\ | \sin \eta | = 0\ $, which is always the case at the South Pole or whenever the instrument is pointed straight to the South or the North, the width of the trapezoid in declination is zero.

These patches, narrow in declination and elongated in azimuth, are degenerate (or ill-conditioned) modes. In the following we discuss this statement in detail for both temperature and polarization and investigate possible degeneracy-breaking effects.

\subsection{Total intensity measurements.}
Let us start with the total intensity measurements and consider a data subset that has the same ground contribution. It is described by~\eqref{eq:dataScan}, with the pointing matrix, $\b{A}$, merely composed of ones and zeros. For this single scan the sky modes, $\b{\tilde s}$, that are degenerate with the ground template signal have to fulfil the following relation, stemming from~\eqref{eq:degModes},
\begin{eqnarray} 
\left.\b{A}\right|_{d, s, \psi}\,\b{\tilde s} = \left.\b{G}\right|_{d, s, \psi}\,\b{\tilde g} = \left.\b{1}\right|_{d, s, \psi} g_\psi.
\end{eqnarray}
Thus for this subset of measurements, the ground template can only give rise to a constant offset in the time-domain for all samples of the scan. Given that $\b{A}$ here simply assigns the pixel amplitudes to the respective time samples without changing their values, there is only one sky mode that reproduces this behavior: a constant offset in the corresponding sky map,
\begin{eqnarray}
\b{\tilde s} \propto
\left[
\begin{smallmatrix}
1\\ \smallskip
\vdots\\
1
\end{smallmatrix}
\right] \equiv \b{1}_{\left.\b{s}\right|_{\left(d, s, \psi\right)}}
\end{eqnarray}
This demonstrates that, as intuitively expected, the absolute offset of the map produced from these measurements is unavoidably lost as it is degenerate with the ground signal, $g_\psi$. 

Let us consider another data subset taken by detector $d'$, during scan $s'$ and with the azimuth coordinate corresponding to bin $\psi'$. The data for this scan can be expressed by a relation analogous to~\eqref{eq:dataScan}. If either the scan, the detector or the azimuthal bin is different between these two data subsets, then the subsets will have independent, and {\em a priori} different, ground-pickup amplitudes $g_\psi$ and $g_{\psi'}$. If the sky observed during these scans overlaps, then the combined data set, $d_{\left\{d, s, \psi\right\} \cup\left\{d', s', \psi'\right\}}$, will have again only one degenerate mode pair, 
\begin{eqnarray}
\b{\tilde s} \propto
\left[
\begin{smallmatrix}
1\\ \smallskip
\vdots\\
1
\end{smallmatrix}
\right] = 
\b{1}_{\left.\b{s}\right|_{\left(d, s, \psi\right)} \cup \left.\b{s}\right|_{\left(d', s', \psi'\right)}}
\ \ \
\hbox{\rm and}
\ \ \ 
\b{\tilde g} \propto
\left[
\begin{smallmatrix}
1\\ \smallskip
\vdots\\
1
\end{smallmatrix}
\right] = 
\b{1}_{\left.\b{g}\right|_{\left(d, s, \psi\right)} \cup \left.\b{g}\right|_{\left(d', s', \psi'\right)}}.
\end{eqnarray}
This reflects the fact that the relative offset of the two sky maps recovered from each of the subsets can be constrained internally owing to the fact they overlap on the sky and have to recover the same sky signal in each common pixel. 

Otherwise, if no overlap exist, the data set made of two subsets will have two degenerate pairs of modes, which can be cast as either the absolute offsets of both of the sky patches or as an absolute offset of both of them and a relative offset between them. 

This latter situation can happen if the two subsets correspond to the same detector, $d$, and the same scan, $s$, but to two different though adjacent azimuthal bins. As described above, their corresponding sky areas will be indeed strictly speaking disjoint. However, as the sky maps are necessarily pixelized, there will be some pixels on the border between two patches which will straddle both of them, constraining their relative offset. The constraint will not be very strong though, and the corresponding degeneracy only weakly broken. An upshot of this is that a map produced from the single constant elevation scan data of a single detector will have as many ill-constrained modes as there are azimuthal bins used to represent the ground pick-up. These modes will become even more ill-constrained if the number of bins increases because the relative offset uncertainty for sky patches corresponding to two
extreme bins increases in proportion to $\sqrt n_\psi$, while the uncertainty on the offset of two adjacent bins remains roughly unchanged. This scaling reflects the fact that the relative offset between two non-overlapping sky patches with $n$ intermediaries is a result of a random walk of the adjacent patches' offsets, each subject to the same uncertainty~\citep[][]{StomporWhite2004}. Consequently ill-constrained modes can be suppressed if fewer azimuthal bins are used. Similarly, the uncertainty on the offset of two adjacent bins can be decreased if larger sky pixels are adopted, and more samples from both azimuth bins fall into them. However, the pixel size is typically set by the beam size, while the size of the azimuthal bins is driven by our preconceptions about the ground signal and the structure of the far side lobes. Consequently, whatever freedom is left should be used with care, as potential improvements
in statistical uncertainty can be translated into increased ground pick-up residual.

There are two reasons why these ill-constrained modes may be further suppressed in the final maps, combining the data of all the detector and all of the scans.First, for a single scan the additional constraints on the relative offsets of these patches typically also come from the data collected by different detectors. This is because the azimuthal bins are often defined differently for different detectors so the sky patch corresponding to an azimuthal bin of one detector will often overlap with two sky patches corresponding to two different bins of the other detector, thus providing an extra leverage on their relative offset. Second, the sky patches corresponding to fixed azimuthal bins of different scans can have different width because the parallactic angle is different in the two scans (see previous section). This introduces additional overlaps between the patches of different scans, helping to constrain their relative offsets.

In general the global offset of the final map is expected to be the only truly degenerate mode in the total intensity maps derived from data that are contaminated by ground pickup that requires explicit filtering. Notwithstanding this, the constraints that can be set between the relative offsets of different patches with the same ground pick up are usually inferior to those between different parts of the same patch, and some ill-constrained large scale modes should be expected, predominantly in the declination direction. 

We note that these conclusions apply qualitatively to any observational site on Earth, including the poles, with the difference that as the single ground bin patches become very narrow in declination the relative offset degeneracies in this direction are broken only by the pixel effects. By contrast, the bin size only plays a role in breaking the degeneracies in the azimuthal direction.

It is important to appreciate the role of the assumptions in breaking these (near) degeneracies. The choices made about the sky
pixel size, the pixelization itself, the binning, and the size of the bins, all impact the degeneracies and can be used, or abused, to break them. In addition, the offset degeneracies can be broken if a less flexible model for the ground signal is used, for instance if we impose a prior constraint on the relative change of the ground signal from one bin to the next. The key parameter in such cases would be the assumed coherence length for the ground signal.

\subsection{Polarization-sensitive measurements.}
The situation for polarization-sensitive observations is potentially more complex due to more complex form of the pointing matrix. However, it is qualitatively similar to the total intensity case. For concreteness, we discuss the case with three Stokes parameters contributing to the measurements, and thus with the pointing matrix as defined in~\eqref{eq:data3stokes}.  
Since the polarization
orientation may change during the operations because of some polarization modulator, we introduce a different ground pick-up amplitude not only for each azimuthal orientation of the instrument but also for each different position of the polarizer, as defined in the instrument coordinates. For simplicity, we will however keep on using a single azimuthal bin number, $\psi$, to distinguish between the ground signal amplitudes.

We again focus on a single constant elevation scan and a single detector. Our data model is then again given by~\eqref{eq:dataScan}, where the same ground pick up is added to each measurement. The major qualitative difference from the total intensity case is that in the polarization-sensitive case the pointing matrix elements may be different from sample to sample, even for samples falling into the same sky pixel on a single crossing, as could be the case if fast rotating half-wave plate were employed to modulate the signal. However, for the data subset selected above the angle of the polarizer is fixed in the instrument frame, so the change of the pointing matrix elements, defined by the polarizer orientation but with respect to the sky coordinates, can be only related to the parallactic angle change with the azimuth of the observation. Typically the angle change within a range of azimuths corresponding to a single sky pixel can be safely neglected and we may assign a single polarizer angle as measured with respect to the sky coordinates for each pixel observed with the data subset. These angles may be somewhat different for two different pixels if these are observed at different azimuths, but as the latter have to fall within a single ground template bin, the bins would need to be rather broad to make such an effect important. Nonetheless, henceforth we assume that for the data subset as defined earlier and characterized by the same ground-pick up amplitude, the polarizer's angle in the sky coordinates, and thus the mixing matrix elements, may at most depend on the observed sky pixels and will have a unique value for all observations falling within the same pixel.  We note that such small angle variations do not appear if the observations are conducted from the Earth's poles.

As in the total intensity case, the degenerate sky modes have to be able to mimic an offset in time-domain data. This can be the case for three linearly independent sky defined as,
\begin{eqnarray}
\b{\tilde s_I} \propto 
\left[
\begin{smallmatrix}\smallskip
1\\ \smallskip
0\\
0\\ \smallskip
\vdots\\ \smallskip
1\\ \smallskip
0\\
0
\end{smallmatrix}
\right], 
\ \ \ \ \ \ 
\b{\tilde s_Q} \propto 
\left[
\begin{smallmatrix} \smallskip
0\\ \smallskip
c_0^{-1}\\
0\\ \smallskip
\vdots\\
0\\ \smallskip
c_{n_p-1}^{-1}\\
0
\end{smallmatrix}
\right], \ \ \ \ \ \ 
\b{\tilde s_U} \propto 
\left[
\begin{smallmatrix}\smallskip
0\\ \smallskip
0\\
s_0^{-1}\\\smallskip
\vdots \\
0\\ \smallskip
0\\ \smallskip
s_{n_p-1}
\end{smallmatrix}
\right],
\label{eqn:polDegCES}
\end{eqnarray}
where, 
\begin{eqnarray}
s_p & \equiv & \sin 2\varphi_p\\
c_p & \equiv & \cos 2\varphi_p,
\end{eqnarray}
and $\varphi_p$ stands for a polarizer angle in the sky coordinate in pixel $p\, (=0,\dots, n_p-1)$. Each of these sky modes is a vector
of $n_p$ triples where the elements of each triple correspond to the $I$, $Q$, and $U$ Stokes parameters. We note that if the sky rotation is negligible across the sky patch covered by the scan these three modes correspond to map offsets of the maps of the respective Stokes parameters.

Within our data subset, each pixel is observed with only a single orientation of the polarizer and we thus cannot estimate all three Stokes parameters separately, but merely their linear combination, $I + Q \cos 2\varphi_p + U \sin 2\varphi_p$, even if no ground pick up is considered. The corresponding two-dimensional degeneracy space is a sub-space of the three-dimensional space spanned by the vectors defined in Eq.~(\ref{eqn:polDegCES}). Consequently, adding the ground pick-up merely adds one degenerate vector to the map-making problem, corresponding to the total offset of the linear combination of the Stokes parameters, $I + Q \cos 2\varphi_p + U \sin 2\varphi_p$\footnote{We also note that even if one of the cosines, $c_p$, or sines, $s_p$, happens to be zero, and therefore only two of the three modes in~\eqref{eqn:polDegCES} are indeed degenerate, the latter statement remains true and the loss of information is the same in all these cases.}.

To recover all the Stokes parameters from data modeled as in~\eqref{eq:data3stokes}, we need at least three visits to each pixel with a different orientation of the polarizer. These can be provided by other detectors in the focal plane during the same or different
constant elevation scans, or come from the same detector if its polarizer direction is modulated either on short or long timescales. In all these cases the new data will have not only a different polarization angle but also potentially a different ground-pickup. Each of these extra data sets can likewise have up to three degenerate sky modes, which for data subset $i\, (=0,1,2)$, we denote as 
$\b{\tilde s_I}^{\left( i\right)}$, $\b{\tilde s_Q}^{\left( i\right)}$ and $\b{\tilde s_U}^{\left( i\right)}$, respectively. For these data sets considered together, however only the intensity offset, $\b{\tilde s_I} = \b{\tilde s_I}^{\left(0\right)} = \b{\tilde s_I}^{\left(1\right)} = \b{\tilde s_I}^{\left(2\right)} $, always leads to degeneracy, while $\b{\tilde s_Q}^{\left(i\right)}$ and $\b{\tilde s_U}^{\left(i\right)}$ will only do so if the polarizer angles for each data subset are effectively the same for all observed pixels, and therefore  
\begin{eqnarray}
\b{\tilde s_Q}^{\left(i\right)} \propto  \b{\tilde s_Q} \equiv
\left[
\begin{smallmatrix} \smallskip
0\\ \smallskip
1\\
0\\ \smallskip
\vdots\\
0\\ \smallskip
1\\
0
\end{smallmatrix}
\right], \ \ \ \ \ \ \ \ \ \ 
\b{\tilde s_U}^{\left( i\right)} \propto 
\b{\tilde s_U} \equiv
\left[
\begin{smallmatrix}\smallskip
0\\ \smallskip
0\\
1\\\smallskip
\vdots \\
0\\ \smallskip
0\\ \smallskip
1
\end{smallmatrix}
\right], \hskip 20pt \hbox{\rm for $i = 0, 1, 2$}.
\label{eqn:polDegCESspecial}
\end{eqnarray} 
In this case each of the recovered maps of the Stokes parameters will have an arbitrary offset corresponding to three degenerate vectors, $\b{\tilde s_I}$,  $\b{\tilde s_Q}$, and $\b{\tilde s_U}$, as defined.

If the angles do change somewhat from pixel to pixel within a single data subset
(i.e., when the change of the parallactic angle within the azimuthal bin is not negligible) only the total intensity map will have an arbitrary offset. This is because in this case $\b{A}^{\left(i\right)}\, \b{\tilde s_Q}^{\left(j\right)}$ and $\b{A}^{\left(i\right)} \,\b{\tilde s_U}^{\left(j\right)}$ are time-domain vectors with elements which depend on time in a non-trivial way. Here, $\b{A}^{\left(i\right)}$ is a pointing matrix specific to subset $i$, while the combined pointing matrix for the three subset is given by,
\begin{eqnarray}
\b{A} \equiv 
\left[
\begin{array}{c}
{\displaystyle \b{A}^{\left(0\right)}}\\
{\displaystyle \b{A}^{\left(1\right)}}\\
{\displaystyle \b{A}^{\left(2\right)}}
\end{array}
\right].
\end{eqnarray}
We therefore also see that the time-domain vectors, $\b{A}\, \b{\tilde s_Q}^{\left(j\right)}$ and $\b{A} \,\b{\tilde s_U}^{\left(j\right)}$, are non-trivial and therefore cannot typically be mimicked by three ground template offsets and the degeneracy condition in~\eqref{eq:degModes} can not be fulfilled. 

However, as the angle change due to the sky rotation is typically small, $\b{\tilde s_Q}$ and $\b{\tilde s_U}$ may be potentially ill-constrained, even if not strictly singular, and the offsets of the $Q$ and $U$ maps may be very uncertain.

The offsets between sky patches corresponding to adjacent ground template bins during the same constant elevation scan can be further constrained as in the case of the total intensity only measurements. The potential degeneracies can then be suppressed with the help of data from the other detectors and/or different scans, although again a natural expectation is that there will be long sky modes in the declination direction which may be ill-constrained.

If the observation is taken from the Earth's poles, the maps recovered from the three subsets of the data taken at the same elevation will have all three degenerate offsets, which will propagate to the final maps combining all the data. In addition, the relative offsets between the sky patches taken at different elevation will only be set by the presence of pixels common to both patches and therefore will lead to long modes in declination which will be ill-constrained.

\end{document}